\documentclass{article}
\usepackage{arxiv}
\usepackage[utf8]{inputenc} 
\usepackage[T1]{fontenc}    
\usepackage{hyperref}       
\usepackage{url}            
\usepackage{booktabs}       
\usepackage{amsfonts}       
\usepackage{nicefrac}       
\usepackage{microtype}      
\usepackage{lipsum}
\usepackage{tabto}
\usepackage{graphicx}
\usepackage{caption}
\usepackage{float}
\usepackage{subfigure}
\usepackage{xcolor}
\usepackage{amssymb}
\usepackage{amsmath}
\usepackage{latexsym}
\usepackage{algorithm}
\usepackage{algpseudocode}
\usepackage[mathscr]{euscript}
\usepackage{url}
\usepackage{bm}
\usepackage{url}
\usepackage{textcomp}
\usepackage{placeins}
\usepackage[superscript,biblabel]{cite}
\newcommand\norm[1]{\left\lVert#1\right\rVert}

\usepackage[title]{appendix}

\title{Reduced-order modeling of large-scale turbulence \\ using Koopman $\beta$-variational autoencoders}

\author{
 Rakesh Halder\thanks{Corresponding author: rakesh.halder@bsc.es} \\
  Barcelona Supercomputing Center\\
  Barcelona, Spain 08034 \\
   \And
 Benet Eiximeno \\
  Barcelona Supercomputing Center\\
  Barcelona, Spain 08034 \\
  \And
 Oriol Lehmkuhl \\
  Barcelona Supercomputing Center\\
  Barcelona, Spain 08034 \\
}

\begin{document}
\maketitle

\begin{abstract}
Reduced-order models (ROMs) are very popular for surrogate modeling of full-order computational fluid dynamics (CFD) simulations, allowing for real-time approximation of complex flow phenomena. However, their application to CFD models including large eddy simulation (LES) and direct numerical simulaton (DNS) is limited due to the highly chaotic and multi-scale nature of resolved turbulent flow. Due to the large amounts of noise present in small-scale turbulent structures, error accumulation becomes a major issue, making long-term prediction of unsteady flow infeasible. While linear subspace methods like dynamic mode decomposition (DMD) can be used to pre-process turbulent flow data to remove small-scale structures, this often requires a very large number of modes and a non-trivial mode selection process. In this work, a ROM framework using Koopman $\beta$-variational autoencoders ($\beta$-VAEs) is introduced for reduced-order modeling of large-scale turbulence. The Koopman operator captures the variation of a non-linear dynamical system through a linear representation of state observables. By constraining the latent space of a $\beta$-VAE to grow linearly using a Koopman-inspired loss function, small-scale turbulent structures are filtered out in reconstructions of input data and latent variables are denoised in an unsupervised manner so that they can be sufficiently modeled over time. Combined with a long short-term memory (LSTM) ensemble for time series prediction of latent variables, the model is tested on LES flow past a Windsor body at multiple yaw angles, showing that the Koopman $\beta$-VAE can effectively denoise latent variables and remove small-scale structures from reconstructions while acting globally over multiple cases.
\end{abstract}

\section{Introduction}

Reduced-order models (ROMs) have become an indispensable tool in engineering design processes, allowing for accurate and rapid real-time approximations of full-order models (FOMs) of high-fidelity complex physical systems~\cite{lucia2004reduced}. ROMs create a low-dimensional surrogate model of the FOM using a limited number of high-fidelity training data. ROMs are trained during a computationally expensive \emph{offline stage}, where both the FOM is run to collect training data for use with the surrogate model and the ROM is trained. An inexpensive \emph{online stage} involves rapidly evaluating the ROM at non-evaluated designs and/or timesteps. Approaches used to develop the surrogate model involve linear subspace methods like the proper orthogonal decomposition (POD)~\cite{BerkoozStructure}, dynamic mode decomposition (DMD)~\cite{tu2013dynamic}, as well as deep neural networks, particularly convolutional autoencoders (CAEs)~\cite{Lee2020ModelRO}. CAEs allow for a non-linear relationship between the full-order states and a low dimensional \emph{latent space}, which can drastically lower the number of variables required for reasonable accuracy when compared to linear methods, particularly for highly non-linear problems.
~\cite{halder2022non,Lee2020ModelRO,maulik2021reduced}.

When applied to fluid dynamics, ROMs have been shown to offer very good accuracy when applied to laminar flow~\cite{HESTHAVEN201855,halder2024reduced} and problems utilizing Reynolds-averaged Navier–Stokes (RANS) turbulence models~\cite{carlberg2011efficient,he2021efficient}. However, ROMs face difficulty in producing accurate results for scale-resolving turbulent simulations such as large eddy simulation (LES)~\cite{arnold2022large, quaini2024bridging}. Turbulent flow involves an energy cascade~\cite{leonard1975energy} where energy is transferred from large, coherent structures like vortices to increasingly smaller eddies that are chaotic in nature. At larger scales, the flow is organized and driven by external forces. As energy cascades down to smaller scales, flow interactions become increasingly disordered until viscous forces dominate and energy is dissipated. The highly non-linear, unsteady, multi-scale, and chaotic nature of scale-resolved turbulent flows make them difficult to model accurately in a low-dimensional framework when using both physics-based and data-driven models. When applied to LES, ROMs suffer from instability, a loss of accuracy over time, and limited generalizability. Predictions become less accurate over time due to error accumulation, making ROMs unsuitable for long-term prediction of unsteady flow.

When using surrogate models in computational fluid dynamics (CFD), it is often the case that obtaining predictions of the large-scale flow behavior that is organized and recurring is sufficient to guide design optimization processes. The large-scale turbulent structures are coherent, contain most of the flow's kinetic energy, and define the macroscopic properties of the flow. Conversely, the highly chaotic nature of small-scale turbulent structures can make them intractably difficult to model over time. This is particularly problematic when using time series models like long short-term memory (LSTM) or transformer neural networks~\cite{vinuesa2022enhancing} as the latent variables contain large amounts of noise, making it difficult learn patterns even when using large models. While turbulent flow data can be pre-processed using DMD by selecting modes corresponding to large-scale structures, this process is often cumbersome due to the large number of modes required for reasonable accuracy and the difficulty in formulating a selection criteria for modes. Furthermore, DMD cannot be applied to multiple turbulent flow datasets simultaneously, hindering its use in design optimization.

The Koopman operator~\cite{koopman1931hamiltonian, otto2021koopman} is a linear, infinite-dimensional operator that captures the evolution of a non-linear dynamical system using a linear representation of state observables. The Koopman operator is also instrumental in DMD~\cite{williams2015data}, allowing for the extraction of spatio-temporal modes from data which are found as approximate eigenfunctions and eigenvalues of the Koopman operator. By constraining state observables to evolve linearly, the Koopman operator can allow for filtering out small-scale turbulent fluctuations while retaining large-scale coherent flow structures. In the context of autoencoders, the Koopman operator has previously been implemented to forecast sequential data such as time series~\cite{azencot2020forecasting,naiman2023generative}, utilizing the ability to model complex non-linear dynamics in a linearly evolving latent space. Koopman autoencoders have also been used for dynamical systems governed by physics. Lusch et al.~\cite{lusch2018deep} utilize the Koopman autoencoder in deep neural networks to extract eigenfunctions of dynamical systems including fluid flow by identifying non-linear coordinates on which the dynamics are globally linear through the use of Koopman-based loss functions. Otto et al.~\cite{otto2019linearly} impose a linear dynamics constraint on the latent space through the use of a simple loss function in order to extract Koopman eigenfunctions. A work by Nayak et al.~\cite{nayak2023koopman} uses Koopman autoencoders with physics-informed constraints for reduced-order modeling of kinetic plasmas. Each of these works directly model the subsequent state using the autoencoder rather than reconstructing the input and do not focus on highly chaotic problems. To the best of our knowledge, there are no works that have used Koopman autoencoders for the purpose of filtering out small-scale structures from turbulent flow.

In this work, we introduce the use of Koopman $\beta$-variational autoencoders ($\beta$-VAEs) for unsupervised filtering of small-scale turbulent structures from input fields and denoising latent variables. This allows for the extraction of only large-scale coherent structures which correspond to a smoothly varying latent space that is able to be used for reduced-order modeling. $\beta$-VAEs are a probabilistic formulation of autoencoders that encourage latent variables to follow a target distribution, most often Gaussian, a useful property for attaining latent variables that are similar in both magnitude and structure. A Koopman-inspired loss function is used to constrain the latent variables to grow linearly in time, which also denoises them and as a result filters out small-scale chaotic fluctuations in reconstructions of input fields. While Koopman operator theory involves infinite-dimensional spaces spanned by infinitely many possible observables, the loss function in this work is a linear dynamics constraint using only the latent variables as observables in a finite-dimensional setting. A training procedure is also detailed that avoids overfitting to temporal patterns present within the training data. For time series forecasting of latent variables, LSTM ensemble models are used as they lead to improved robustness and stability over long time horizons~\cite{halder2024reduced}. The ROM is tested on LES flow past a Windsor body~\cite{page2022towards}, a standardized automotive benchmark case, at 5 different yaw angles, highlighting the ability of the Koopman $\beta$-VAEs to act globally over multiple cases to smooth latent variables.

\section{Methods}
This section presents the methods used for the ROM implementing Koopman $\beta$-VAEs. Some background information of $\beta$-VAEs is given and the implementation of the Koopman loss function is described in addition to the training procedure for the neural network. The time series prediction component of the ROM, an LSTM ensemble utilizing bootstrap aggregation introduced by Halder et al.~\cite{halder2024reduced} is described briefly.

\subsection{Koopman $\beta$-Variational Autoencoders}
\subsubsection{$\beta$-Variational Autoencoders}

Standard convolutional autoencoders reconstruct the high-dimensional input state $\bm{x} \in \mathbb{R}^{n_x \times n_y}$ using a combination of two individual feedforward neural networks, the encoder $f_{\text{enc}}$ and decoder $f_{\text{dec}}$,

\begin{equation}
f: \bm{\hat{x}} = f_{\text{enc}} \circ f_{\text{dec}}(\bm{x}),
\end{equation}

where $f_{\text{enc}}$ maps from the input to the latent vector $\bm{z} \in \mathbb{R}^{k}$ and $f_{\text{dec}}$ takes the latent vector as an input and provides an approximate reconstruction $\bm{\hat{x}}$. Standard autoencoders impose no constraints or regularization on the latent space, allowing the encoder to map inputs to arbitrary regions without encouraging properties such as smoothness or continuity. The loss function typically used for standard CAEs in ROM applications is the mean square error (MSE),

\begin{equation}
\mathcal{L}_{\text{CAE}} =  \mathcal{L}_{\text{MSE}}.
\end{equation}

Variational autoencoders, first introduced by Kingma and Welling~\cite{kingma2022autoencodingvariationalbayes}, incorporate a probabilistic framework. Rather than the encoder mapping inputs to a fixed latent vector, a probabilistic encoder $q_{\phi}(\bm{z|x})$ maps the inputs to the parameters of a latent distribution, most often a multivariate Gaussian,

\begin{equation}
  q_{\phi}(\bm{z|x}) = \mathcal{N}(\bm{z}; \bm{\mu}_{\phi}(\bm{x}), \bm{\sigma}^2_{\phi}(\bm{x})),
\end{equation}

where $\bm{\mu}_{\phi}$ and $\bm{\sigma}^2_{\phi}$ are the mean and variance respectively. Instead of mapping to a single latent vector, the encoder in a VAE consists of two neural networks that output the mean and variance of the approximate posterior distribution over the latent variables. The decoder $p_{\theta}(\bm{x|z})$ in VAEs remains deterministic, and latent vectors are sampled from the encoder outputs $\bm{z} \sim \mathcal{N}(\bm{\mu}_{\phi}(\bm{x}), \bm{\sigma}^2_{\phi}(\bm{x}))$. Since sampling from the encoder's distribution is non-differentiable and prevents backpropagation, a \emph{reparameterization trick} is employed, expressing the sampled latent vector as

\begin{equation}
\bm{z} = \bm{\mu}_{\phi}(\bm{x}) + \bm{\sigma}_{\phi}(\bm{x})\epsilon,
\end{equation}

where $\epsilon \sim \mathcal{N}(0, I)$ is a source of randomness that allows for indirect sampling of $\bm{z}$. The original work by Kingma and Welling can be consulted for further detail. By incorporating a probabilistic framework, VAEs allow for regularizing the latent variables to closely follow a target distribution, typically an isotropic Gaussian. Some desirable properties of isotropic Gaussians are that the variables are independent of each other and within the same numerical range. To encourage this, the loss function is augmented by the Kullback–Leibler (KL) divergence ($D_{\text{KL}})$, a measure of distance between two distributions, between the encoder and prior function $p(\bm{z}) = \mathcal{N}(0, I)$,

\begin{equation}
  \mathcal{L}_{\text{VAE}} =  \mathcal{L}_{\text{MSE}} + D_{\text{KL}}(q_{\phi}(\bm{z|x}), p(\bm{z})).
\end{equation}

To control the impact of the KL divergence term and subsequently the level of regularization, $\beta$-VAEs were introduced by Higgins et al.~\cite{higgins2017beta}, which simply add a regularization term $\beta$ to the KL divergence term of the loss function,

\begin{equation}
  \mathcal{L}_{\text{$\beta$-VAE}} =  \mathcal{L}_{\text{MSE}} + \beta D_{\text{KL}}(q_{\phi}(\bm{z|x}), p(\bm{z})).
\end{equation}

In practice, $\beta$ is often slowly and linearly increased from 0 to a maximum value $\beta_{\text{max}}$ over a number of epochs after which it remains constant to allow for better stability during training~\cite{bowman2015generating}. $\beta$-VAEs have been used for a wide number of tasks, including learning interpretable generative latent factors for images~\cite{higgins2017beta}, anomaly detection~\cite{ulger2021anomaly}, and non-linear modal analysis of fluid flow~\cite{eivazi2022towards}.

\subsubsection{Koopman Loss Function}

To retain only the large-scale coherent structures in the autoencoder reconstructions given raw turbulent flow data, the goal is to regularize the latent variables in a manner that filters out the small-scale choatic fluctuations. If small-scale flow structures are retained, this leads to highly noisy latent variables which are difficult to model. Encouraging the latent variables to evolve linearly according to a trainable Koopman operator matrix $\bm{A} \in \mathbb{R}^{k \times k}$ can lead to latent variables that exhibit a smooth evolution over time, resulting in reconstructions that do not include noise from small-scale structures and are able to be sufficiently modeled. The Koopman operator as used works with the finite-dimensional latent variables and constrains them to evolve linearly in time,
\begin{equation}
  \bm{z}_{t+1} = \bm{A}\bm{z}_{t},
\end{equation}

where $\bm{z}_{t}$ are the latent variables for a given temporal snapshot. The $\beta$-VAE loss function is augmented by a Koopman loss term $\mathcal{L}_{\text{Koop}}$,

\begin{equation}
  \mathcal{L}_{\text{Koop}} = \dfrac{\norm{\bm{z}_{t+1} - \bm{Az}_{t}}^{2}} {\norm{\bm{z}_{t+1}}^2}.
\end{equation}

The loss term is the relative $L^2$ error between the sampled latent vector at $t+1$ and its Koopman approximation. A relative error is used so that the magnitudes of the latent variables are not arbitrarily driven to zero in order to minimize the loss. Augmenting the $\beta$-VAE loss function with the Koopman loss results in $\mathcal{L}_{\text{K$\beta$-VAE}}$ given as

\begin{equation}
  \mathcal{L}_{\text{K$\beta$-VAE}} = \mathcal{L}_{\text{MSE}} + \beta D_{\text{KL}}(q_{\phi}(\bm{z|x}), p(\bm{z})) + \alpha \mathcal{L}_{\text{Koop}},
\end{equation}

where $\alpha$ is a regularization term for the Koopman loss. The Koopman operator matrix $\bm{A}$ is initialized as the identity matrix and all of its elements are trainable. While a vanilla CAE can be used for our desired purpose, in practice they lead to poor stability of the Koopman operator during training. Since vanilla CAEs impose no regularization on the latent space, it is often the case that the magnitudes of the latent variables vary drastically and are sensitive to the initialization of the neural network. As a result, the elements of $\bm{A}$ tend to overfit to these trends in order to minimize the loss and fail to denoise and smooth the latent variables. By using a $\beta$-VAE and constraining the latent variables to be similar in magnitude, the training process is more stable and the desired result is more easily attained. Independence of latent variables, which is often why $\beta$-VAEs are preferred, cannot be easily attained when implementing the Koopman loss. This would require that $\bm{A}$ remain diagonal, which is too rigid of a constraint when attempting to encourage linear growth over time. 

\begin{figure}[!htpb]
  \centering
  \includegraphics[width=1.0\textwidth]{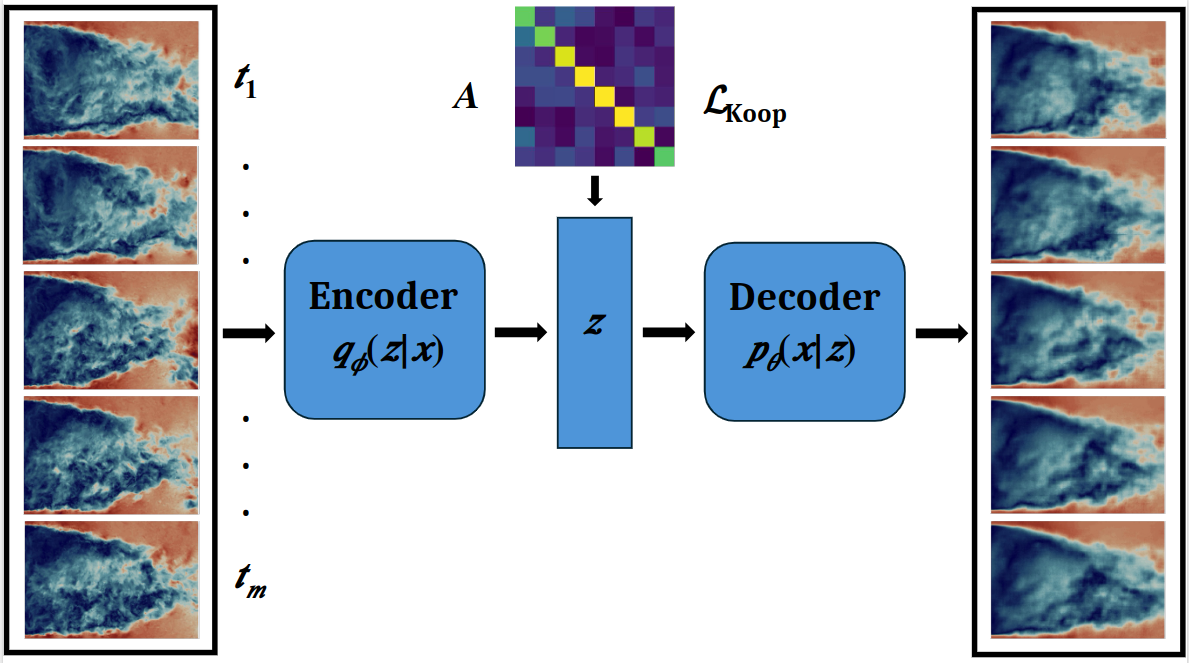}
  \caption{Schematic of the Koopman $\beta$-VAE, where small-scale structures are filtered out from turbulent input data through the use of the Koopman loss function.}
  \label{fig:vae}
  \end{figure}

For proper implementation of the Koopman loss, training mini-batches containing temporally ordered data must be used. However, this can lead to lower reconstruction accuracy as the network overfits to temporal patterns in the data. Temporally consecutive data snapshots also exhibit some degree of correlation, which violates the independent and identically distributed (i.i.d.) assumption that VAEs make about the training data. To mitigate this, the model is pre-trained using only the standard $\beta$-VAE loss ($\alpha = 0$) with randomly shuffled mini-batches for a prescribed number of $e_{\text{pre}}$ epochs so that the network can better fit the overall trends of the training data and learn a more structured latent space. After pre-training, temporally ordered mini batches are used to train the network for $e_{\text{Koop}}$ epochs as $\alpha$ is slowly and linearly raised from 0 to a maximum value $\alpha_{\text{max}}$ over a number of epochs. In practice, we find that allocating approximately 2/3 of the total number of epochs to $e_{\text{pre}}$ and 1/3 to $e_{\text{Koop}}$ leads to optimal performance. A schematic of the Koopman $\beta$-VAE is given in Figure~\ref{fig:vae}.

\subsection{LSTM Ensembles}

Neural network architectures such as transformers have been used in non-intrusive ROMs for time series prediction of low-dimensional coefficients~\cite{solera2024beta,wu2022non} as they offer a more powerful and expressive architecture compared to LSTMs. However, like with all time series models, error accumulation is a major issue when using transformers. Small errors made in early predictions can compound over time, leading to large inaccuracies as the prediction horizon grows. The model's performance becomes very sensitive to the initialization of its parameters, an undesirable property for practical use. A common approach used to mitigate this issue is ensemble learning~\cite{wichard2004time}, where multiple individual models, or \emph{weak learners}, are combined and have their predictions aggregated to create a more stable and robust model. While the individual models exhibit high variance, their combined results provide more accurate and stable predictions. Additionally, this significantly diminishes the effect of the model's initial parameters.

\begin{figure}[!h]
  \centering
  \includegraphics[width=1.0\textwidth]{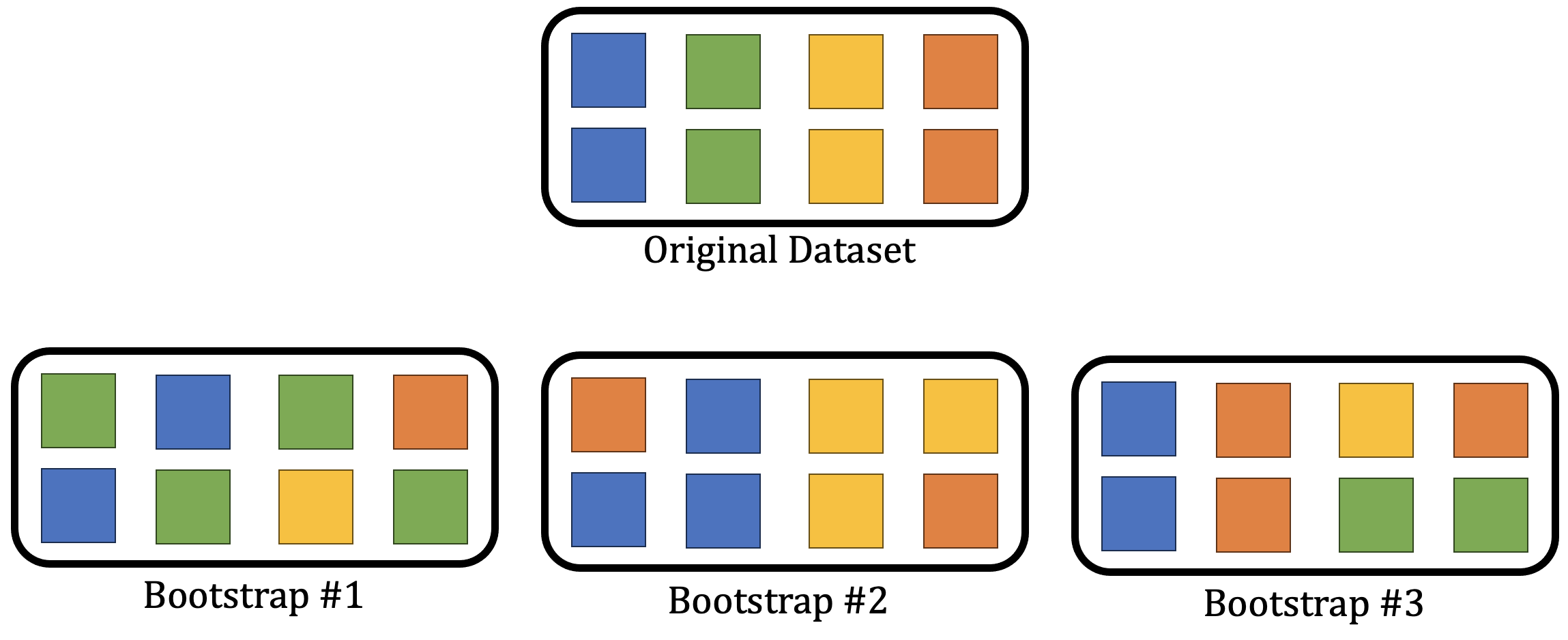}
  \caption{An example of bootstrapping, where random subsets of the original dataset are chosen through sampling with replacement.}
  \label{fig:bagged}
  \end{figure}

Bootstrap aggregating, often referred to as bagging~\cite{breiman1996bagging}, is a widely used ensemble learning approach that simply averages the results of multiple weak learners that are trained on different subsets of the training data chosen through random selection with replacement, as shown in Figure~\ref{fig:bagged}, where each subset has the same number of data points as the original dataset. By training multiple models on different subsets of the data, the weak learners become diverse and learn different patterns of the training data. As a result, aggregating their outputs leads to a large reduction in variance, leading to more stable and accurate results over time. Bagging also allows models to be trained in parallel as they are independent of each other, making it more computationally efficient than other ensemble methods like boosting~\cite{freund1999short}, which requires sequential training of weak learners. As an increasing number of weak learners are added to the ensemble, the reduction in variance pleateaus, leading to diminishing returns in model performance. Beyond a certain point, the added computational expense of including more weak learners in the ensemble exceeds the marginal gains in performance. The optimal number of weak learners to use is highly problem dependent and difficult to know \emph{a priori}. LSTMs are chosen to be used over transformers due to having a significantly lower training cost as well as less hyperparameters to tune, while still being powerful time series prediction models. A many-to-one LSTM architecture is used to predict $\bm{z}$ a single temporal snapshot ahead given a window of the previous $w$ latent variables. During inference, predictions are re-incorporated into the current window, leading to autoregressive forecasting.

Although in theory the latent variables can be predicted using the Koopman operator $\bm{z}_{t+1} = \bm{A}\bm{z}_{t}$, this is infeasible for long-term predictions. While the Koopman operator does impose a linear dynamics constraint on the latent variables, in practice the latent variables do not follow a perfectly linear evolution. The residual errors that exist can compound very quickly over time, leading to states $\bm{z}_{t}$ that diverge greatly from the structure of the training data. As a result, using only the Koopman operator to evolve the latent variables will result in rapid error accumulation and states that are not appropriately approximated by $\bm{A}$. Furthermore, a benefit of using an LSTM ensemble is that there is feedback from previous states to stabilize predictions.

\subsection{Reduced-Order Model}

The ROM implemented in this work involves a computationally expensive offline stage where high-fidelity snapshots are computed by solving the FOM and training both the Koopman $\beta$-VAE and LSTM ensemble. A number of $T_i$ initial snapshots are first assembled into a snapshot matrix $\bm{X} \in \mathbb{R}^{(n_d \times T_i) \times n_x \times n_y \times n_c}$, where $n_d$ is the number of parameters present in the training data, $n_x$ and $n_y$ are the number of grid points in each direction, with $n_x \times n_y = N$, and $n_c$ is the number of channels for each velocity component. The Koopman $\beta$-VAE is first pre-trained on randomly shuffled mini-batches from $\bm{X}$ for $e_{\text{pre}}$ epochs without the Koopman loss function ($\alpha = 0$) to allow the network to learn a set of well-structured latent variables in addition to avoid overfitting to temporal patterns in the data. Next, the model is trained for $e_{\text{Koop}}$ epochs on temporally ordered mini-batches to smooth and denoise the latent variables so small-scale turbulent structures are filtered out in reconstructions. The encoder means $\bm{\mu}$ from each sample are used to construct a matrix of latent variables $\bm{Z} \in \mathbb{R}^{T_i \times k}$. The means are used as they represent the most likely latent representations of each sample, providing a deterministic way of describing the flow dynamics. Using the latent variables, $l$ individual LSTMs with the same initial weights and biases are trained on different subsets of sequences of length $w$ generated from $\bm{Z}$ through selection with replacement.

During the online stage, the last $w$ temporal snapshots of $\bm{Z}$ are used as the initial LSTM window $\bm{Z}_t$ for autoregressive forecasting of latent variables until the end of the desired time horizon at time $T$. The average value taken over each LSTM is used to compute $\bm{z}_{t+1}$, which is incorporated into the current window. Finally, the full-order states are predicted using the decoder $\bm{\tilde{X}} = p_{\theta}\left(\bm{{Z}}\right)$. The ROM is outlined in Algorithm~\ref{alg:rom}.

\begin{algorithm}[!htbp]
  \caption{Offline and online stages of Koopman $\beta$-VAE ROM}\label{alg:rom}
  \begin{algorithmic}[1]
  \Function{KOOPMAN\_ $\beta$-VAE\_ OFFLINE}{$k, l, w$}
  \State Compute high-fidelity snapshots by solving FOM and assemble into $\bm{X}$.
  \State Pre-train Koopman $\beta$-VAE with latent dimension $k$ for $e_{\text{pre}}$ epochs ($\alpha$ = 0) using randomly shuffled mini-batches.
  \State Train Koopman $\beta$-VAE for $e_{\text{Koop}}$ epochs using temporally ordered mini-batches.
  \State Calculate latent variables for training data by using encoder means $\bm{Z} = q_{\phi}\left(\bm{X} \right)$.
  \State Train $l$ LSTMs $\mathcal{E} = [\text{LSTM}_{1}(\bm{Z}), \text{LSTM}_{2}(\bm{Z}), \cdots \text{LSTM}_{l}(\bm{Z})]$ using a window size of $w$ on sequences selected through replacement from $\bm{Z}$.
  \State \Return $\left(p_{\theta}, \bm{Z}, \mathcal{E} \right)$
  \EndFunction
  \end{algorithmic}
  \[\]
  \begin{algorithmic}[1]
  \Function{KOOPMAN\_ $\beta$-VAE\_ ONLINE}{$p_{\theta}, \bm{Z}, \mathcal{E}$}
  \State Use the last $w$ temporal snapshots of $\bm{Z}$ as the current LSTM window $\bm{Z}_{t}$.
  \For{$t \in \{T_{i}, T_{i+1},\dots,T-1\}$}
  \State Compute $\bm{z}_{t+1} = \mathcal{E}(\bm{Z}_{t})$ by using an average of the $l$ individual LSTM predictions.
  \State Incorporate $\bm{z}_{t+1}$ into the current window $\bm{Z}_{t}$.
  \State $\bm{Z}[t+1] = \bm{z}_{t+1}$
  \EndFor
  \State Predict full-order fields $\bm{\tilde{X}} = p_{\theta}\left(\bm{{Z}}\right)$
  \State \Return $\bm{\tilde{X}}$
  \EndFunction
  \end{algorithmic}
  \end{algorithm}

\section{Results}
\begin{figure}[!h]
   \centering
   \includegraphics[width=0.99\textwidth]{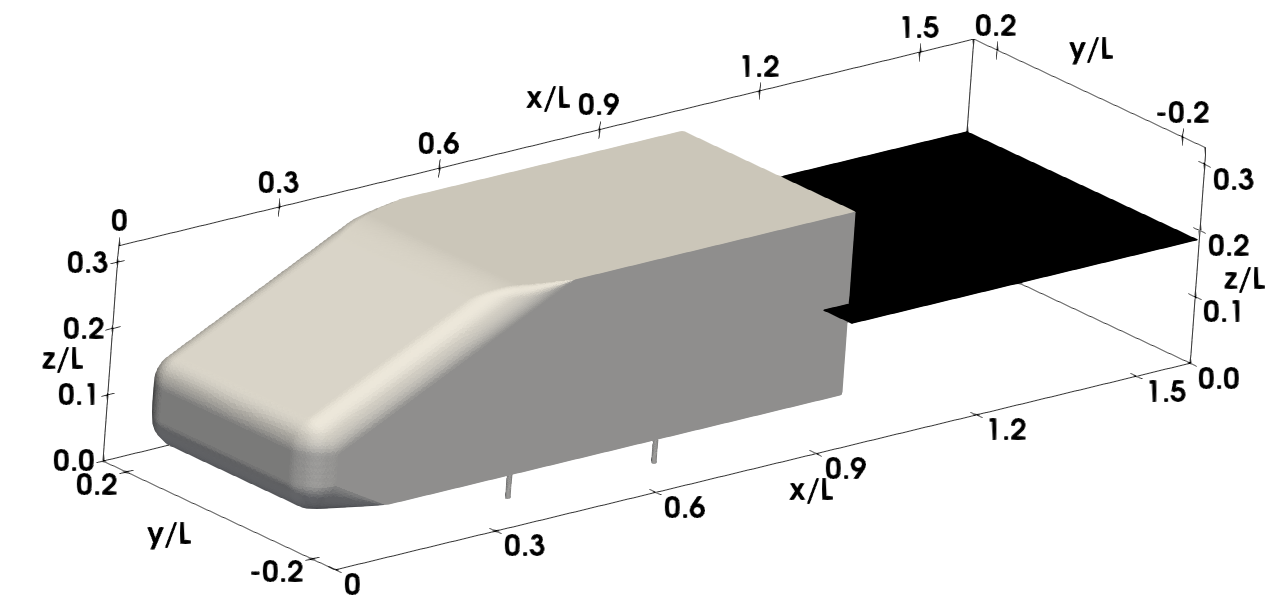}
   \caption{Geometry of the Windsor body (gray) and the plane in the wake on which flow data are interpolated onto (black).}
   \label{fig:windsor}
\end{figure}

The test case used in this work involves large eddy simulations of flow past a Windsor body, a simplified square-back vehicle shown in Figure~\ref{fig:windsor}, at 5 different yaw angles $\delta$ = [2.5, 5, 7.5, 10, 12.5] degrees. Data are interpolated onto the plane in the turbulent wake depicted in black. The flow is simulated at a Reynolds number $Re_L = U_\infty L/ \nu = 2.9 \times 10^6$, where $U_\infty$ is the freestream velocity, $L$ is the body length, and $\nu$ is the kinematic viscosity. SOD2D (Spectral high-Order coDe 2 solve partial
Differential equations), a low-dissipation GPU-based spectral element method (SEM) code~\cite{gasparino2024sod2d}, is used to solve the spatially filtered Navier-Stokes equations,

\begin{equation}
   \frac{\partial \bar{u}_i}{\partial x_i} = 0,
   \label{eq:continuity}
\end{equation}

\begin{equation}
   \frac{\partial \bar{u}_i}{\partial t}
   + \frac{\partial  \bar{u}_i \bar{u}_j }{\partial x_j}
   - \nu \frac{\partial^2  \bar{u}_i }{\partial x_j x_j}
   + \frac{1}{\rho} \frac{\partial \bar{p}}{\partial x_i}
   = \frac{\partial \tau_{ij}}{x_j},
   \label{eq:momentum}
\end{equation}

where $x_i$ are the spatial coordinates ($x, y$, and $z$), $u_i$ are the velocity components ($u, v$, and $w$), $p$ is the pressure, and $\rho$ is the density. The filtered variables are represented using a bar. The right-hand side of Equation~\ref{eq:momentum} represents the subgrid stresses, with the anisotropic part represented as

\begin{equation}
   \tau_{ij} - \frac{1}{3}\tau_{kk}\delta_{ij} = -2\nu_{\text{sgs}}\bar{\mathcal{S}}_{ij},
\end{equation}

where the large-scale strain rate tensor $\bar{\mathcal{S}}_{ij}$ is evaluated as $\bar{\mathcal{S}}_{ij} = \frac{1}{2}\left(g_{ij} + g_{ji}  \right),$ $g_{ij} = \partial \bar{u}_i / \partial x_j$, and $\delta_{ij}$ is the Kronecker delta function. The unresolved flow scales are modeled using a local formulation of the integral length scale approximation presented in a work by Lehmkuhl et al.~\cite{lehmkuhl2019extension}. The near-wall region is modeled using the Reichardt wall-law~\cite{reichardt1951vollstandige} using an exchange location in the 5th node~\cite{lehmkuhl2018large}.

The velocity components $u$ and $v$ are interpolated onto a uniform grid measuring 384 $\times$ 256 points on the x-y plane at $z/L = 0.186$ with bounds $x/L \in [1, 1.6]$ and $y/L \in [-0.2, 0.2]$. 

\begin{figure}[!h]
   \centering
   \includegraphics[width=0.99\textwidth]{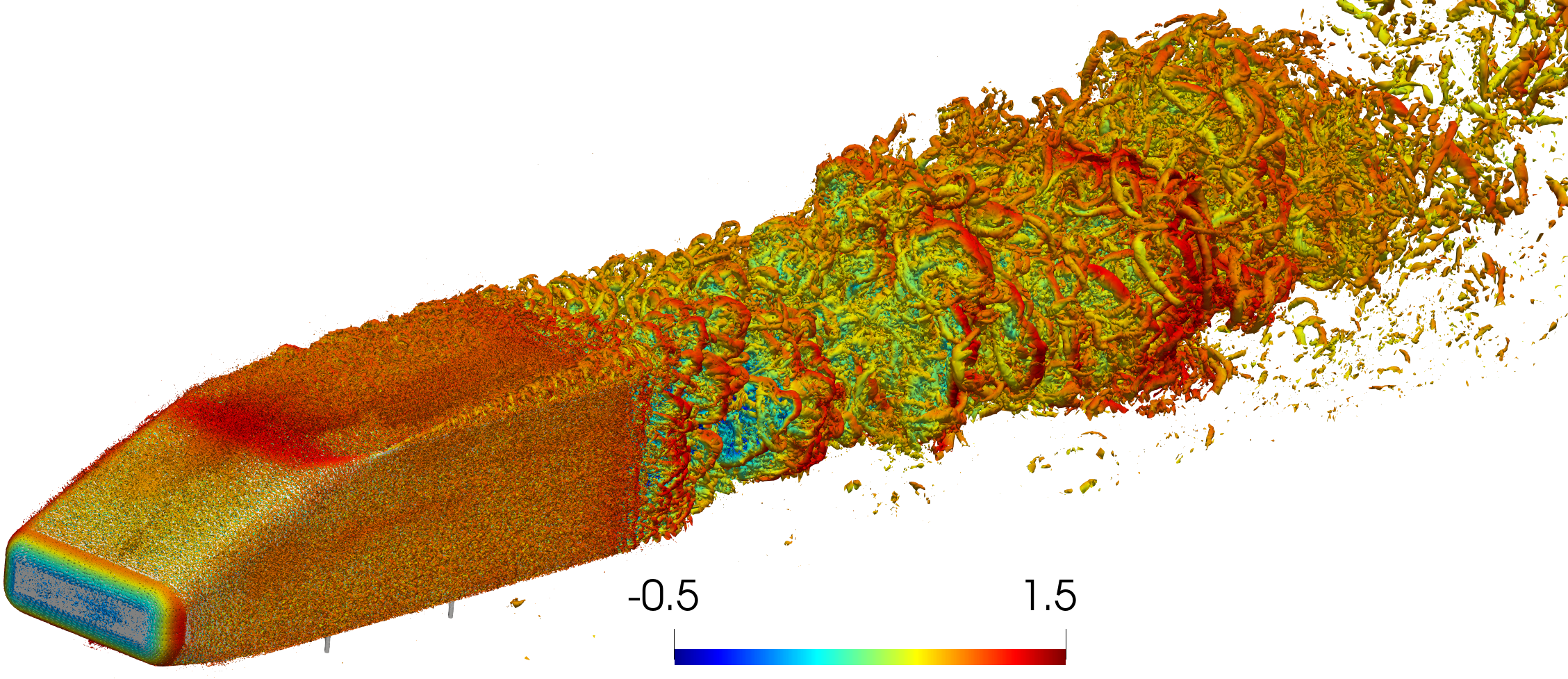}
   \caption{Instantaneous $Q$ criterion isocontours at $Q=100$ and $\delta=12.5$ portraying the streamwise velocity magnitude.}
   \label{fig:windsorqs}
\end{figure}

640 snapshots spanned along 58 convective time units, $t=L/U_{\infty}$ are generated at each angle and split into 80\% training and 20\% test data, corresponding to $T_i = 512$ training snapshots and 128 test snapshots, resulting in a total of 2560 training and 640 test samples. In order to provide an example of the flow topology, Figure~\ref{fig:windsorqs} presents an instantaneous representation of the flow for the case at $\delta=12.5$. More details on the case description, flow evaluation, and validation of the numerical methodology can be found in the AC-1.12 entry of the ERCOFTAC Knowledge Wiki \cite{Eiximeno2025AC112}.

\begin{figure}[!h]
  \centering
  \includegraphics[width=0.6\textwidth]{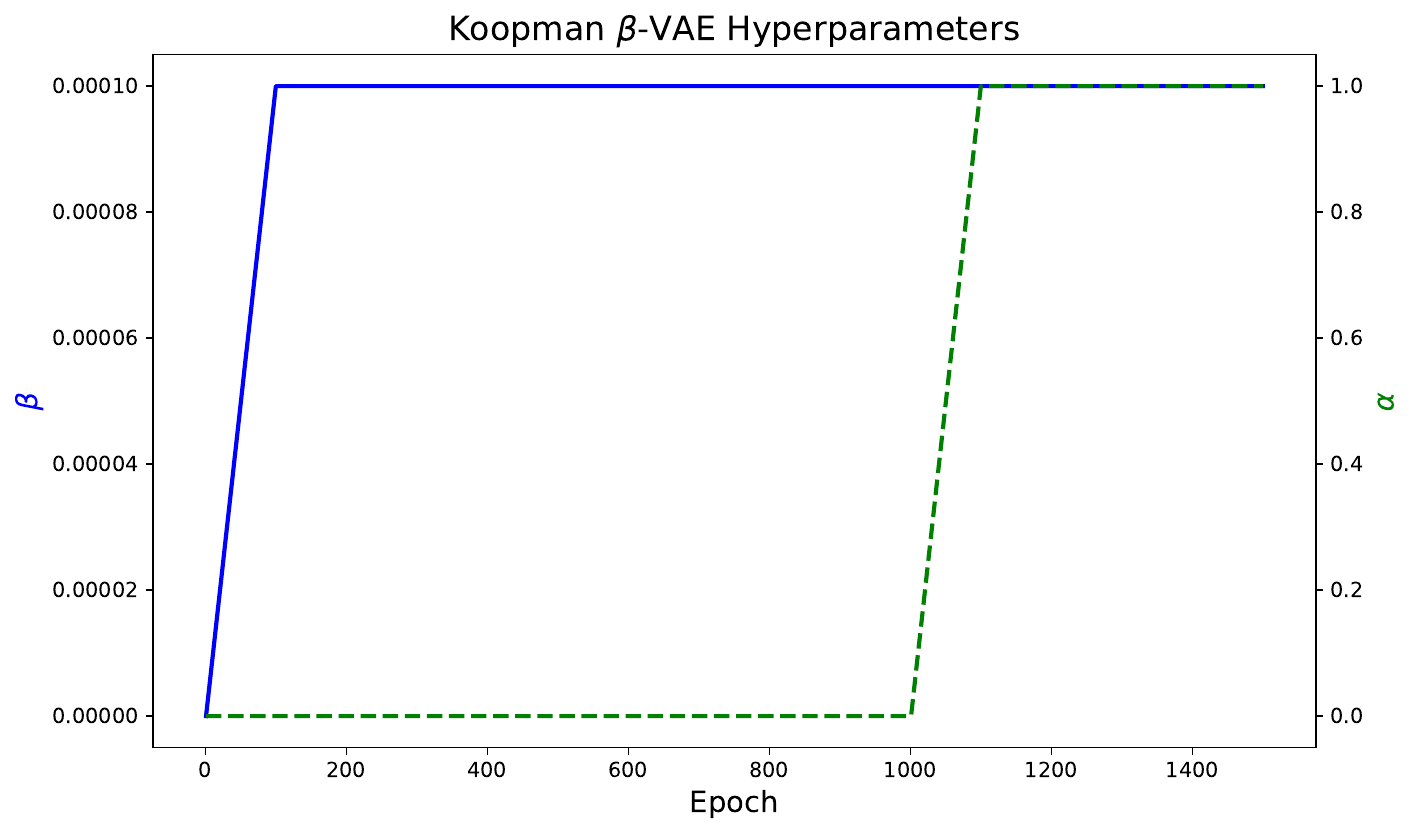}
  \caption{Koopman $\beta$-VAE hyperparameters.}
  \label{fig:params}
\end{figure}

Before training the Koopman $\beta$-VAE, the snapshot data are pre-processed using feature-wise min-max scaling between a range of [0, 1]. Data normalization improves model performance and allows for learning the optimal network parameters at a faster rate~\cite{Sola1997ImportanceOI}. A sigmoid activation function is used in the output layer of the model to constrain the range of the outputs to match the range of the inputs, after which the data is unscaled to the original range. The model is trained for 1500 epochs, with $e_{\text{pre}}$ = 1000 and $e_{\text{Koop}}$ = 500. A batch size of 64 is used for both training stages. The full architecture of the Koopman $\beta$-VAE can be found in Appendix~\ref{appendix:arch}.
A comparison to a non pre-trained model is given in Appendix~\ref{appendix:pre}. When trained on an NVIDIA H100 GPU, the model takes approximately 1200 seconds to train. Values of $\beta_{\text{max}} = 1 \times 10^{-4}$ and $\alpha_{\text{max}} = 1$ are used. The variation of both parameters is shown in Figure~\ref{fig:params}. $\beta$ is increased from 0 to $\beta_{\text{max}}$ linearly over 100 epochs, in the first 10\% of the pre-training procedure, after which it remains constant. $\alpha$ remains at 0 for the pre-training procedure to retain the vanilla $\beta$-VAE loss, after which it is increased linearly from 0 to $\alpha_{\text{max}}$ over 100 epochs in the first 20\% of $e_{\text{Koop}}$. Both values are increased linearly from 0 to add stability to the training process and avoid sudden over-regularization of latent variables. $\beta_{\text{max}}$ was chosen to maximize the reconstruction accuracy whilst retaining a latent space that remains well-structured (similar latent variable magnitudes) after the implementation of the Koopman loss. A comparison to a Koopman CAE ($\beta_{\text{max}}$ = 0) is given in Appendix~\ref{appendix:cae}. The effect of different values of $\alpha_{\text{max}}$ is given in Appendix~\ref{appendix:alpha}. 
A bidirectional LSTM architecture~\cite{schuster1997bidirectional} with three hidden layers each consisting of 96 neurons is used for the LSTM ensemble. A dropout rate~\cite{srivastava2014dropout} of 0.2 is used in each hidden layer. Again, the data are pre-processed using min-max scaling and a sigmoid layer is used in the output layer. Bidirectional LSTMs process data in both forward and backward directions, allowing the model to leverage both past and future context. A window size of $w$ = 64 is used. $l = 128$ weak learners are used and trained consecutively in parallel across 4 NVIDIA H100 GPUs, with a single model taking approximately 60 seconds to train. Training sequences are generated at each yaw angle and the ensemble model is trained on all of them simultaneously using randomly shuffled mini-batches of size 64. All hyperparameters were chosen through a trial-and-error process. At each yaw angle, inference over the 128 test snapshots takes approximately 24 seconds, which is negligible when compared to the time required for simulating the FOM. Results from a CAE-LSTM ROM using a vanilla CAE and LSTM ensemble for time series prediction are given in Appendix~\ref{appendix:caelstm} to provide a baseline for comparison.

\subsection{Reconstruction Accuracy}

The reconstruction accuracy of the Koopman $\beta$-VAE is compared against a vanilla CAE using the same architecture and latent dimension that is trained for 1000 epochs using only the MSE loss function ($\beta = 0$) on randomly shuffled mini-batches. We expect the reconstruction accuracy to be lower as small-scale turbulent structures are filtered out from the training data. The metric used to assess performance is the relative $L^2$ error between the raw data and reconstruction $\epsilon$,

\begin{equation}
  \epsilon =   \dfrac{\norm{\bm{x} - \hat{\bm{x}}}^{2}} {\norm{\bm{x}}^2}.
\end{equation}

\begin{figure}[!htpb]
  \centering
  \includegraphics[width=0.7\textwidth]{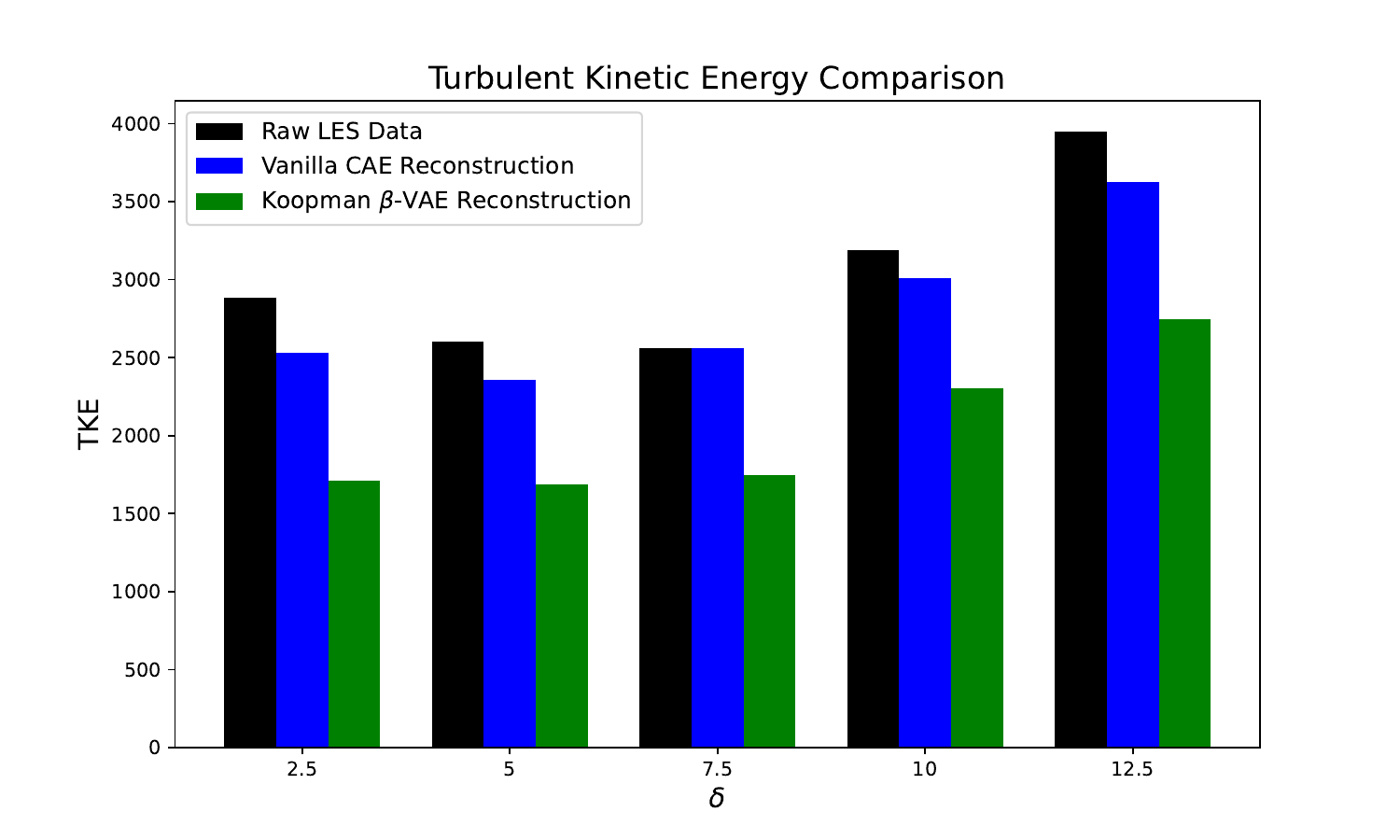}
  \caption{Turbulent kinetic energy comparison for the training dataset.}
  \label{fig:tke}
\end{figure}

The turbulent kinetic energy $TKE$ is also measured to quantify the reduction in velocity fluctuations from the mean flow, which are dominated by small-scale turbulent structures,

\begin{equation}
  TKE = \frac{1}{2} \sum_{t=1}^{T_i} \left( (u^t - \bar{u})^2 +  (v^t - \bar{v})^2    \right),
\end{equation}

where $\bar{u}$ and $\bar{v}$ are the mean x and y velocity respectively. Table~\ref{table:recon} shows both the overall and component-wise relative errors averaged over all of the training data. As expected, the vanilla CAE reconstructs the snapshots much more accurately due to the inclusion of small-scale turbulent structures. Both models exhibit significantly higher errors in reconstructing $v$ compared to $u$; structures corresponding to the spanwise velocity generally correspond to smaller and more chaotic scales, while the streamwise velocity is more spatially and temporally coherent. Using the Koopman $\beta$-VAE results in lower reconstruction accuracy due to the exclusion of small-scale structures. As a result, there is a significant decrease in the average turbulent kinetic energy. The turbulent kinetic energy averaged over all yaw angles of the raw LES training data is equal to $3.04 \times 10^3$; the vanilla CAE reconstructions retain approximately 93.5\% of this, whereas those from the Koopman $\beta$-VAE retain approximately 67.2\%, an expected and desired result. An angle-wise comparison of $TKE$ between the training data and model reconstructions is given in Figure~\ref{fig:tke}, where it is shown that the vanilla CAE retains more kinetic energy than the Koopman $\beta$-VAE at each angle and matches the raw LES data almost exactly at $\delta = 7.5$.

\begin{figure}[!htpb]
  \centering
  \includegraphics[width=0.7\textwidth]{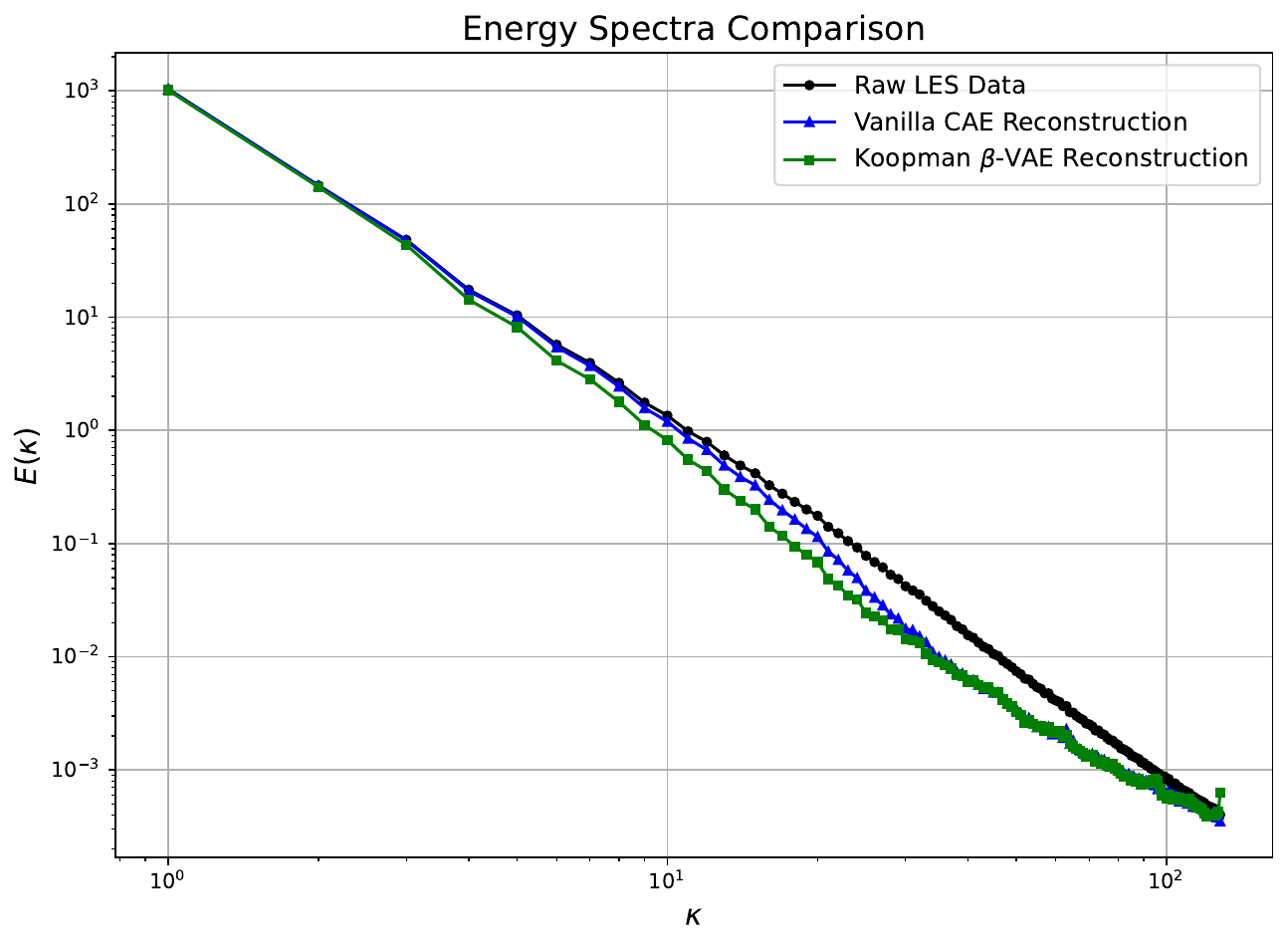}
  \caption{Energy spectra comparison of the training dataset.}
  \label{fig:fft}
\end{figure}

Spectral analysis is a useful method for visualizing the energy cascade of turbulent flow, or how kinetic energy is distributed among turbulent eddies of different sizes. Given the velocity components, a two-dimensional fast Fourier transform (FFT) $\mathcal{F}$ is applied to each velocity component, converting them from the physical space to the wavenumber space $\hat{u}(\bm{\kappa}), \hat{v}(\bm{\kappa})$,

\begin{equation}
  \hat{u}(\bm{\kappa}) = \mathcal{F}(u(x,y)),
\end{equation}
where $\bm{\kappa} = (\kappa_x, \kappa_y)$ is the wave vector. The magnitude of $\kappa = \vert \bm{\kappa} \vert$ is the wavenumber, which is inversely proportional to the eddy size. The kinetic energy $E(\kappa)$ in the wavenumber space is computed as 

\begin{equation}
E(\kappa) = \frac{1}{2} \left( \vert \hat{u}(\bm{\kappa}) \vert ^2 + \vert \hat{v}(\bm{\kappa}) \vert ^2 \right).
\end{equation}
Absolute values are used as the FFT produces complex numbers. To obtain a one-dimensional energy spectrum, values of $\kappa$ are binned linearly from 1 to $\text{max}(\kappa)$ in intervals of 1. Within these bins, the corresponding energy values are averaged. Figure~\ref{fig:fft} shows an energy spectra comparison between the raw LES data and both models. At lower wavenumbers, which correspond to larger scales, the kinetic energy of the flow is distributed similarly between both models and the raw LES data. For intermediate wavenumbers ($\kappa \in [4, 40]$ ), where smaller scales start to dominate the flow, both models exhibit lower amounts of energy compared to the raw LES data. The Koopman $\beta$-VAE exhibhits a larger decrease than the vanilla CAE, showing that the addition of the Koopman loss function results in filtering of small-scale turbulent structures. At larger wavenumbers, both models are similar due to the limitations of convolutional layers in capturing very fine-scale features.

\begin{table}[]
  \centering
  \begin{tabular}{|l|l|l|l|l|l|}
  \hline
  \textbf{Model} & \textbf{$\epsilon$} & \textbf{$\epsilon_u$} & \textbf{$\epsilon_v$} &  \textbf{$TKE$} \\ \hline
  Vanilla CAE    &          0.147              & 0.114               & 0.254                   & 2.84e3               \\ \hline
  Koopman $\beta$-VAE    & 0.252                      & 0.188                   & 0.443                            & 2.04e3               \\ \hline
  \end{tabular}
  \caption{Reconstruction error and turbulent kinetic energy comparison for training data.}
  \label{table:recon}
  
\end{table}

\begin{figure}[!htpb]
  \centering
  \includegraphics[width=1.0\textwidth]{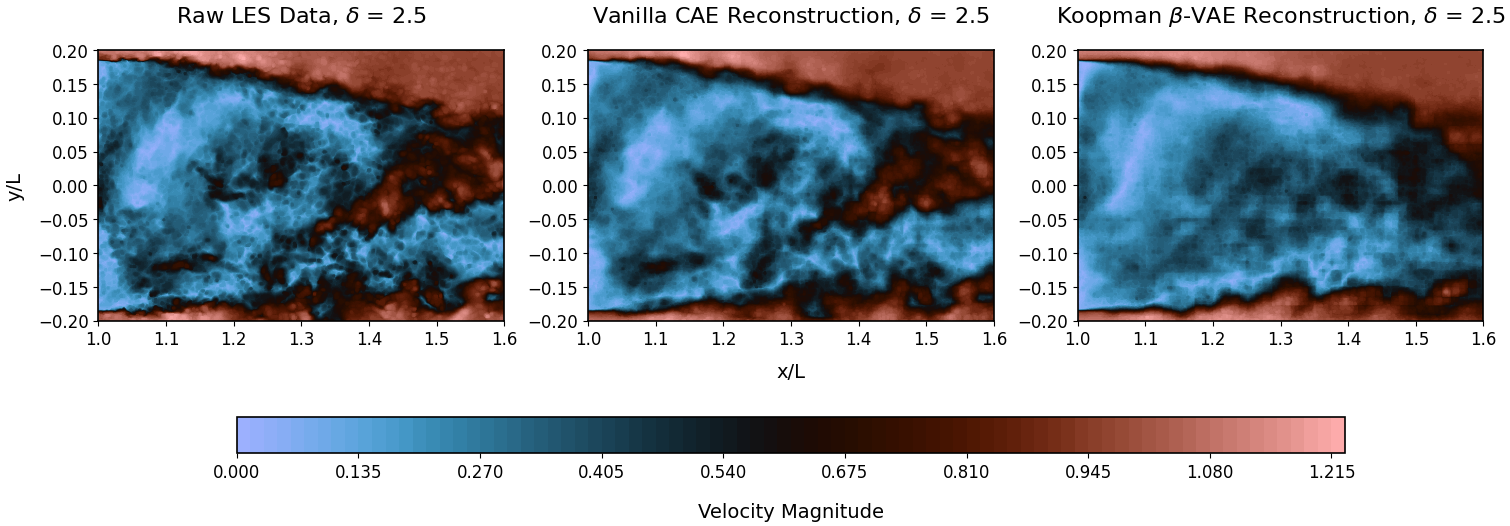}
  \caption{Velocity magnitude comparison at $t = 256$ and $\delta = 2.5$.}
  \label{fig:recon_2.5}

  \end{figure}

\begin{figure}[!htpb]
    \centering
    \includegraphics[width=1.0\textwidth]{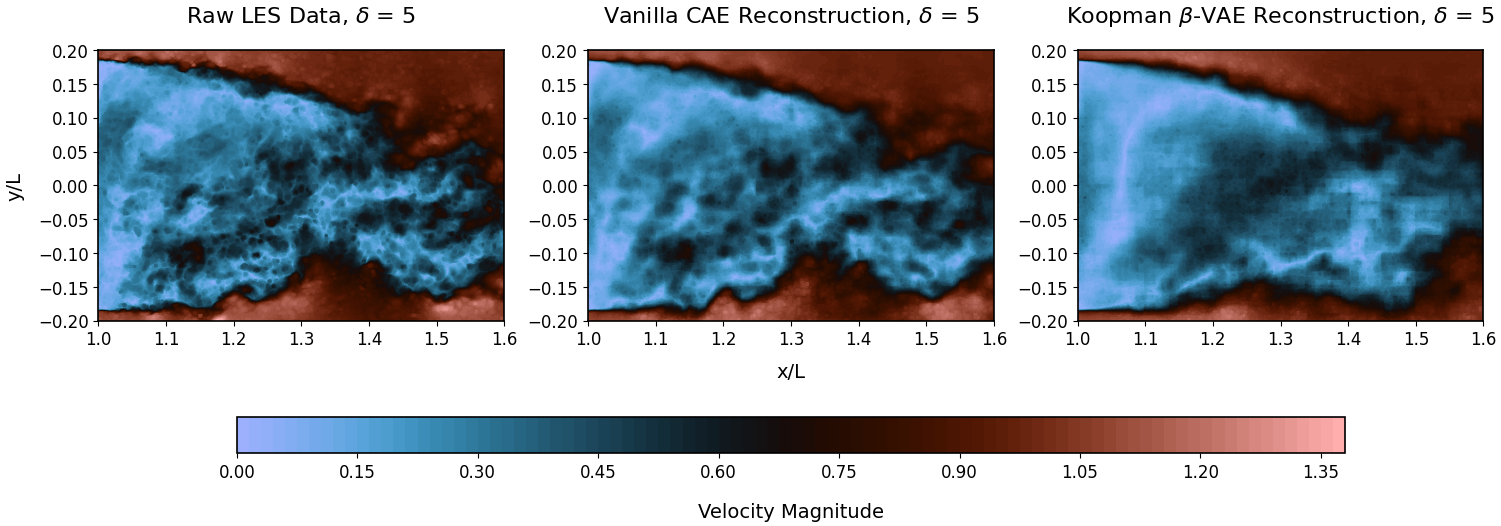}
    \caption{Velocity magnitude comparison at $t = 256$ and $\delta = 5$.}
  \label{fig:recon_5}

\end{figure}

\begin{figure}[!htpb]
  \centering
  \includegraphics[width=1.0\textwidth]{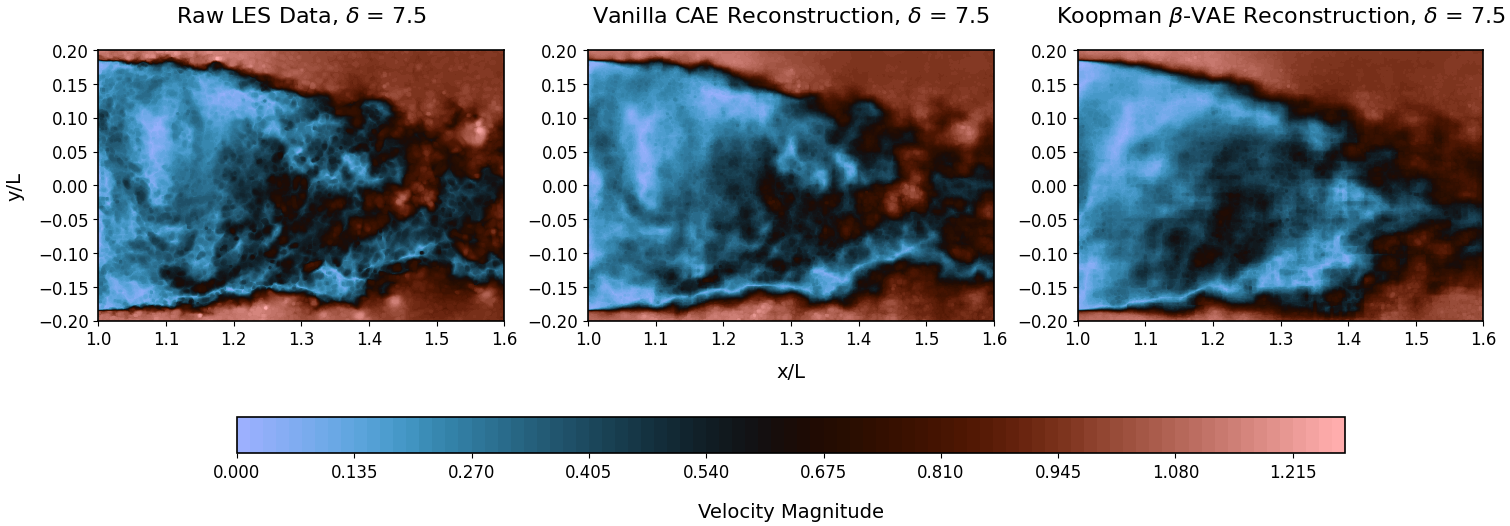}
  \caption{Velocity magnitude comparison at $t = 256$ and $\delta = 7.5$.}
  \label{fig:recon_7.5}
  \end{figure}

\begin{figure}[!htpb]
    \centering
    \includegraphics[width=1.0\textwidth]{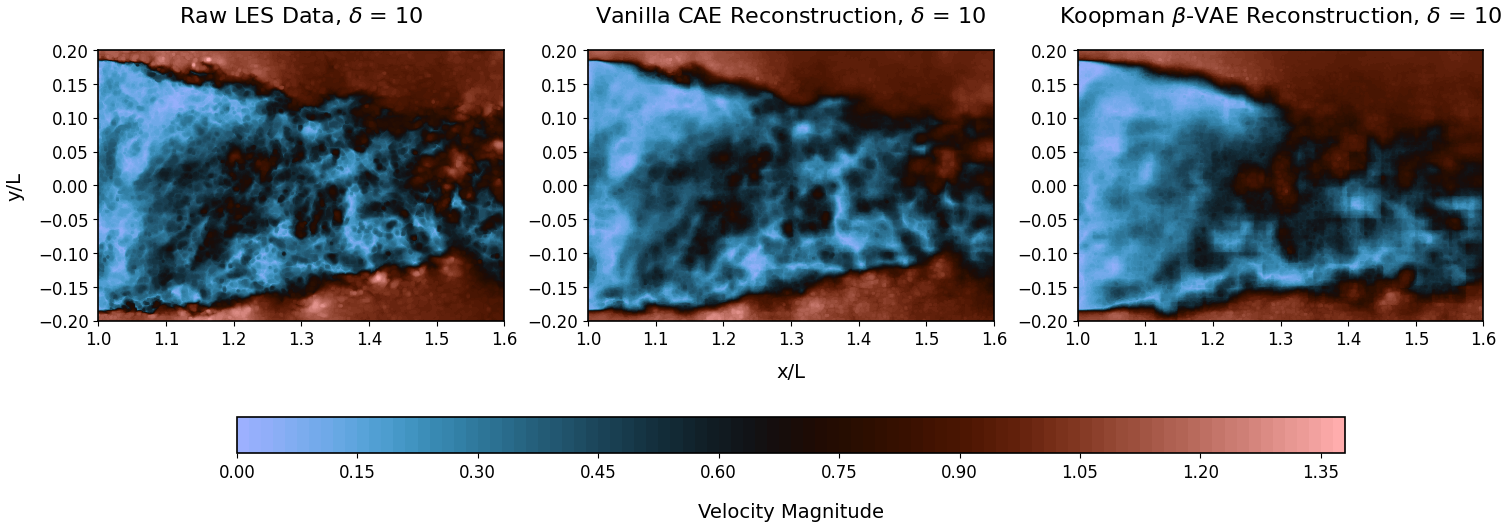}
  \caption{Velocity magnitude comparison at $t = 256$ and $\delta = 10$.}
  \label{fig:recon_10}
\end{figure}

\begin{figure}[!htpb]
  \centering
  \includegraphics[width=1.0\textwidth]{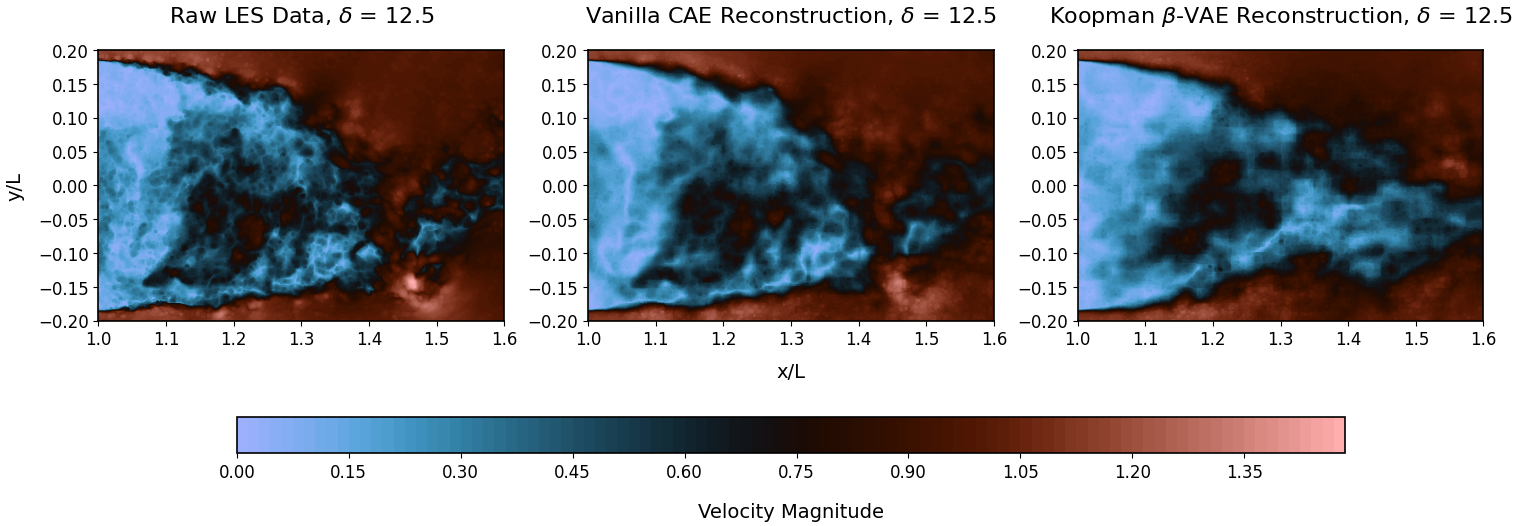}
  \caption{Velocity magnitude comparison at $t = 256$ and $\delta = 12.5$.}
  \label{fig:recon_12.5}
\end{figure}

\begin{figure}[!htpb]
  \centering
  \includegraphics[width=1.0\textwidth]{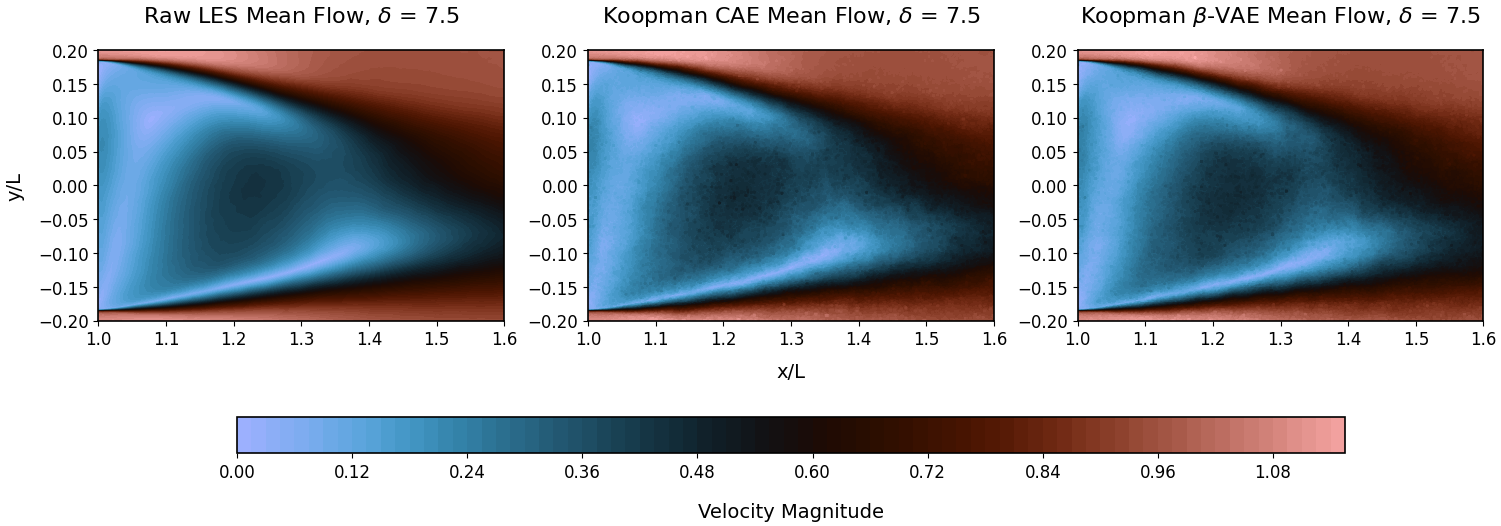}
  \caption{Mean flow comparison for the training data at $\delta = 7.5$.}
  \label{fig:means_train}
\end{figure}

\begin{figure}[!htpb]
  \centering
  \includegraphics[width=0.495\textwidth]{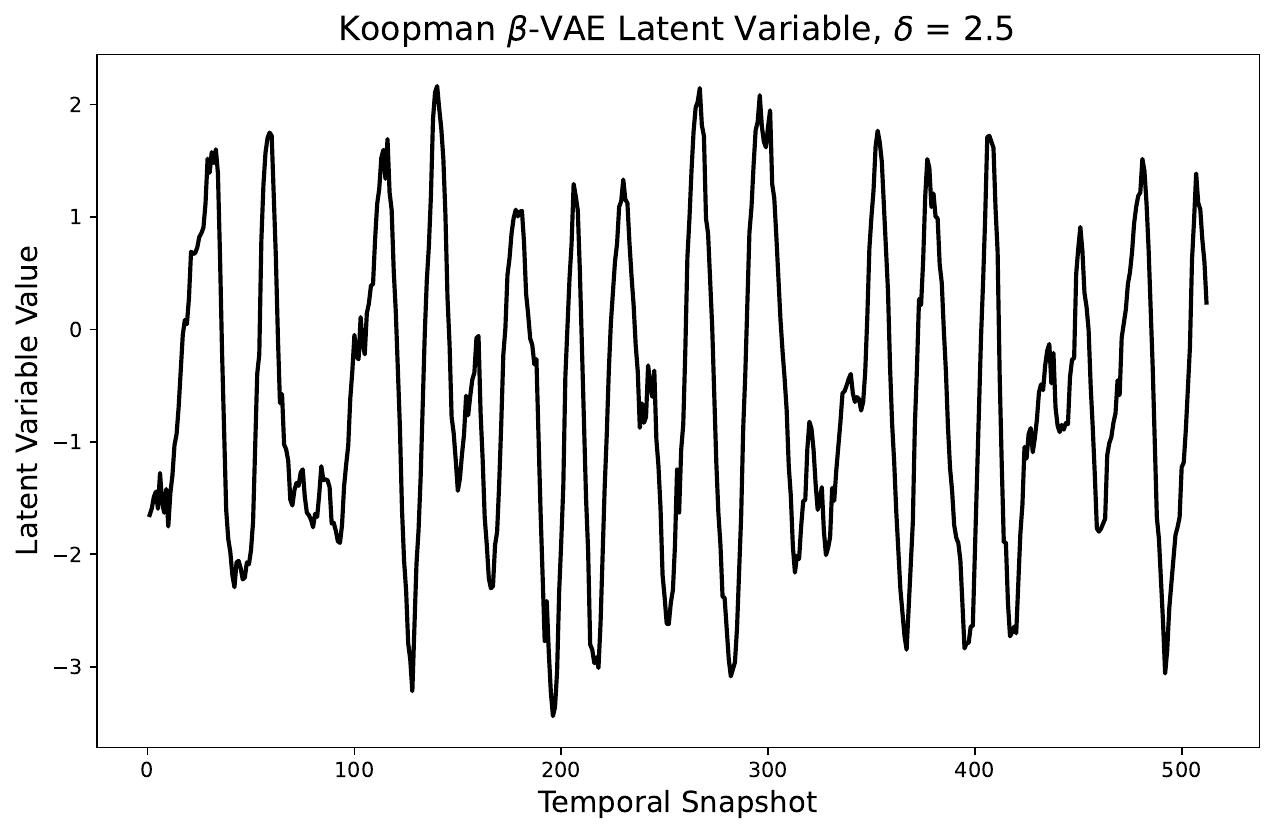}
  \includegraphics[width=0.495\textwidth]{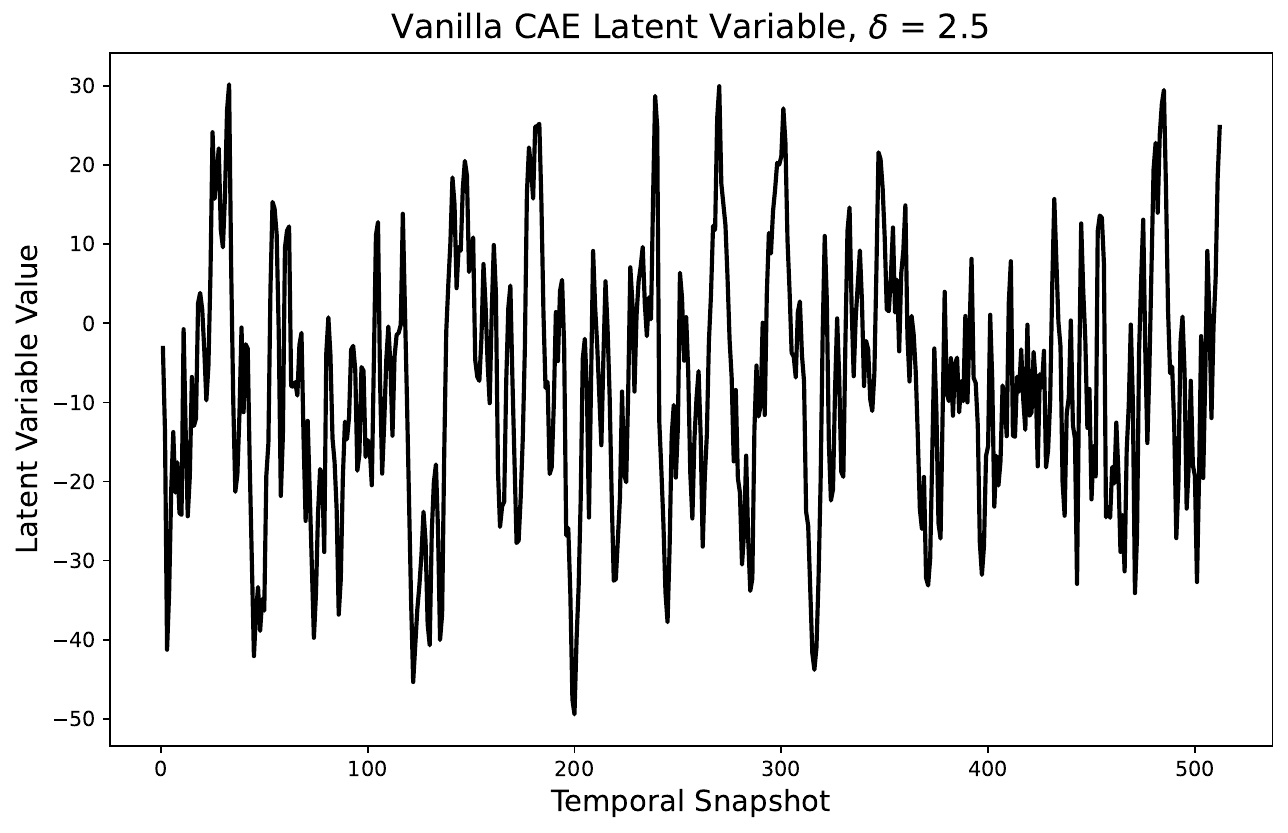}
  \caption{Latent variable comparison at $\delta = 2.5$.}
  \label{fig:latent_2.5}
\end{figure}

\begin{figure}[!htpb]
  \centering
  \includegraphics[width=0.495\textwidth]{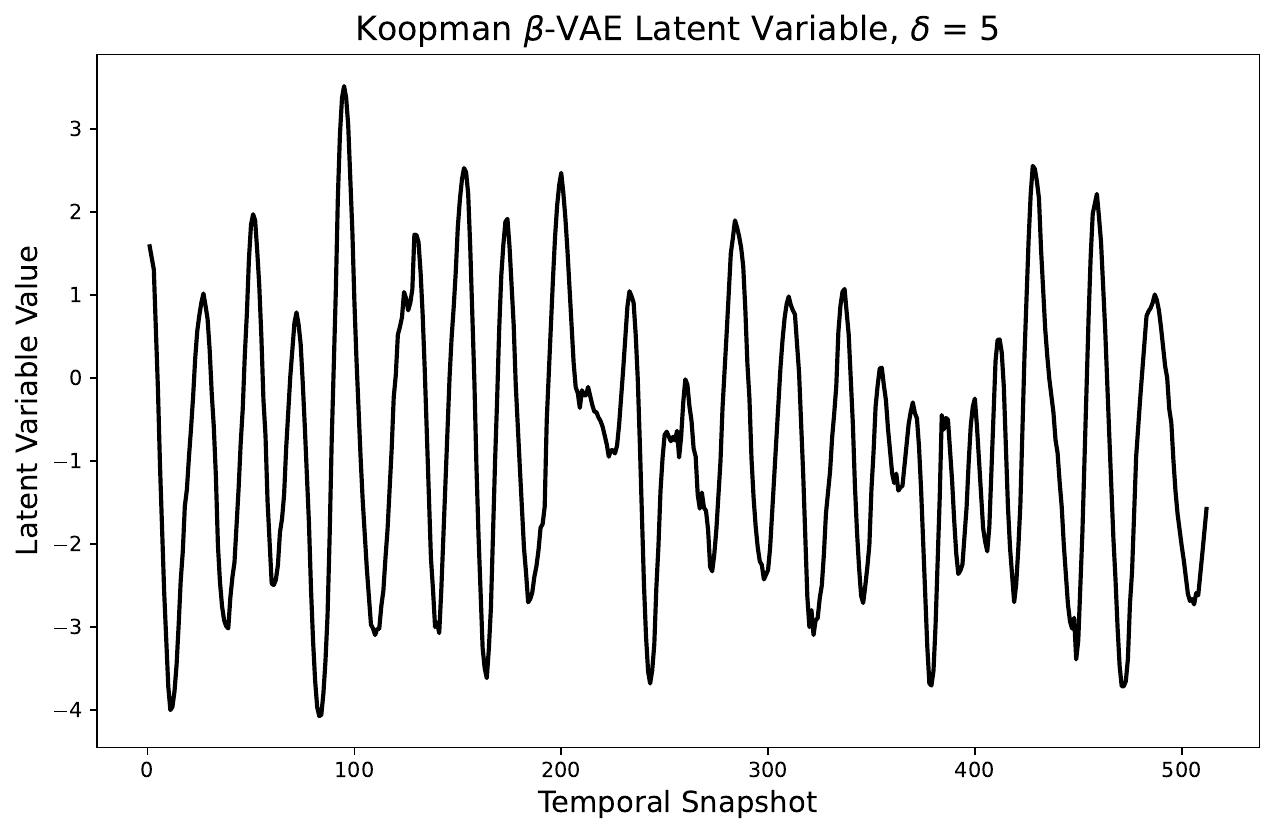}
  \includegraphics[width=0.495\textwidth]{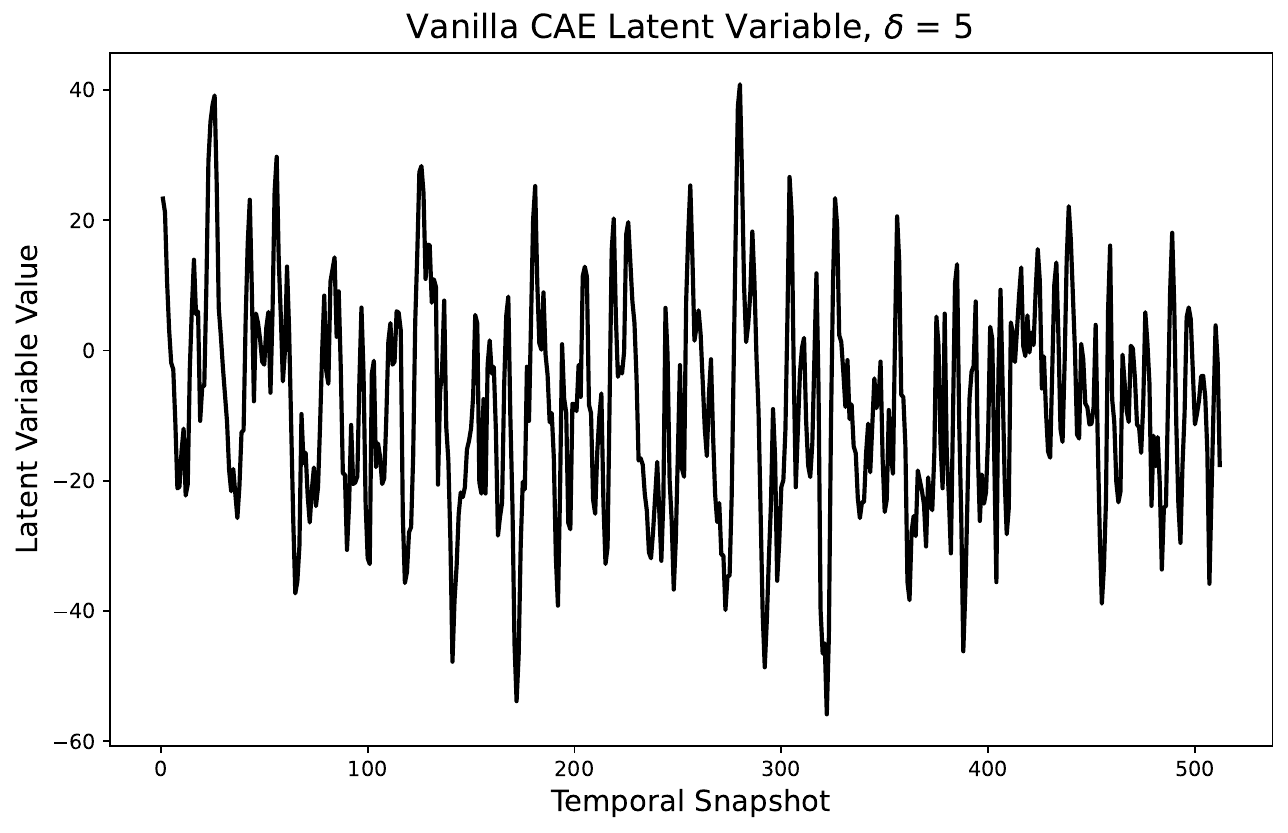}
  \caption{Latent variable comparison at $\delta = 5$.}
  \label{fig:latent_5}
\end{figure}

\begin{figure}[!htpb]
  \centering
  \includegraphics[width=0.495\textwidth]{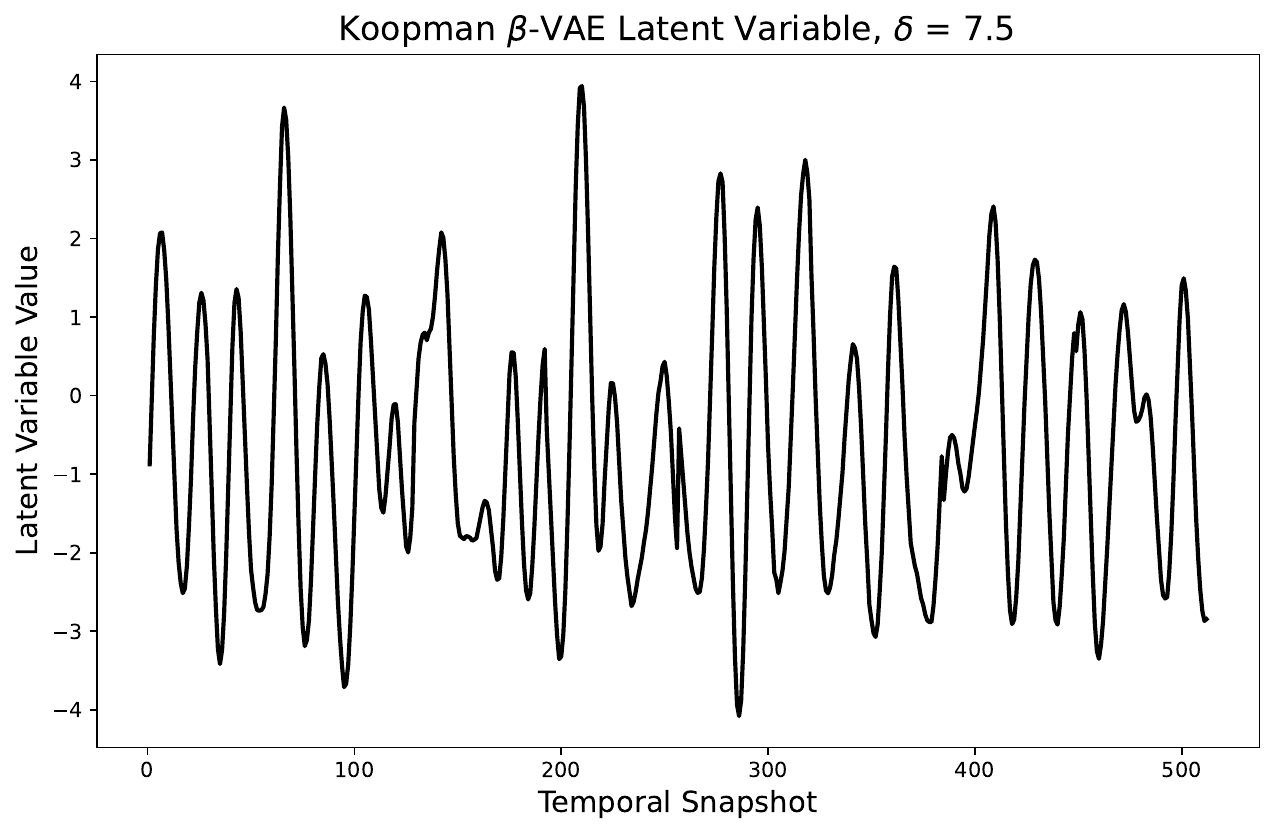}
  \includegraphics[width=0.495\textwidth]{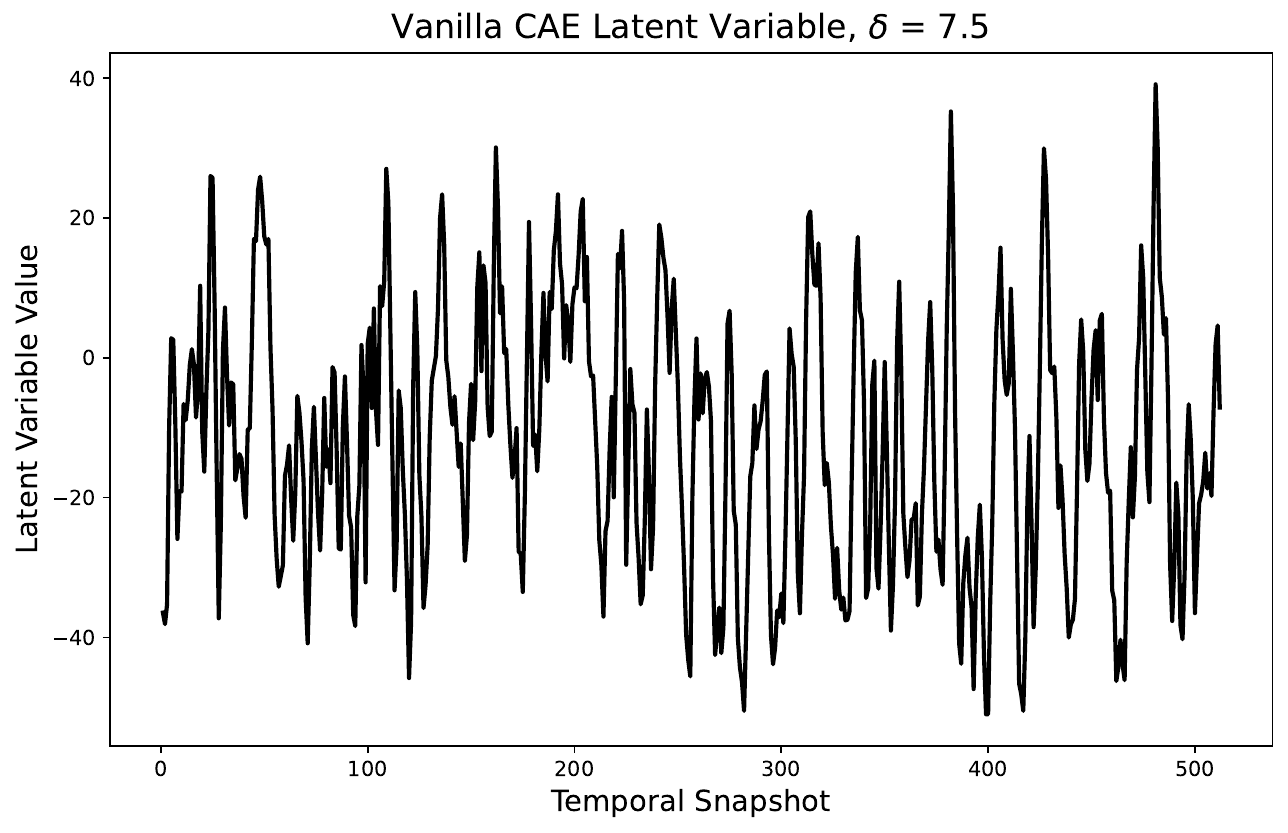}
  \caption{Latent variable comparison at $\delta = 7.5$.}
  \label{fig:latent_7.5}
\end{figure}

\begin{figure}[!htpb]
  \centering
  \includegraphics[width=0.495\textwidth]{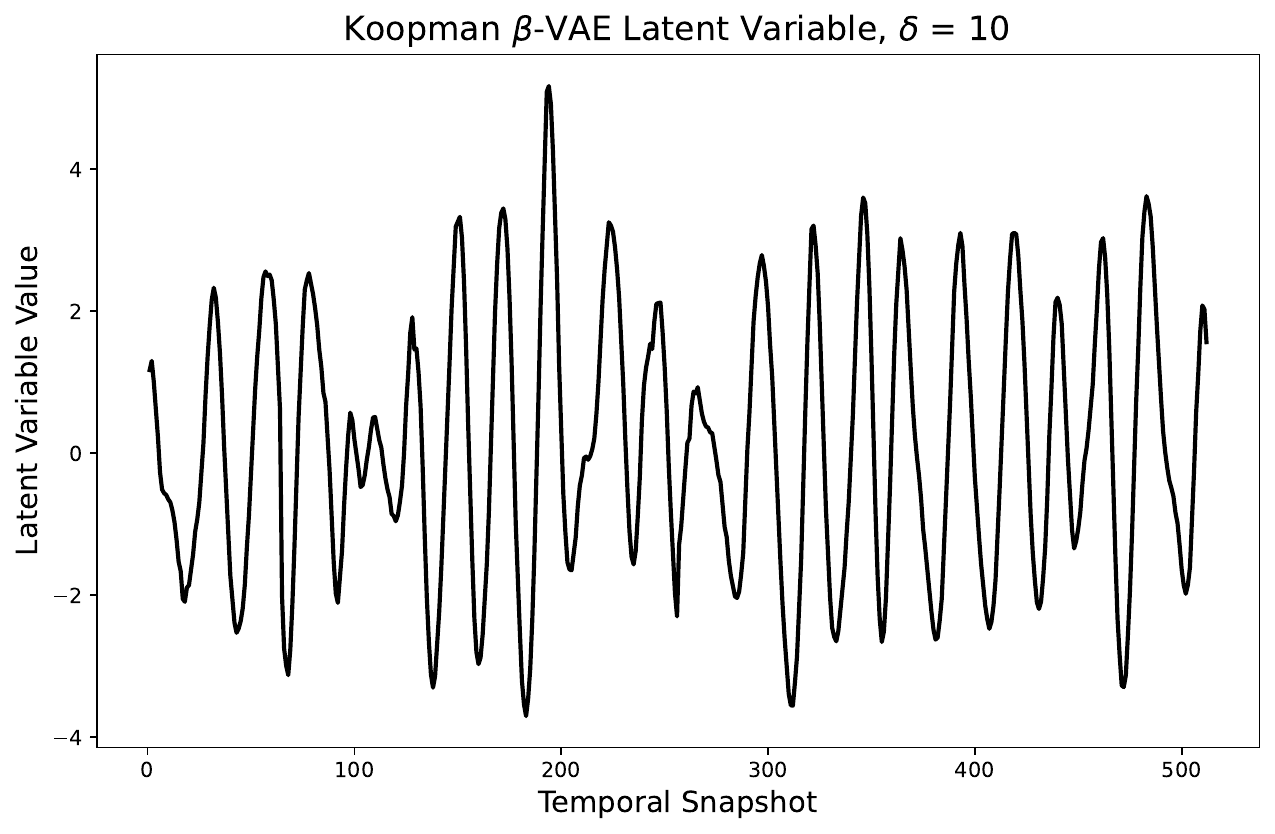}
  \includegraphics[width=0.495\textwidth]{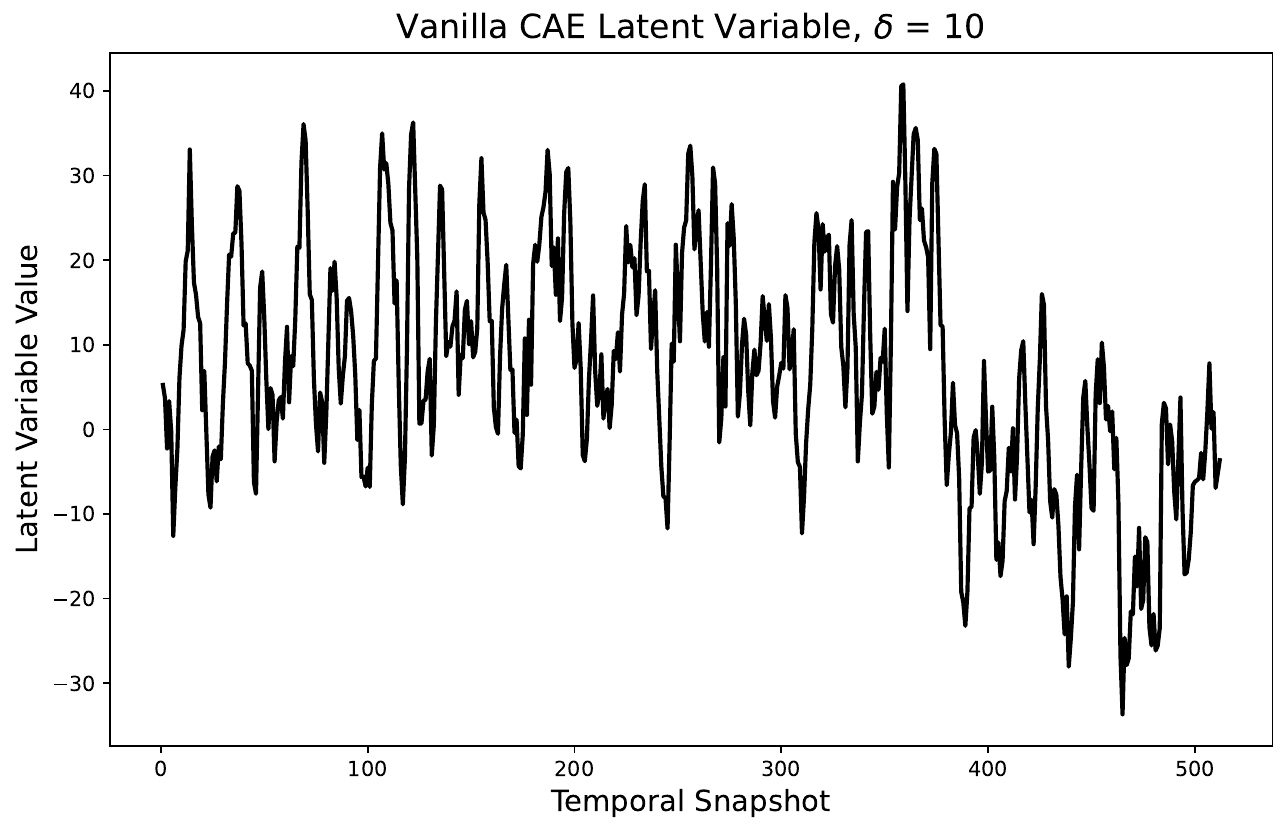}
  \caption{Latent variable comparison at $\delta = 10$.}
  \label{fig:latent_10}

\end{figure}

\begin{figure}[!htpb]
  \centering
  \includegraphics[width=0.495\textwidth]{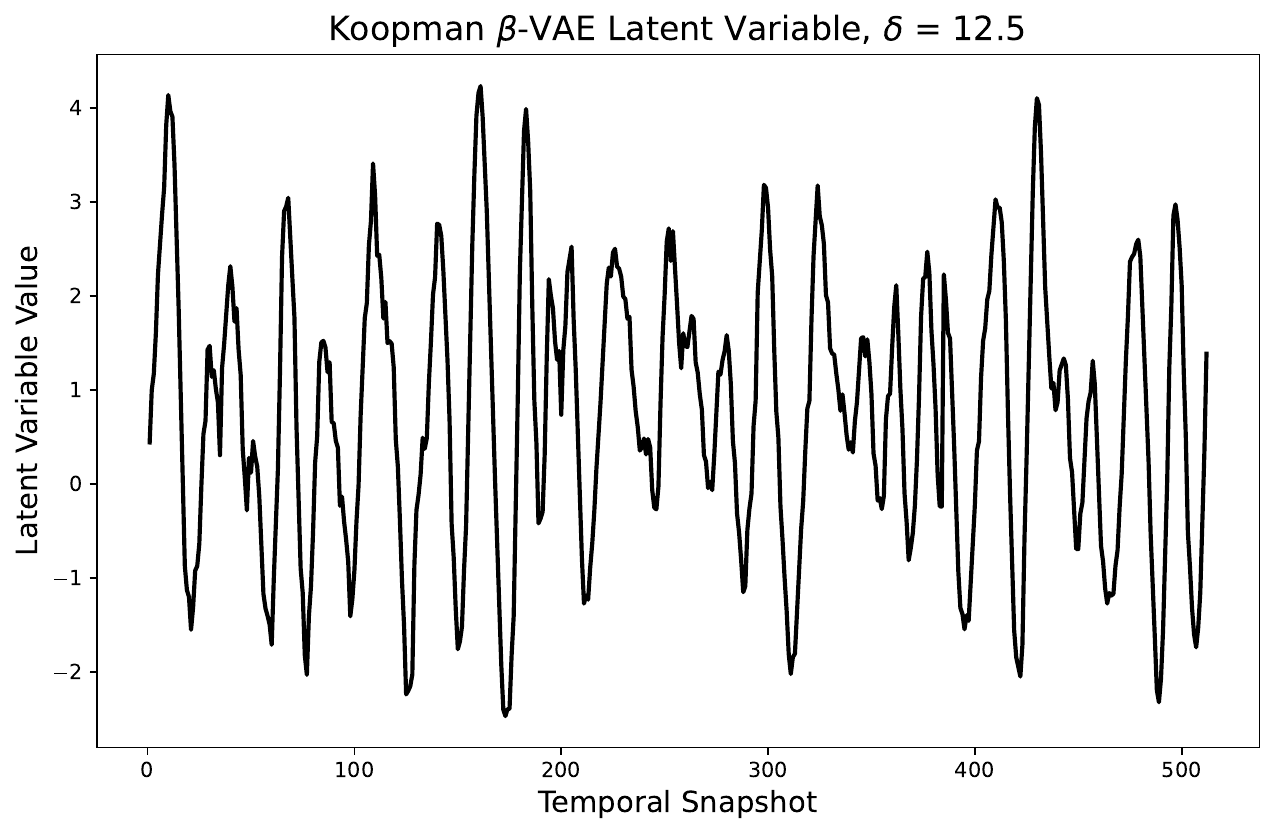}
  \includegraphics[width=0.495\textwidth]{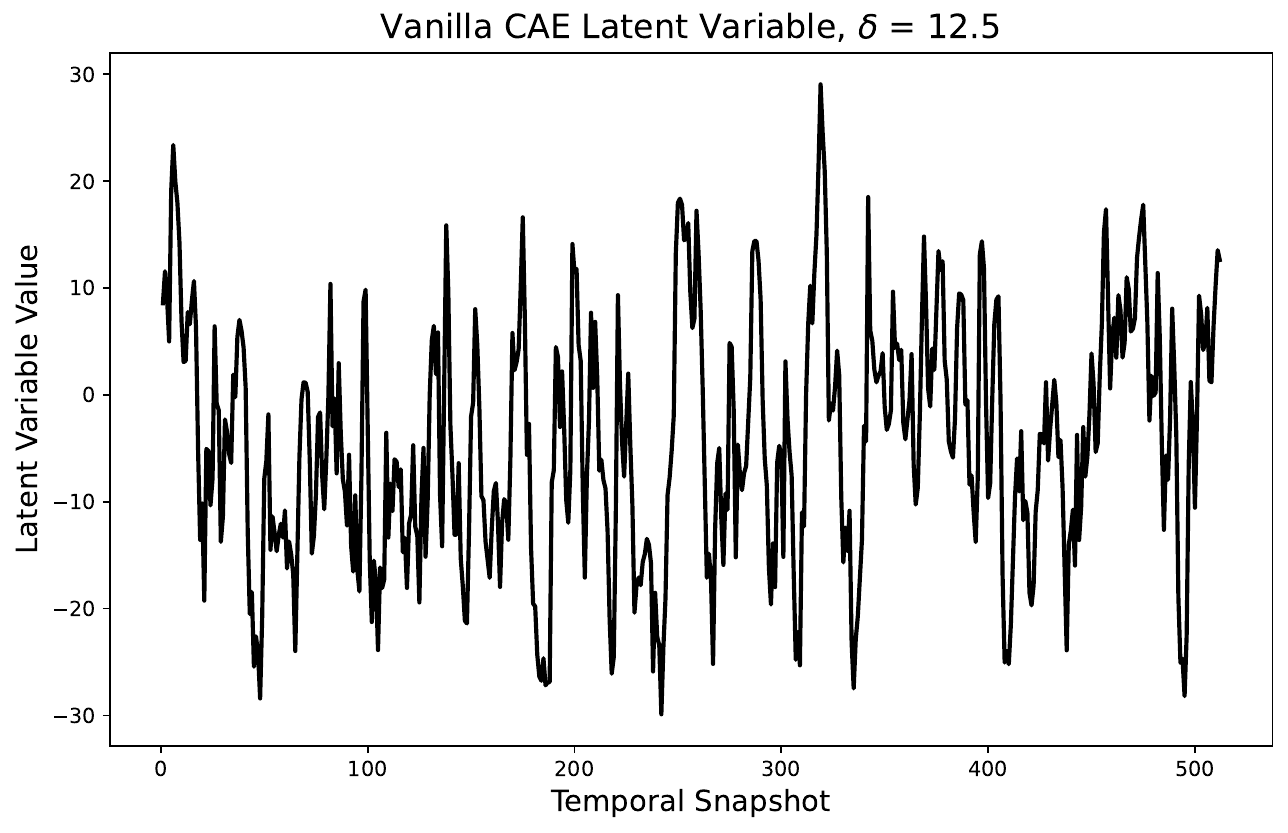}
  \caption{Latent variable comparison at $\delta = 12.5$.}
  \label{fig:latent_12.5}
\end{figure}

Figures~\ref{fig:recon_2.5}-\ref{fig:recon_12.5} show contour plots of the velocity magnitude from the raw LES input data and reconstructions from both autoencoders and all yaw angles at $t = 256$. The vanilla CAE does well at reconstructing most scales of the flow and fine-scale details are well-preserved, although not entirely. Mean flow comparisons of the training dataset and reconstructions given by both models at $\delta = 7.5$ are shown in Figure~\ref{fig:means_train}. Visually, the mean flows are very similar, although there is a loss in sharpness in both models due to the autoencoder not capturing very fine-scale features. Using the Koopman $\beta$-VAE leads to the bulk properties of the flow being retained while finer small-scale structures are largely filtered out. While this leads to lower reconstruction accuracy, the flow is easier to model, as shown by latent variable samples in Figures~\ref{fig:latent_2.5}-\ref{fig:latent_12.5}. The latent variables produced by the Koopman $\beta$-VAE exhibit a much smoother variation over time when compared to the ones produced by the vanilla CAE, which show high levels of noise and are thus difficult to model over time. The numerical range of the latent variables produced by the Koopman $\beta$-VAE remain consistent, following that of an isotropic Gaussian. As the vanilla CAE imposes no constraint on the latent variables, their magnitudes can vary significantly.

\FloatBarrier

\begin{figure}[!htpb]
  \centering
  \includegraphics[width=1.0\textwidth]{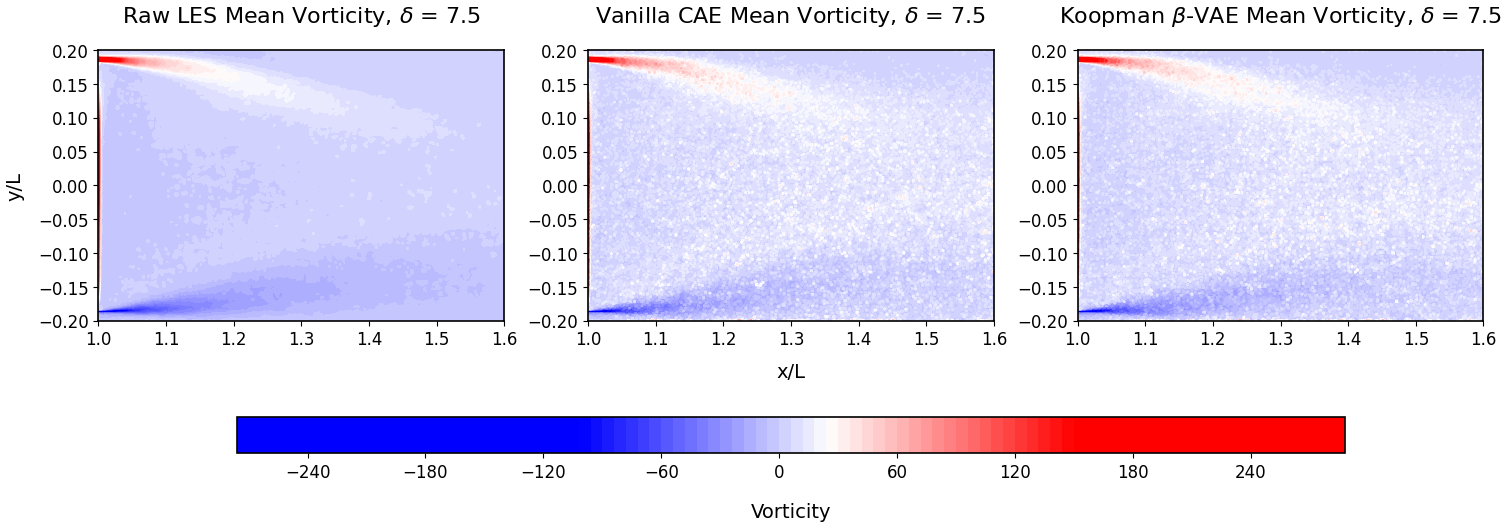}
  \caption{Time-averaged vorticity comparison for the training data at $\delta = 7.5$.}
  \label{fig:vorticity_train}
\end{figure}

To further assess the Koopman $\beta$-VAE's ability to filter out small-scale structures while retaining bulk flow information, a comparison of time-averaged vorticity plots over the training snapshots is given in Figure~\ref{fig:vorticity_train}. The instantaneous vorticity $\omega = \nabla \times (u, v)$ measures the local rotation of fluid elements. Averaged over time, the vorticity field represents regions of constant rotation; high magnitudes correspond to persistent large-scale coherent structures. The vorticity profile from the raw LES data depicts the shear layer profile of the flow, with a layer of positive vorticity on the top of the domain and one of negative vorticity at the bottom. There is also a region of high vorticity at the rear of the Windsor body, which rapidly dissipates downstream. Both models replicate this pattern well, showing that the large-scale coherent flow structures are retained in reconstructions. For both models, there are large amounts of noise present
throughout the domain. This is likely because of lack of smoothness and continuity in reconstructions due to very fine-scale information not being retained. 

\begin{figure}[!htpb]
  \centering
  \includegraphics[width=0.6\textwidth]{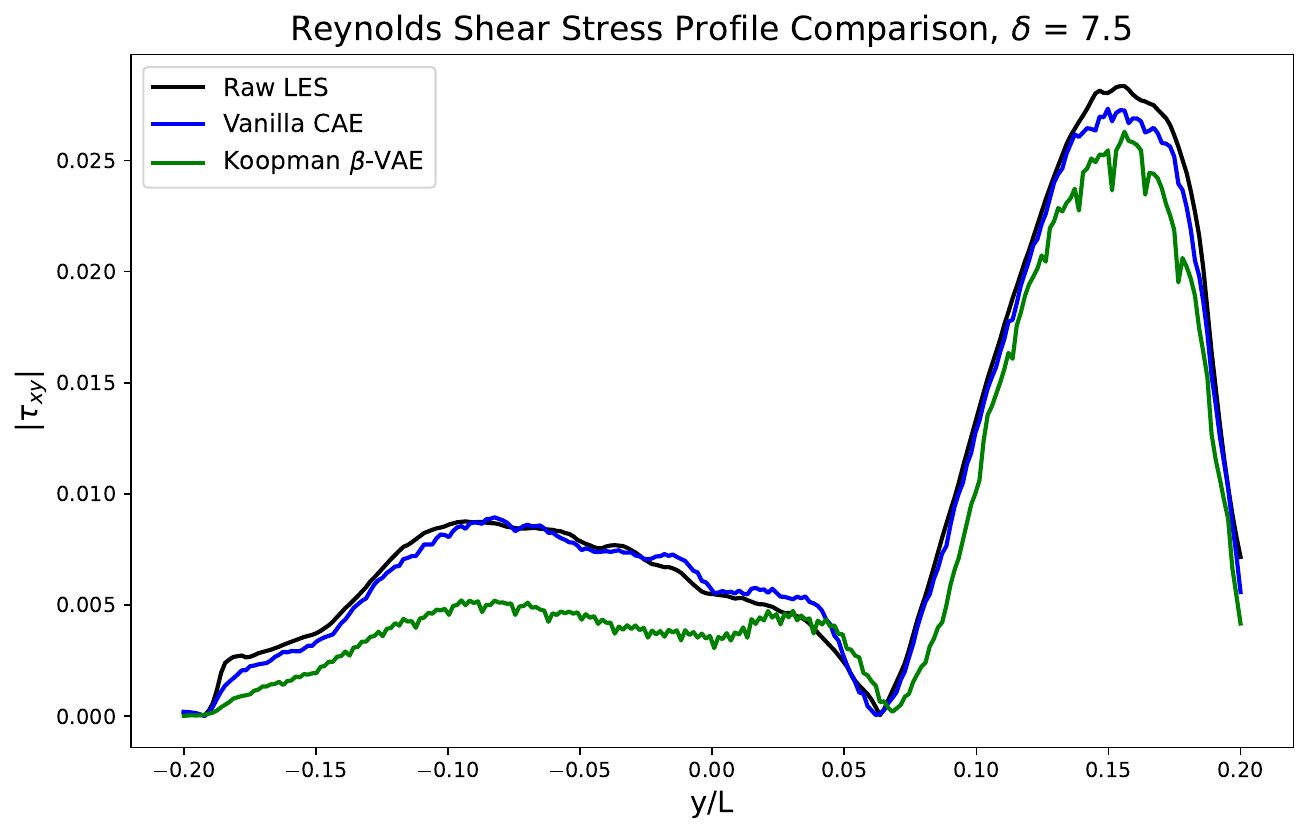}
  \caption{Reynolds shear stress comparison for the training data at $\delta = 7.5$.}
  \label{fig:tau}
\end{figure}

To measure the intensity of turbulent mixing, the Reynolds shear stress $\tau_{xy} = \overline{{u'v'}}$, where $u'$ and $v'$ are the velocity fluctuations is computed for the training data and model reconstructions. As it relates to the intensity of turbulent mixing, the Reynolds shear stress is a good indicator of how active the turbulent energy cascade is and subsequently the multi-scale nature of the flow. Figure~\ref{fig:tau} shows  $\vert \tau_{xy} \vert$ averaged over the x-axis plotted against $y/L$ at $\delta = 7.5$, where it is shown that the intensity of turbulent mixing is similar for the raw LES data and vanilla CAE reconstructions, and significantly lower for the Koopman $\beta$-VAE along the y-axis. As large-scale coherent structures are retained in reconstructions, this is due to the absence of small-scale structures, which are unable to transfer momentum to increasingly smaller scales until they are dissipated, leading to an overall decrease in turbulent mixing. Given the large decrease in turbulent kinetic energy, loss of fineness in reconstructions, smoother latent variables, and lower Reynolds shear stress values, it is shown that the Koopman $\beta$-VAE effectively filters out small-scale turbulent structures from input data. Time-averaged vorticity plots show that large-scale coherent structures are retained in reconstructions, implying that the decrease in reconstruction accuracy is mostly due to small-scale structures being filtered. Additionally, the model can act on multiple datasets simultaneously, a notable advantage over methods like DMD.

\begin{figure}[!htpb]
  \centering
  \includegraphics[width=0.495\textwidth]{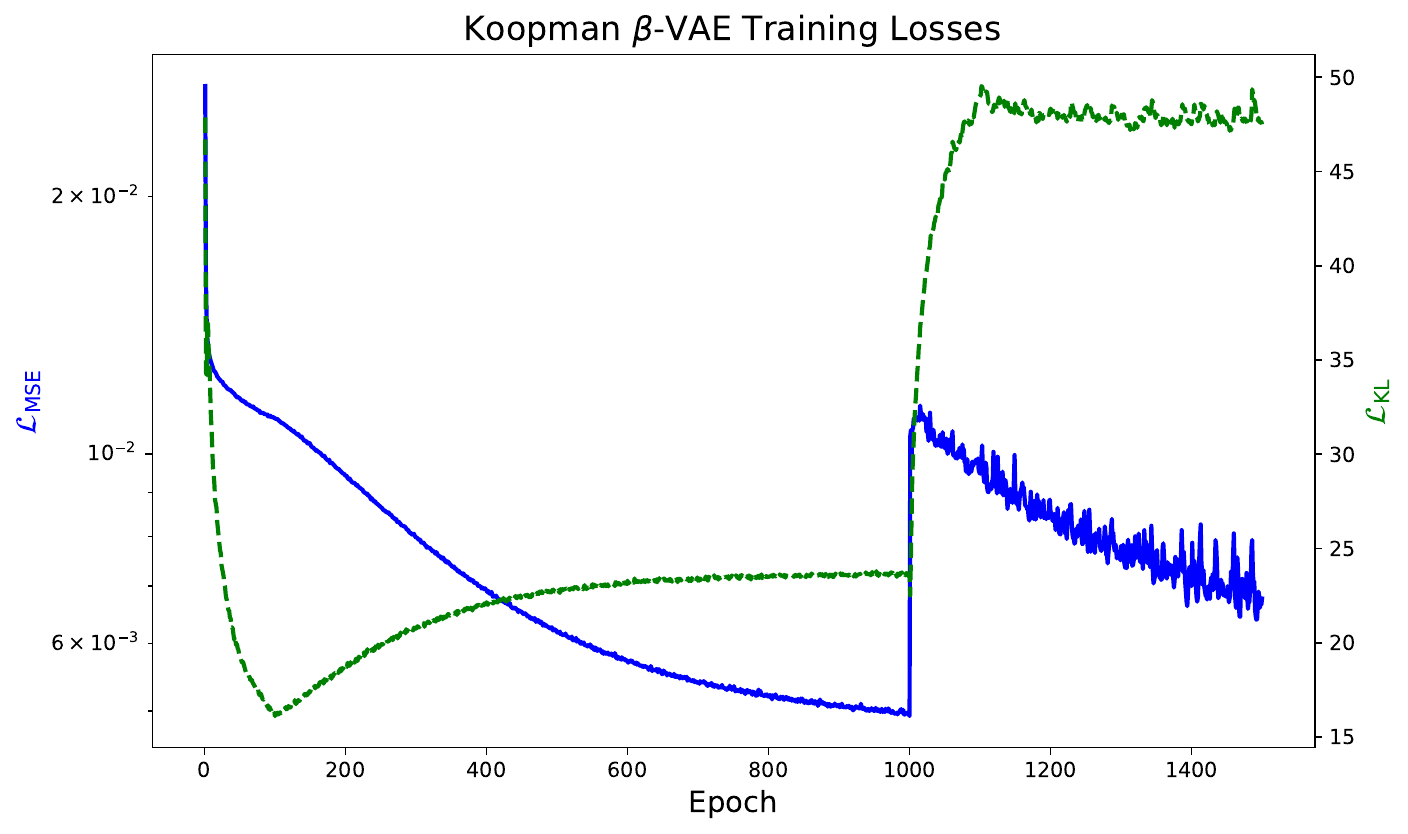}
  \includegraphics[width=0.495\textwidth]{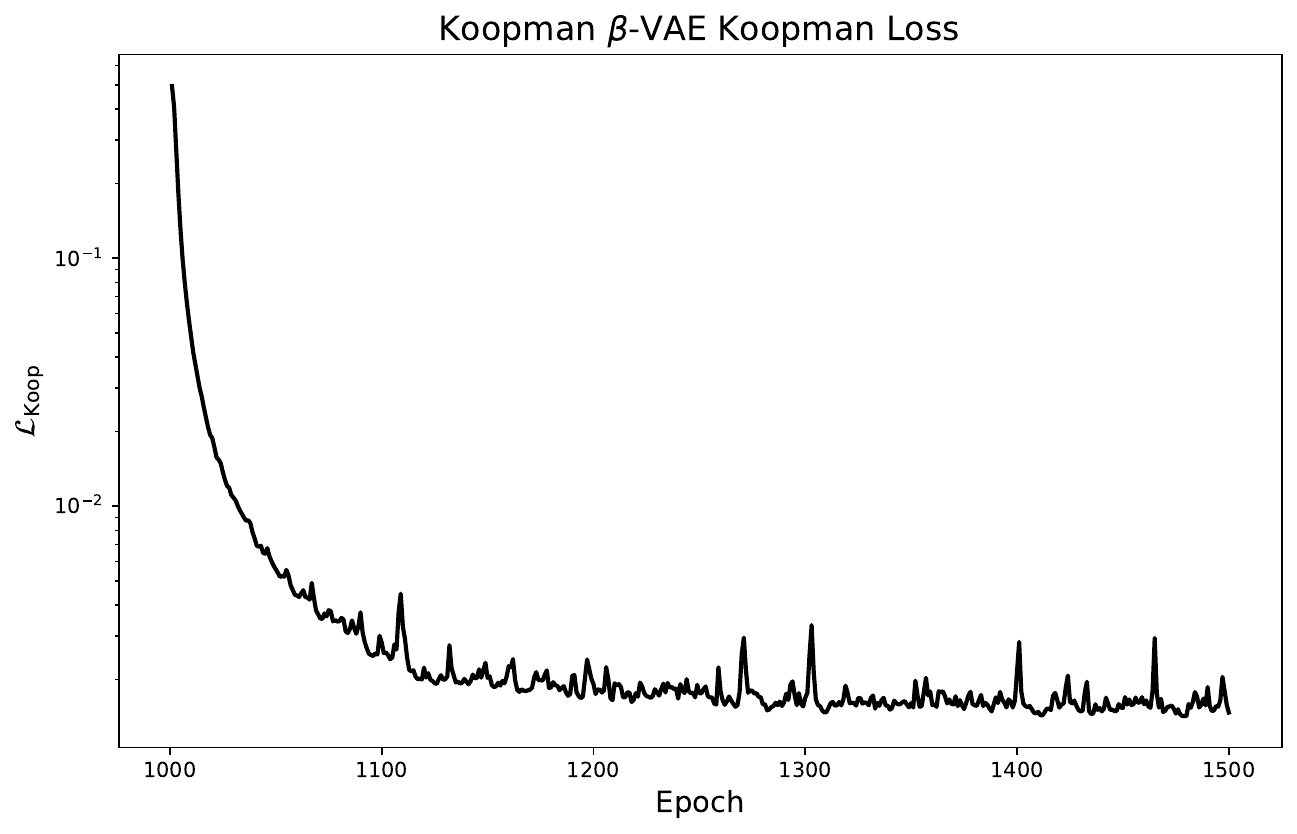}
  \caption{Koopman $\beta$-VAE training losses.}
  \label{fig:losses}
\end{figure}

The individual components of the training loss are shown in Figure~\ref{fig:losses}. $\mathcal{L}_{\text{MSE}}$ exhibits a monotonic decrease until $\mathcal{L}_{\text{Koop}}$ is activated at $e = 1000$, after which it exhibits a sharp increase until it gradually decreases over time, although not to as low as it previously was due to the exclusion of small-scale fluctuations. Although $\mathcal{L}_{\text{MSE}}$ continues to drop during both training stages, the number of epochs is not increased as this leads to lower predictive performance on the test data due to overfitting. $\mathcal{L}_{\text{KL}}$ initially decreases rapidly, until gradually increasing until reaching a near-constant value before $\mathcal{L}_{\text{Koop}}$ is activated. Initially, as $\beta$ increases, the model focuses on reducing $\mathcal{L}_{\text{KL}}$ to regularize the latent space. Once $\beta$ reaches its maximum value and remains constant, the decoder starts to utilize latent information to focus on improving reconstructions, causing $\mathcal{L}_{\text{KL}}$ to slowly increase. After $\mathcal{L}_{\text{Koop}}$ is activated, $\mathcal{L}_{\text{KL}}$ exhibits a sharp increase, after which it slightly fluctuates. This is similar to the behavior of $\mathcal{L}_{\text{Koop}}$, which first decreases rapidly until exhibiting small fluctuations. Due to the implementation of the Koopman operator $\bm{A}$, the latent variables exhibit dependence on each other and are no longer independent of each other after the pre-training stage. However, the marginal distributions are still approximately Gaussian which allows the latent variables to be similar in magnitude. When using the Koopman operator with a vanilla CAE, where magnitudes vary, denoising is not achieved, as shown in Appendix~\ref{appendix:cae}.

\subsection{ROM Results}

\begin{figure}[!htpb]
  \centering
  \includegraphics[width=0.7\textwidth]{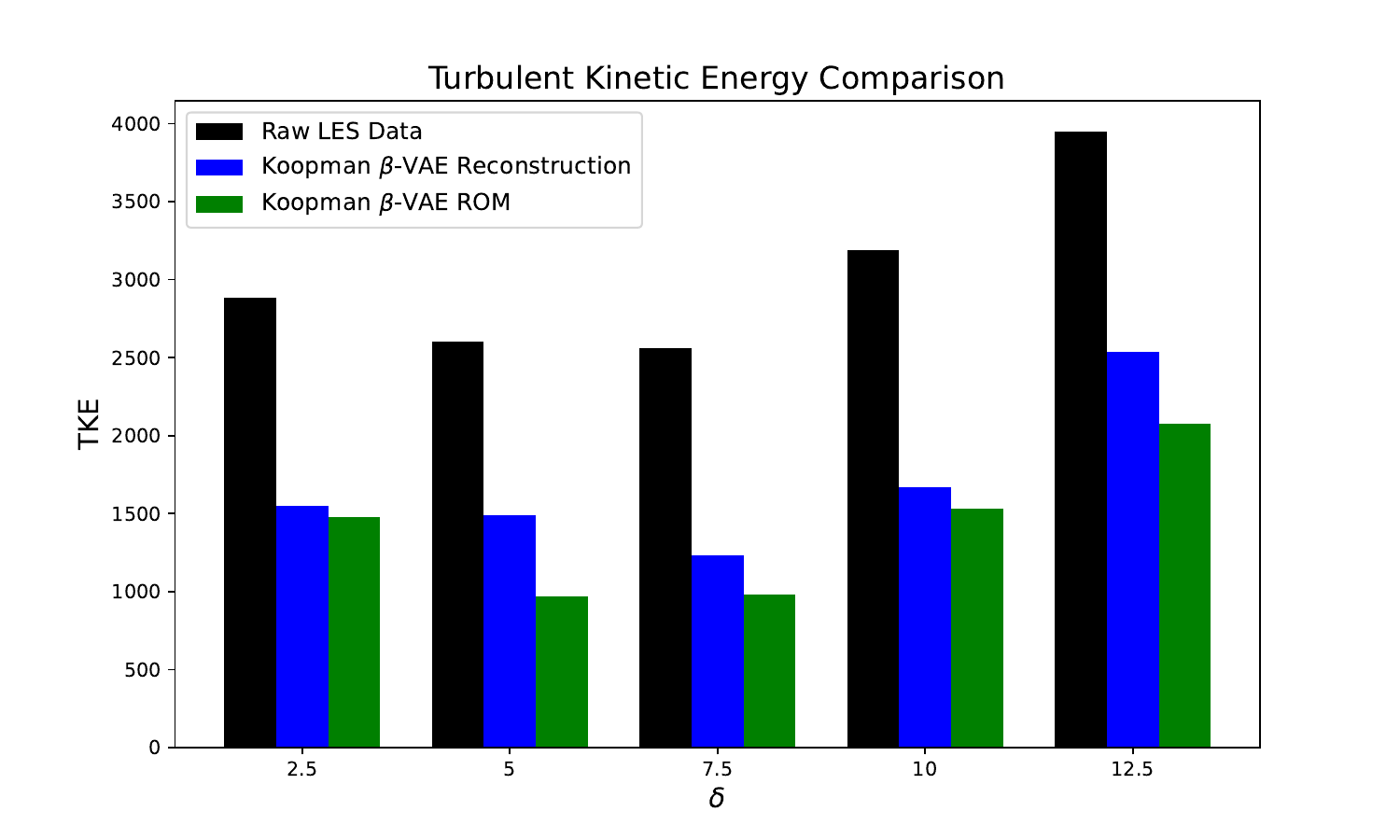}
  \caption{Turbulent kinetic energy comparison for the test dataset.}
  \label{fig:tke_test}
\end{figure}

\begin{figure}[!htpb]
  \centering
  \includegraphics[width=0.7\textwidth]{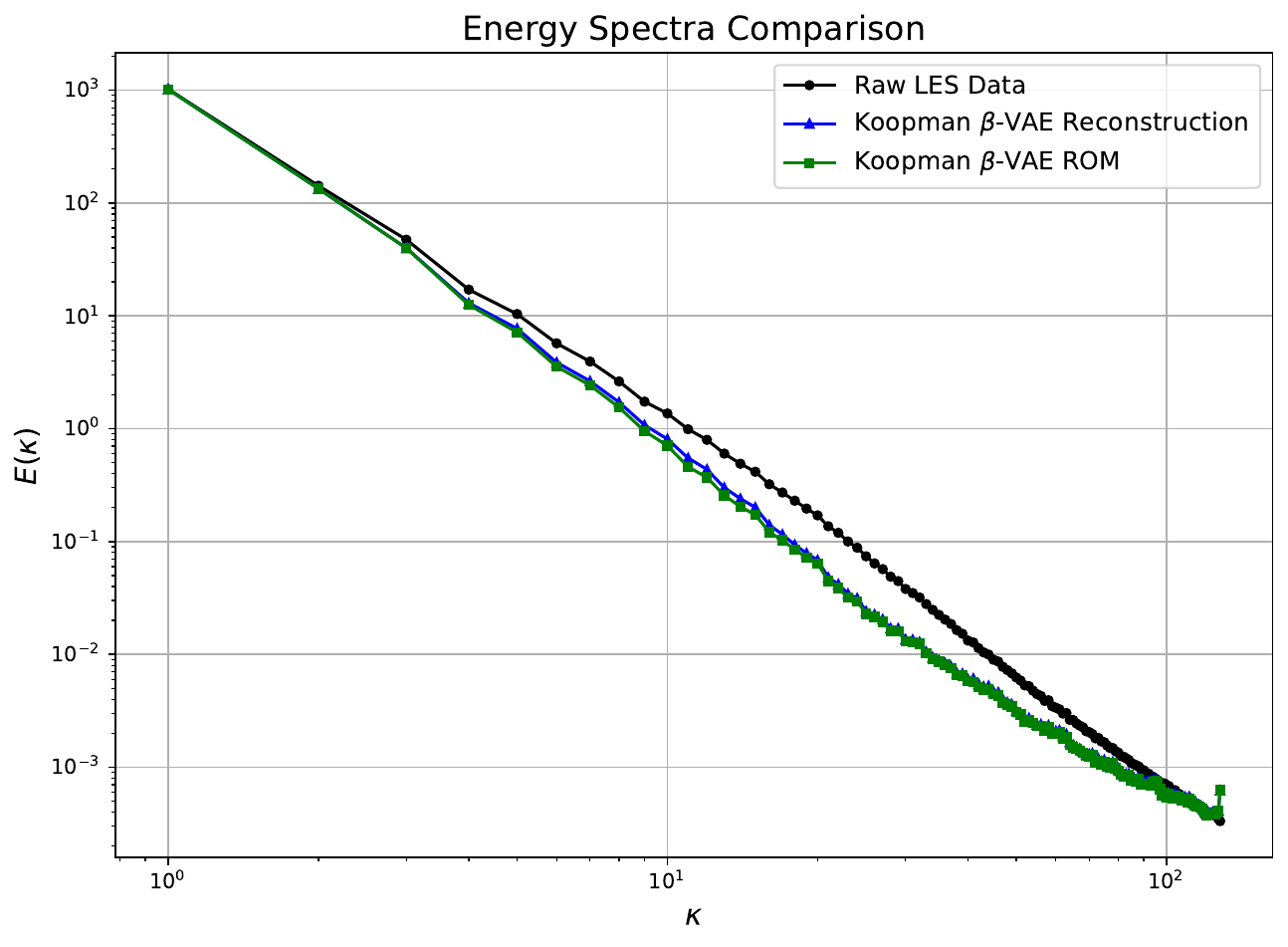}
  \caption{Energy spectra comparison of the test dataset.}
  \label{fig:fft_test}
\end{figure}

\begin{figure}[!htpb]
  \centering
  \includegraphics[width=0.7\textwidth]{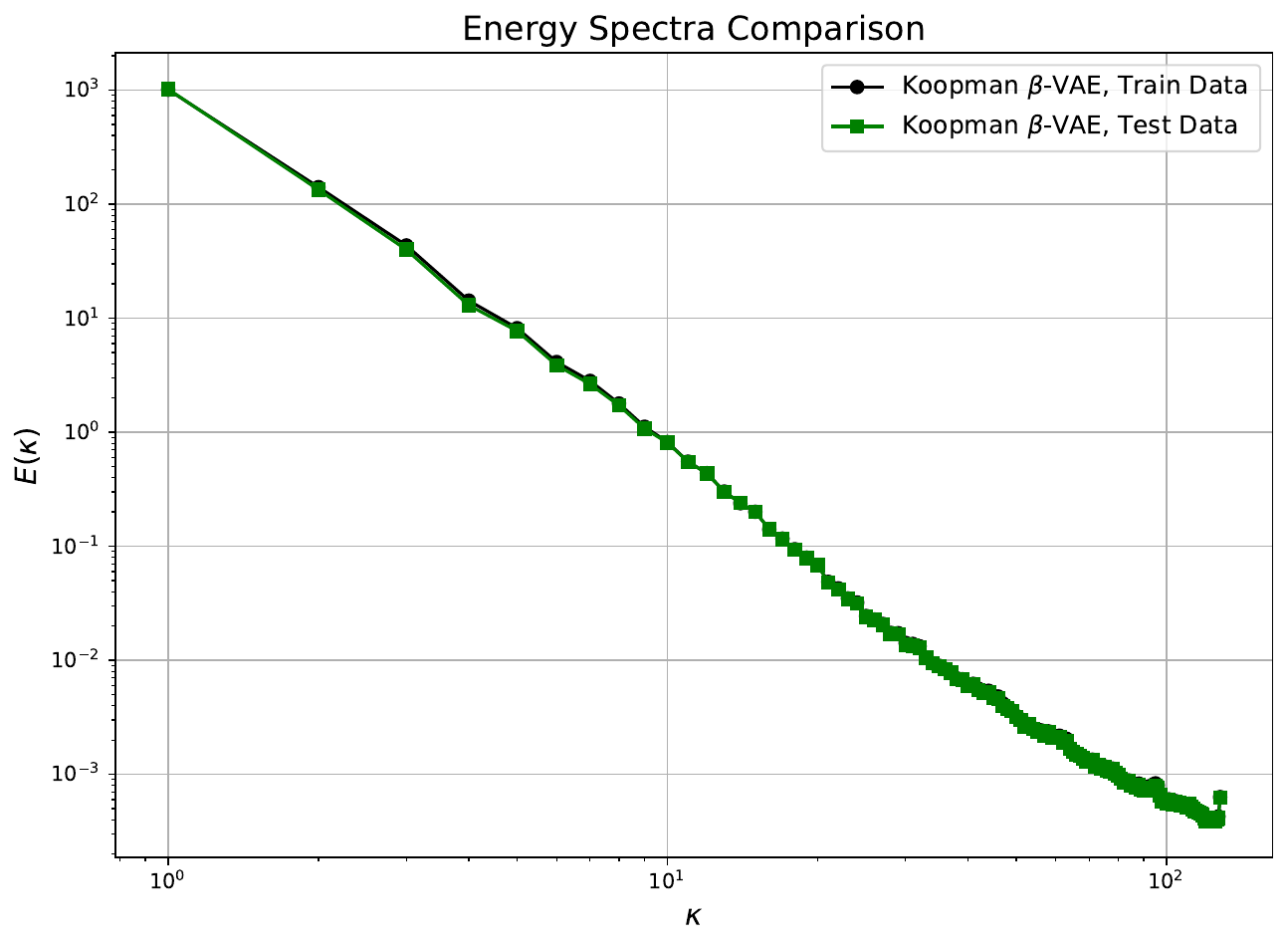}
  \caption{Energy spectra comparison between training and test data reconstructions.}
  \label{fig:fft_diff}
\end{figure}

Table~\ref{table:rom} shows the overall and component-wise predictive accuracy of both the reconstructions and ROM averaged over all of the test data. The errors are significantly higher when compared to the training data, which is expected as machine learning models tend to degrade in performance when extrapolating, especially for highly non-linear data. A similar result is shown in a work by Solera-Rico et al.~\cite{solera2024beta} using $\beta$-VAEs for ROMs of unsteady fluid dynamics. Again, the spanwise errors are significantly greater than the streamwise ones, due to them corresponding to small-scale fluctuations. However, the errors in the streamwise velocity, which is representative of the bulk flow properties, are reasonable. The ROM produces errors that are close to that of the reconstructions, showing that the LSTM ensemble can reasonably predict the evolution of the latent variables over time with a good degree of stability. The turbulent kinetic energy for the raw data averaged over all yaw angles is again equal to $3.04 \times 10^3$; the model reconstructions retain 55.6\% of this. Relative to the TKE of the model training data reconstructions, 83.2\% is retained. The ROM predictions exhibit less turbulent kinetic energy due to additional prediction errors, retaining approximately 44.6\% of the ground truth value and 69\% of the value from the model training data reconstructions. An angle-wise comparison for the test data is given in Figure~\ref{fig:tke_test}; at each angle, the ROM predictions retain less $TKE$ than the reconstructions due to increased error. Figure~\ref{fig:fft_test} shows an energy spectra comparison between the raw test data, model reconstructions, and ROM predictions. The energy spectrum for the test data reconstructions is very similar to that of the ROM predictions, with slight increases at intermediate scales. The energy spectra of the training and test data reconstructions from the Koopman $\beta$-VAE are almost identical, as shown in Figure~\ref{fig:fft_diff}, with very small differences at the smallest wavenumbers corresponding to the largest flow scales. As both test data reconstructions and ROM predictions retain the spectral properties of the training data reconstructions, this suggests that large-scale coherent structures are retained and that the decrease in turbulent kinetic energy arises due to the presence of point-wise errors that decrease the amplitudes of the present structures. While test data reconstruction errors are due to extrapolations of highly non-linear data, additional errors are introduced in ROM predictions from time-series predictions of latent variables.

\begin{figure}[!htpb]
\centering
\includegraphics[width=0.84\textwidth]{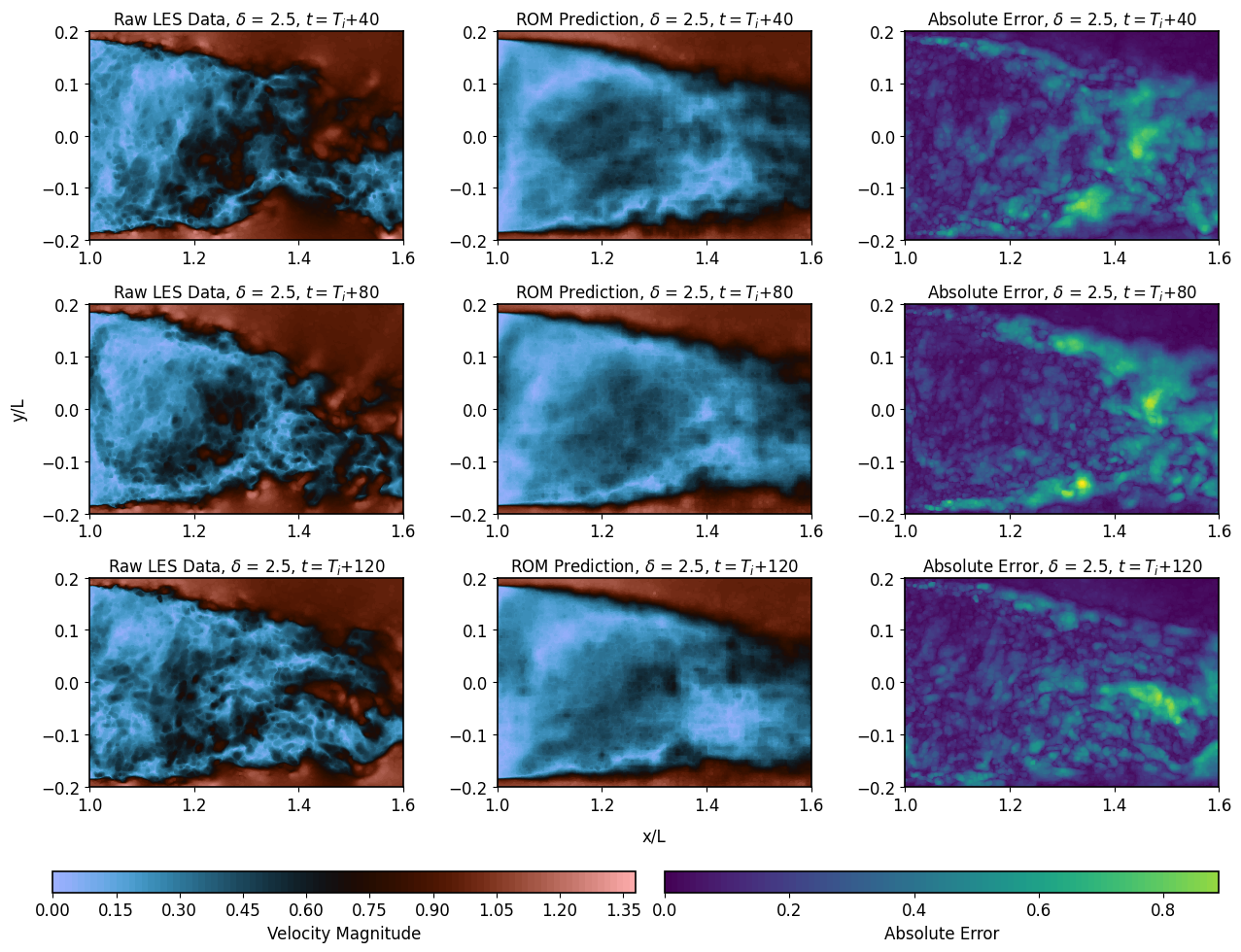}
\caption{ROM predictions and absolute errors at $\delta = 2.5$.}
\label{fig:rom_2.5}
\end{figure}

\begin{figure}[!htpb]
\centering
\includegraphics[width=0.84\textwidth]{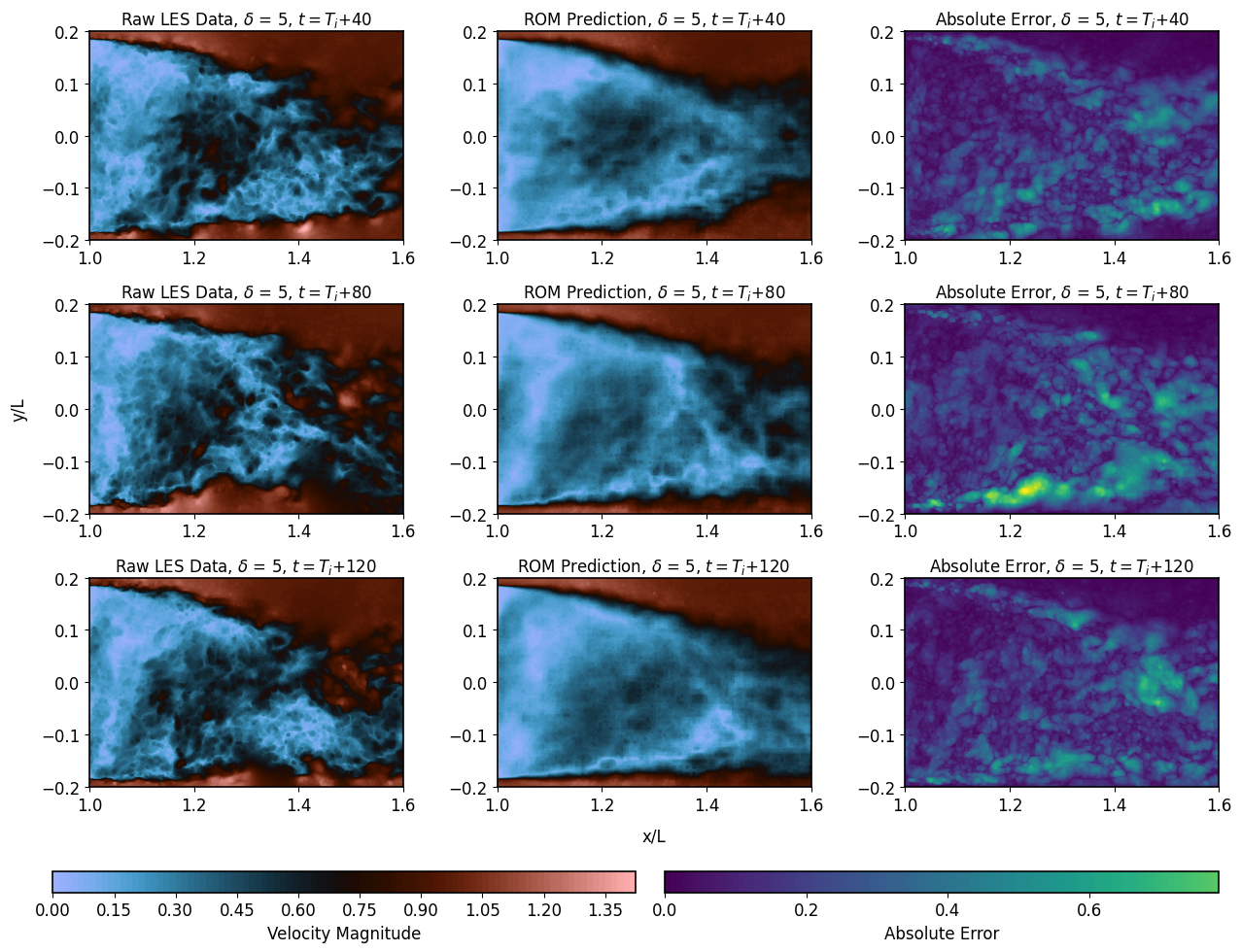}
\caption{ROM predictions and absolute errors at $\delta = 5$.}
\label{fig:rom_5}
\end{figure}

\begin{figure}[!htpb]
\centering
\includegraphics[width=0.84\textwidth]{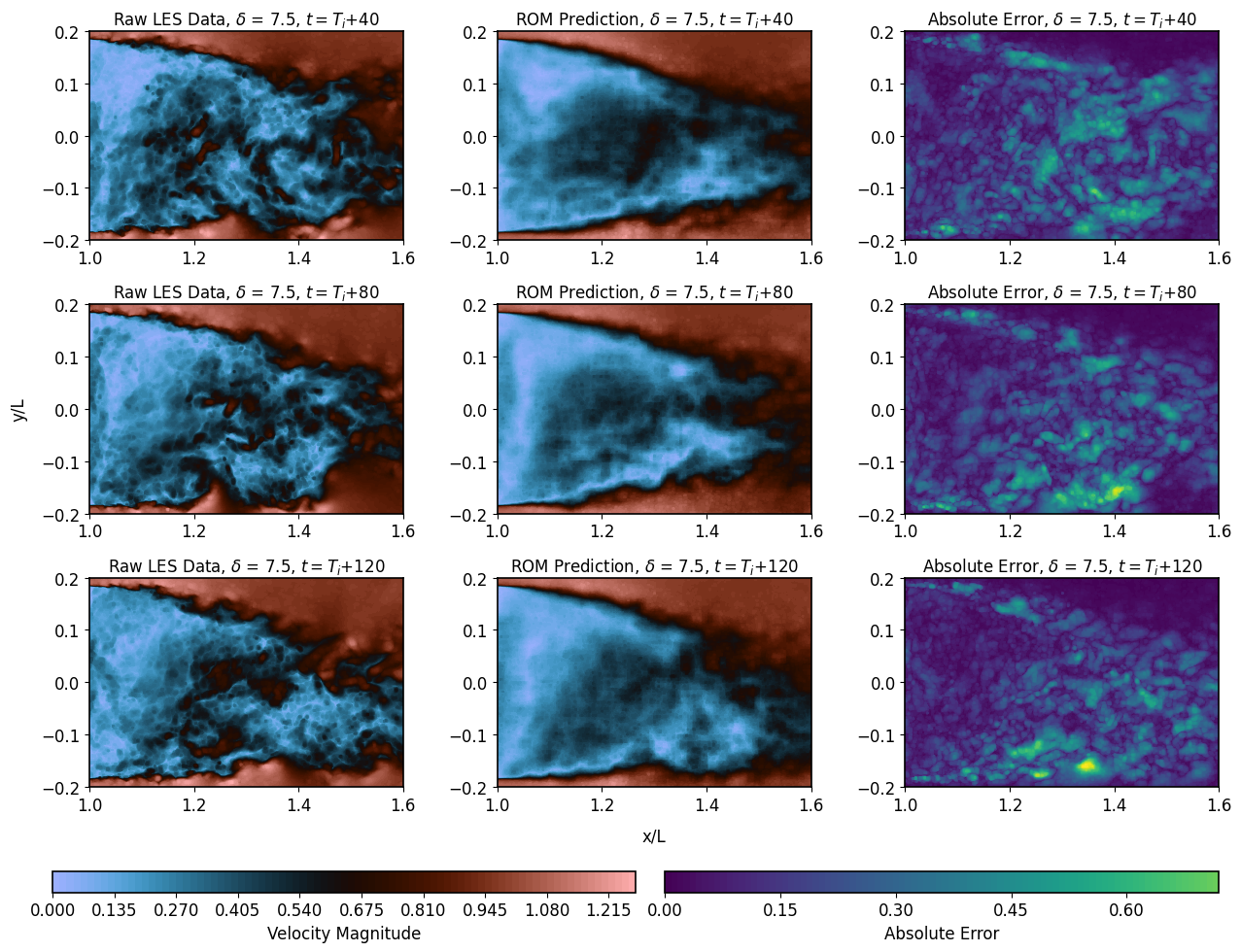}
\caption{ROM predictions and absolute errors at $\delta = 7.5$.}
\label{fig:rom_7.5}
\end{figure}

\begin{figure}[!htpb]
\centering
\includegraphics[width=0.84\textwidth]{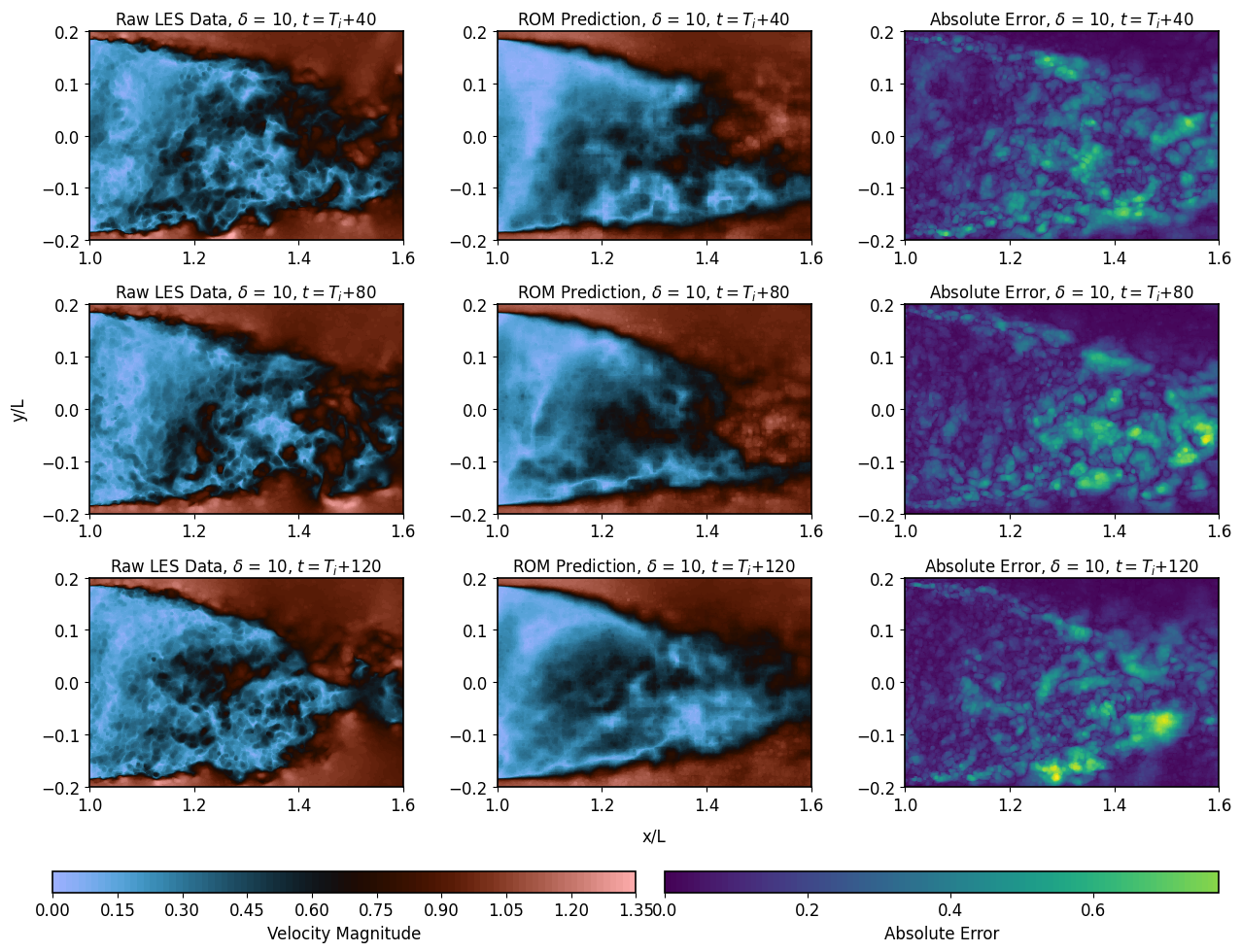}
\caption{ROM predictions and absolute errors at $\delta = 10$.}
\label{fig:rom_10}
\end{figure}

\begin{figure}[!htpb]
\centering
\includegraphics[width=0.84\textwidth]{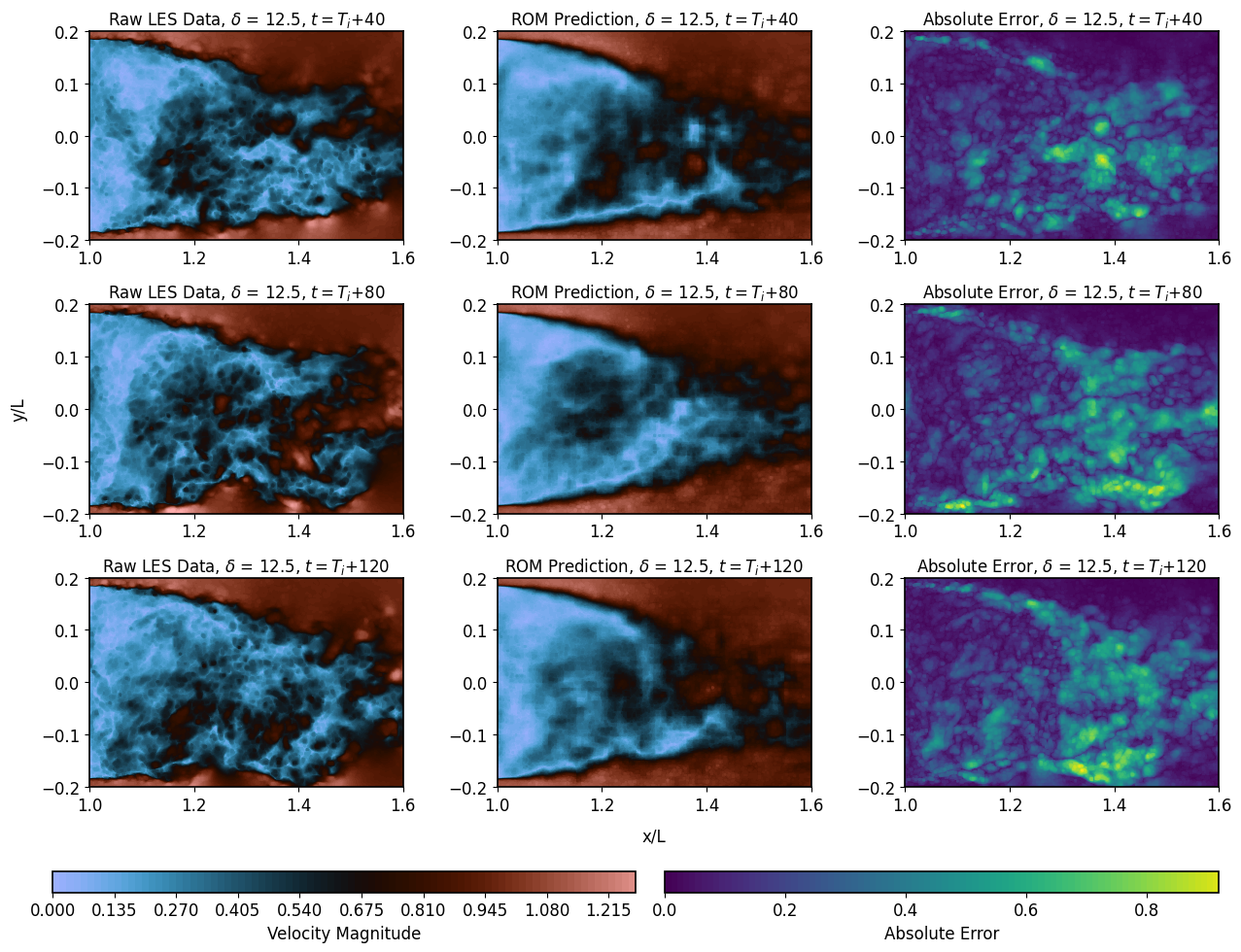}
\caption{ROM predictions and absolute errors at $\delta = 12.5$.}
\label{fig:rom_12.5}
\end{figure}

\begin{figure}[!htpb]
  \centering
  \includegraphics[width=1.0\textwidth]{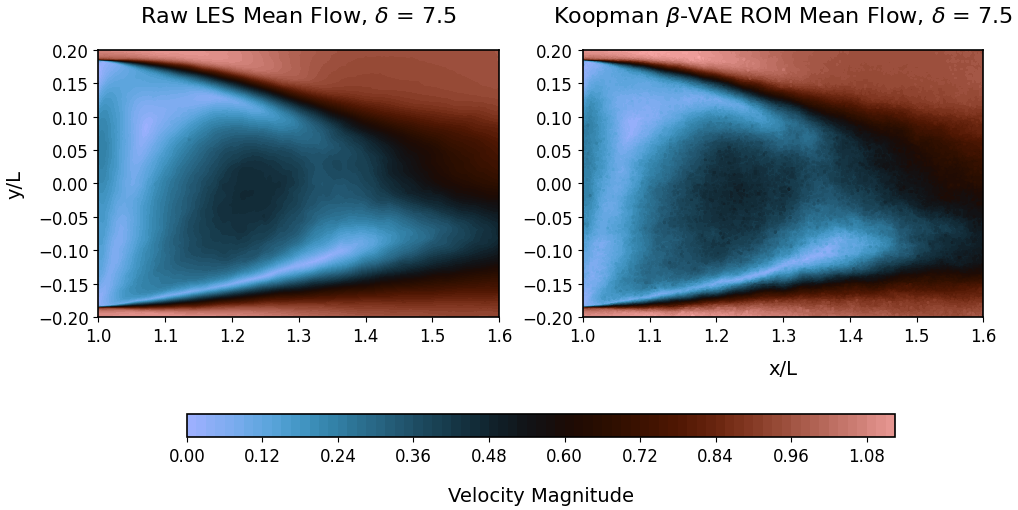}
  \caption{Mean flow comparison between the raw test data and ROM predictions at $\delta$ = 7.5.}
  \label{fig:means_rom}
\end{figure}

Figures~\ref{fig:rom_2.5}-\ref{fig:rom_12.5} show velocity magnitude contours of the raw LES data, ROM predictions, and absolute errors at each yaw angle at $t = T_i + [40, 80, 120]$. The errors are highest at $\delta = [2.5, 12.5]$, which is expected as they are at the extremes of the training data range. At all angles and prediction points, errors are concentrated in the shear layers and near-wake region. Both areas are highly chaotic, especially the near-wake region where both shear layers mix and small-scale structures dominate. The errors don't exhibit a growing trend over time, highlighting the stability of the ROM's predictions which can be attributed to the use of an ensemble model. While the ROM predictions visually differ greatly from the ground truth LES data, the intended goal of the ROM framework is to accurately capture the bulk properties of the flow corresponding to large-scale structures over unseen time horizons in real-time. A comparison between the mean flow from the raw LES data and ROM predictions over the test time horizon is given in Figure~\ref{fig:means_rom}. The mean flow from the ROM predictions closely matches that of the test data, although there is a lack of sharpness and smoothness due to fine-scale information not being retained in reconstructions.

\begin{figure}[!htpb]
  \centering
  \includegraphics[width=0.6\textwidth]{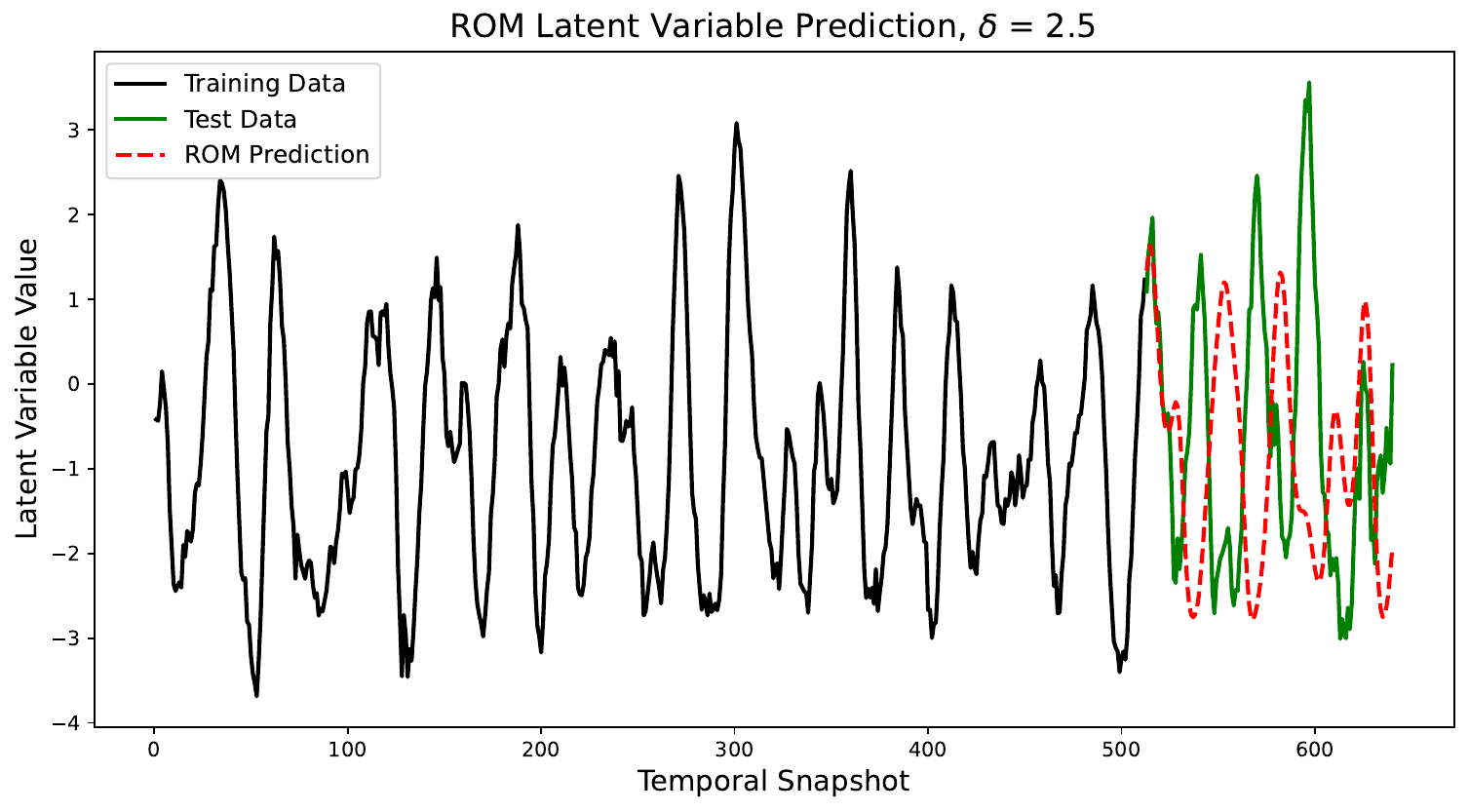}
  \caption{ROM latent variable prediction at $\delta = 2.5$.}
  \label{fig:lat_2.5}
  \end{figure}

\begin{figure}[!htpb]
    \centering
    \includegraphics[width=0.6\textwidth]{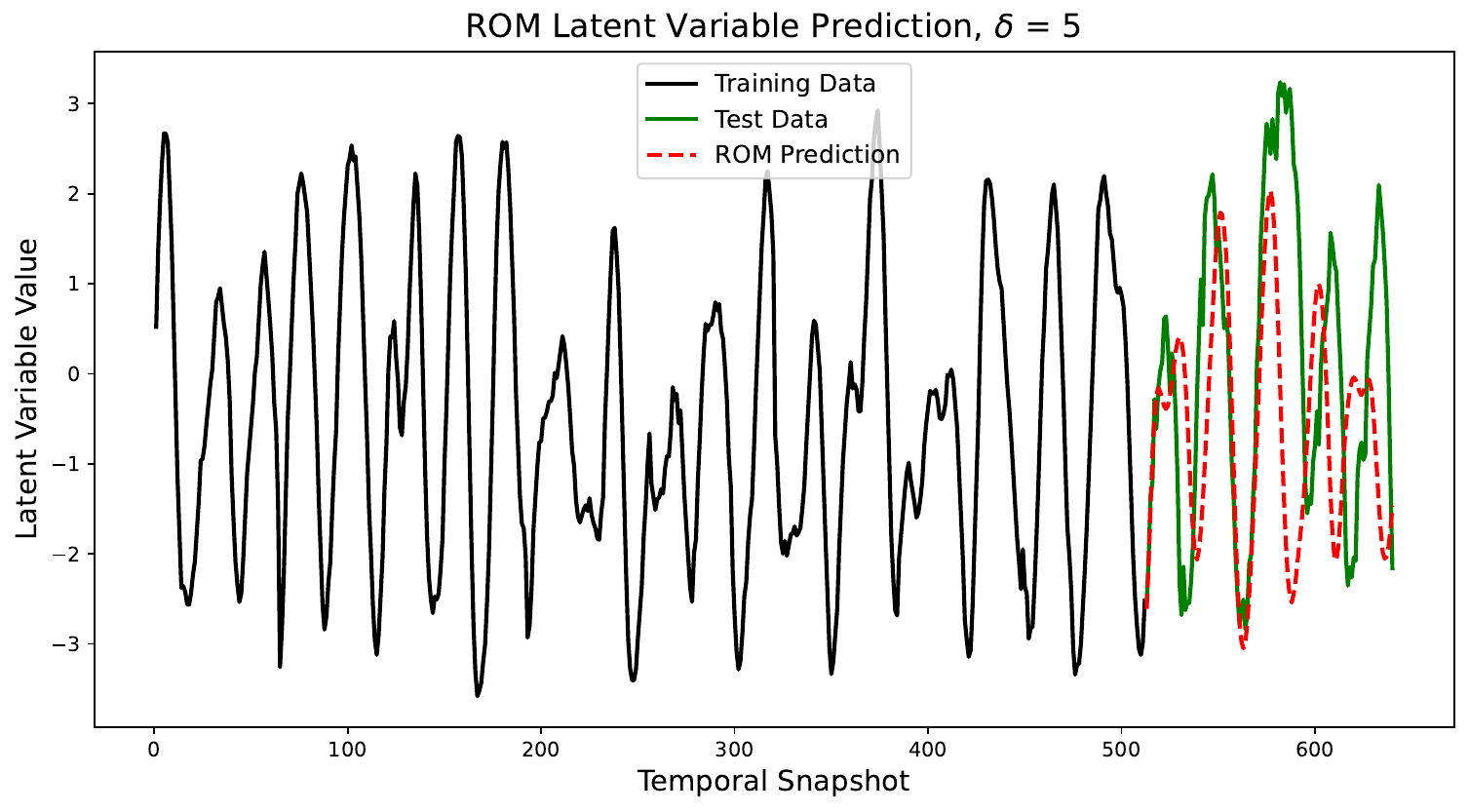}
    \caption{ROM latent variable prediction at $\delta = 5$.}
    \label{fig:lat_5}
    \end{figure}

\begin{figure}[!htpb]
  \centering
  \includegraphics[width=0.6\textwidth]{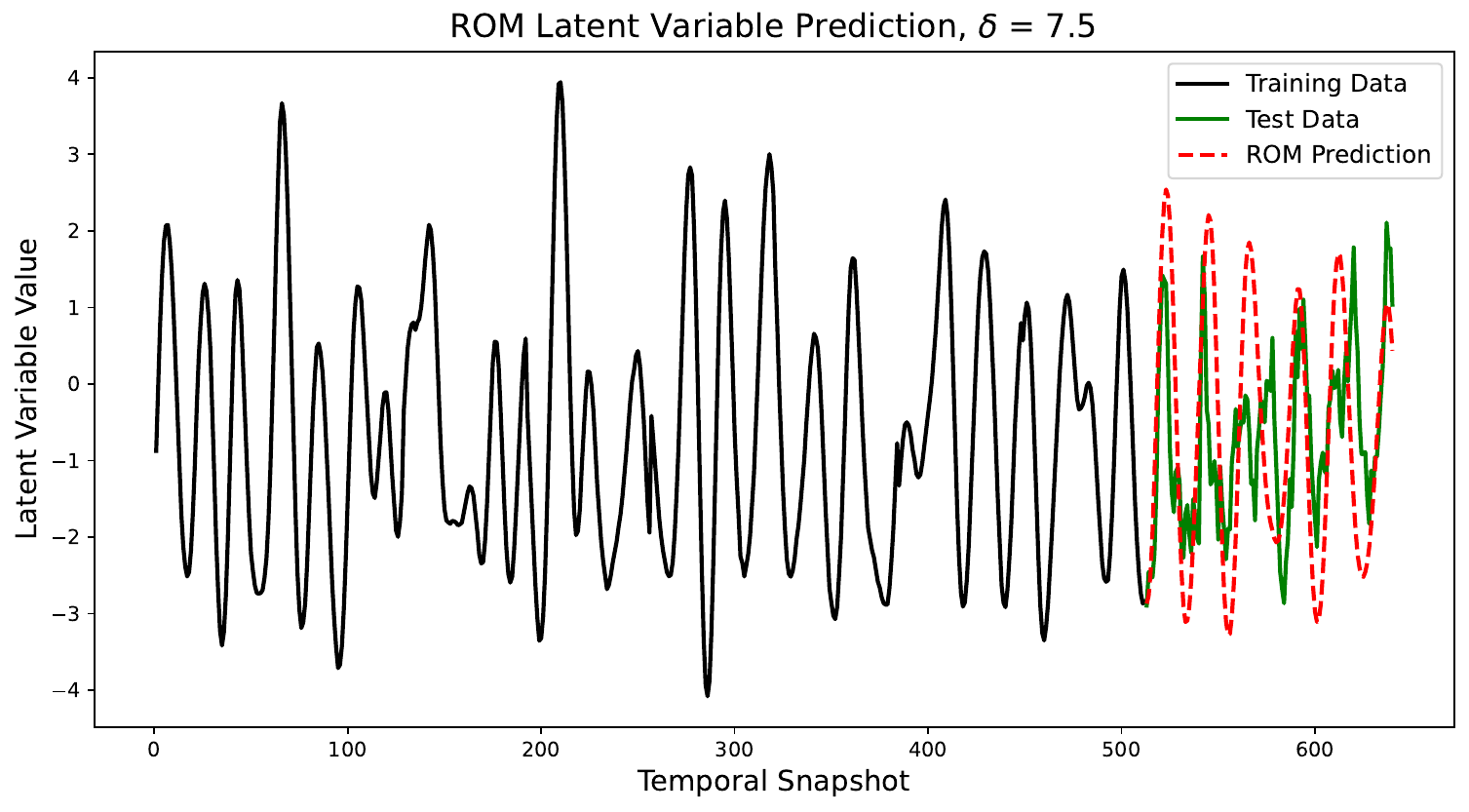}
  \caption{ROM latent variable prediction at $\delta = 7.5$.}
  \label{fig:lat_7.5}
  \end{figure}

\begin{figure}[!htpb]
    \centering
    \includegraphics[width=0.6\textwidth]{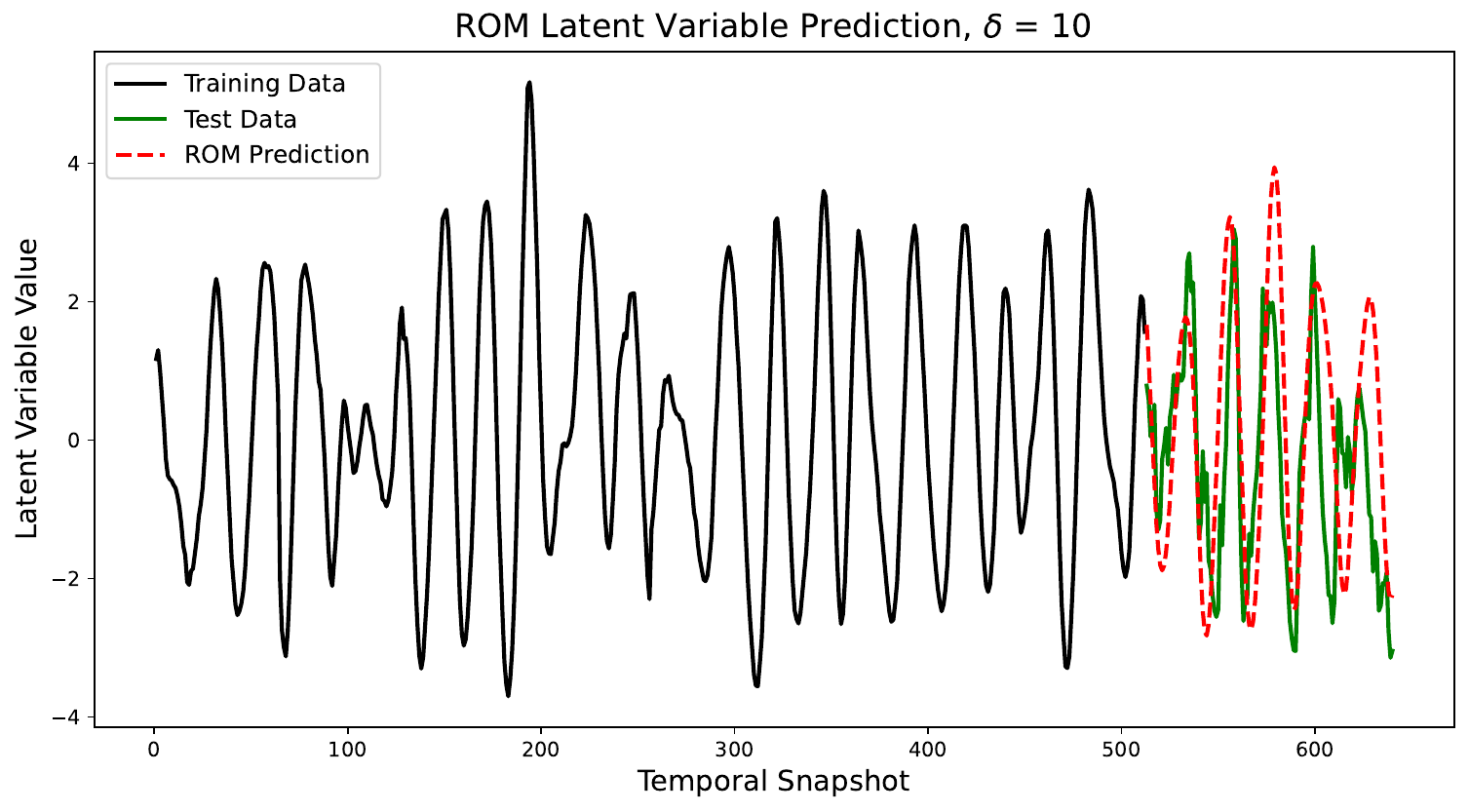}
    \caption{ROM latent variable prediction at $\delta = 10$.}
    \label{fig:lat_10}
    \end{figure}

\begin{figure}[!htpb]
  \centering
  \includegraphics[width=0.6\textwidth]{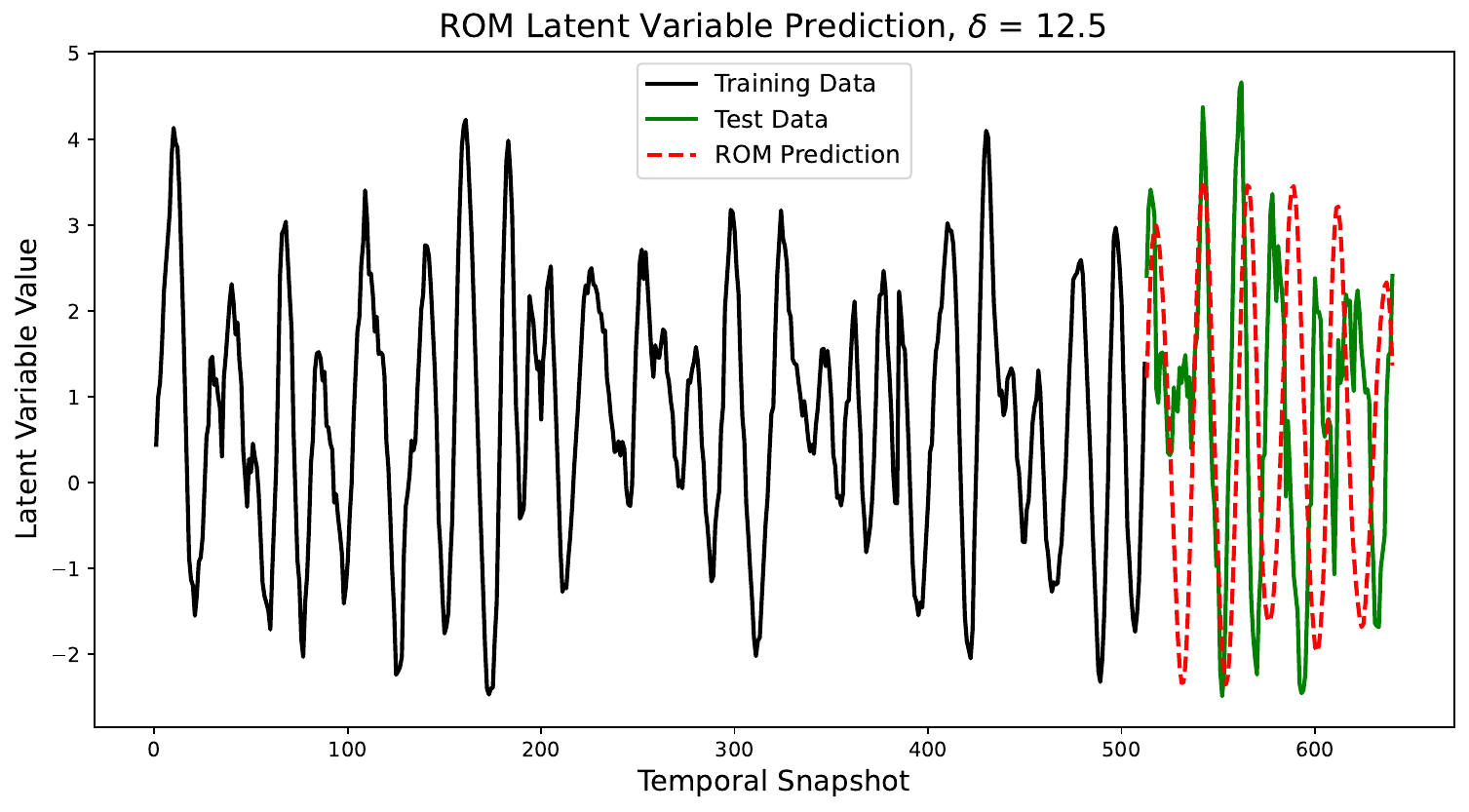}
  \caption{ROM latent variable prediction at $\delta = 12.5$.}
  \label{fig:lat_12.5}
  \end{figure}

Figures~\ref{fig:lat_2.5}-\ref{fig:lat_12.5} contain latent variables from both reconstructions and ROM predictions of the test data at each yaw angle. While the latent variables of the test data are denoised, it is not to the same extent of the training data, which had the Koopman operator directly applied to them. The ROM predicts the overall pattern of the latent variables well, with peaks and troughs being effectively tracked. The predictions at $\delta = [5, 7.5, 10]$ are most accurate, where the patterns of the test data are very well matched. The latent variables are better predicted initially after which there is some divergence from the test data, although this doesn't grow drastically. Again, the predictions are less accurate at $\delta = [2.5, 12.5]$, where the latent variable frequencies vary greatly from the other angles. As an ensemble model was used, the latent variable predictions are largely non-sensitive to the initial weights and biases of the LSTM model and the choice of seed has minimal impact on the predictions. 
\begin{table}[!htbp]
  \centering
  \begin{tabular}{|l|l|l|l|l|l|}
  \hline
  \textbf{Prediction} & \textbf{$\epsilon$} & \textbf{$\epsilon_u$} & \textbf{$\epsilon_v$} &  \textbf{$TKE$} \\ \hline
  Reconstruction    &          0.392              & 0.286               & 0.704         & 1.70e3                       \\ \hline
  ROM    & 0.442                      & 0.323                   & 0.787    & 1.41e3                                       \\ \hline
  \end{tabular}
  \caption{Prediction error and turbulent kinetic energy comparison for test data for the Koopman $\beta$-VAE.}
  \label{table:rom}  
\end{table}

\FloatBarrier

\begin{figure}[!htpb]
  \centering
  \includegraphics[width=1.0\textwidth]{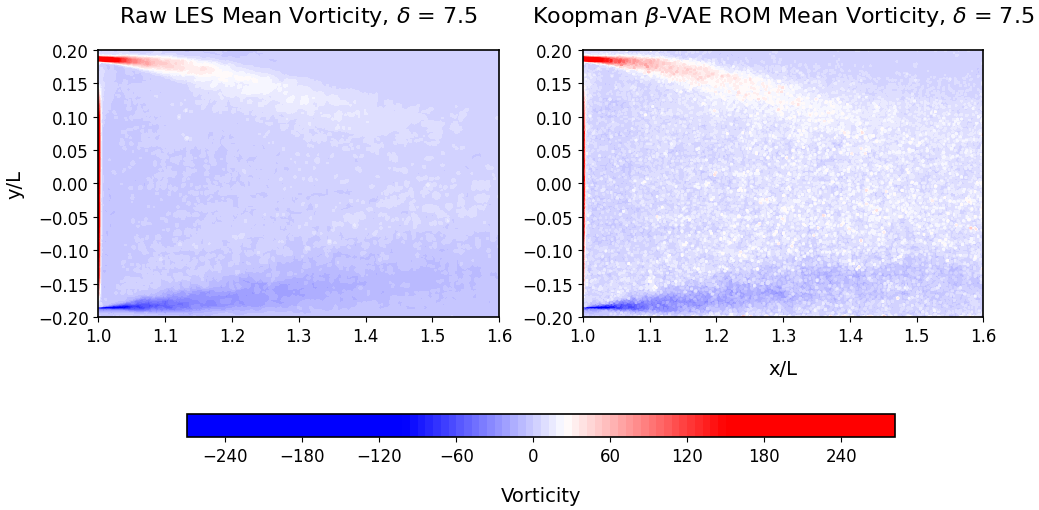}
  \caption{Time-averaged vorticity comparison for the test data at $\delta = 7.5$.}
  \label{fig:vorticity_rom}
\end{figure}

\begin{figure}[!htpb]
  \centering
  \includegraphics[width=0.6\textwidth]{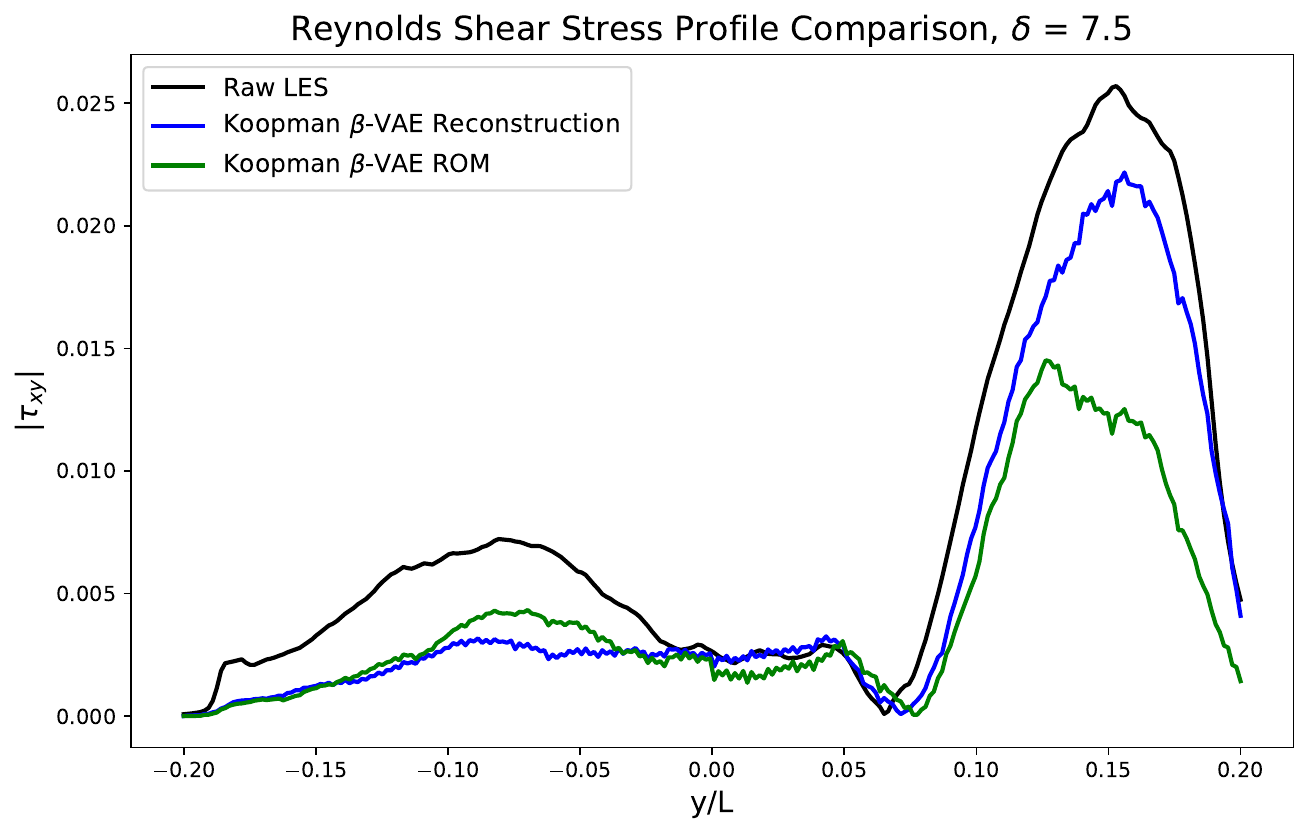}
  \caption{Reynolds shear stress comparison for the test data at $\delta = 7.5$.}
  \label{fig:tau_rom}
\end{figure}

Time-averaged vorticity contours for the raw LES test data and ROM predictions are given in Figure~\ref{fig:vorticity_rom}. The shear layer profile and high rear end vorticity, corresponding to persistent large-scale structures, are retained in the ROM predictions. Again, there is a fair amount of noise present in the ROM predictions due to a lack of smoothness and fine-scale information in the reconstructions. A comparison of the Reynolds shear stress profiles averaged over the x-axis between the raw test data, model reconstructions, and ROM predictions is given in Figure~\ref{fig:tau_rom} at $\delta = 7.5$. Compared to the training data, $\vert \tau_{xy} \vert$ is slightly lower along the y-axis for the model reconstructions, a result of increased reconstruction errors. There is a noticeable decrease in $\vert \tau_{xy} \vert$ for the ROM predictions compared to the reconstructions, which we also believe to be a result of increased prediction errors rather than the disappearance of large-scale flow structures. As $\vert \tau_{xy} \vert$ depends on the product $\overline{{u'v'}}$, increased point-wise errors in both velocity components has a quadratic effect and can significantly alter the shear stress profile. Since the errors are shown to lead to a decrease in turbulent kinetic energy, we can infer that $\overline{{u'v'}}$ is decreased in reconstructions and ROM predictions, leading to attenuation of $\vert \tau_{xy} \vert$. 

\FloatBarrier

\section{Conclusion}

In this work, a ROM framework utilizing Koopman $\beta$-variational autoencoders is introduced for reduced-order modeling of large-scale turbulent flow structures. Surrogate modeling of scale-resolving turbulent flow simulations such as large eddy simulation poses a challenge due to the presence of chaotic small-scale turbulent structures which are difficult to accurately model over time using both physics-based and data-driven methods. Often, the goal of surrogate modeling is to only approximate the macroscopic, bulk flow properties which contain most of the kinetic energy to reasonably inform the design optimization process in real-time. While linear subspace methods like dynamic mode decomposition exist to extract spatio-temporal modes from turbulent flow data, they often require a very large number of modes for satisfactory accuracy and the mode selection process is non-trivial. Additionally, DMD cannot be applied to multiple datasets simultaneously, limiting its use. To this end, Koopman $\beta$-VAEs are implemented as an alternative to vanilla CAEs to filter small-scale turbulent structures from turbulent flow data in an unsupervised manner. Through the use of a Koopman-inspired loss function which encourages a linear evolution of the latent space in time, latent variables are efficiently denoised and are able to be sufficiently modeled over time. $\beta$-VAEs are used as they produce a well-structured latent space that allows for a more efficient implementation of the Koopman loss. To maximize predictive performance and avoid overfitting to temporal patterns, the model is pre-trained on randomly shuffled mini-batches before being trained on temporally ordered mini-batches, which are required for correct implementation of the Koopman loss function.

When applied to a dataset involving turbulent flow past a Windsor body at multiple yaw angles, the Koopman $\beta$-VAE is shown to efficiently filter out small-scale turbulent structures from input data in reconstructions and denoise latent variables. An ensemble LSTM is used as the time series prediction model and can predict latent variables over unseen time horizons with acceptable accuracy and high stability, with ROM errors matching reconstruction errors closely and not growing drastically over time. Mean flow profiles and time-averaged vorticity contours show that both model reconstructions and ROM predictions preserve large-scale coherent structures well.  Results show that although there is a loss in turbulent kinetic energy on test data reconstructions and ROM predictions relative to training data reconstructions, the spectral properties of the flow are retained and that the decrease is due to increased point-wise errors from extrapolation and time-series prediction which lower the amplitudes of the present structures. A limitation of the method is that it is not readily applicable to unstructured meshes and flow data must be interpolated onto a uniform grid. Additionally, the filtering of small-scale structures is unsupervised and a threshold for which flow scales are retained cannot be set explicitly, making it uncontrollable in nature. Future work will focus on extending the method to unstructured meshes using graph neural networks and applying it to three-dimensional flows.

\section*{Acknowledgements}
The research leading to this work has been partially funded by the project TIFON with reference PLEC2023-010251/ AEI/10.13039/501100011033. Benet Eiximeno’s work was funded by a contract from the Subprograma de Ayudas Predoctorales given by the Ministerio de Ciencia e Innovacion (PRE2021-096927). Oriol Lehmkuhl has been partially supported by a Ramon y Cajal postdoctoral contract (Ref: RYC2018- 025949-I). The authors acknowledge the support given by the Departament de Recerca i Universitats de la Generalitat de Catalunya to the Large-scale Computational Fluid Dynamics Research Group (Code: 2021 SGR 00902). We also acknowledge the Barcelona Supercomputing Center for awarding us access to the MareNostrum V machine based in Barcelona, Spain.

\section*{Author Declarations}
\subsection*{Conflict of Interest}
The authors have no conflicts to disclose.

\subsection*{Author Contributions}
\textbf{Rakesh Halder:} Conceptualization (equal), formal analysis, methodology, software, writing - original draft (lead) \\ 
\textbf{Benet Eiximeno:} Data curation, writing – original draft (supporting) \\ 
\textbf{Oriol Lehmkuhl:} Conceptualization (equal), supervision, writing - review and editing  \\

\section*{Data Availability Statement}
The data that support the findings of this study are available from the corresponding author upon reasonable request.

\FloatBarrier

\bibliographystyle{unsrt}
\bibliography{references}

\clearpage

\begin{appendices}

\section{Koopman $\beta$-VAE Architecture}
\label{appendix:arch}

The architecture of the Koopman $\beta$-VAE used in this work is shown in Table~\ref{table:arch}. The encoder and decoder consist of a series of convolutional and convolutional transpose layers with a kernel size of 3 and a stride of 2 followed by batch normalization layers~\cite{ioffe2015batch}, which help reduce internal covariate shift and lead to more efficient gradient flow. In the encoder, the number of filters is doubled in each subsequent convolutional layer while the spatial resolution is reduced by a factor of 2, which allows the network to learn progressively more abstract features, while the opposite is done in the decoder. Two fully connected layers are present in both the encoder and decoder; although this increases the total number of parameters, it also leads to greatly improved reconstruction accuracy. Before the latent space given in the table, the encoder feeds separately into $\bm{\mu}$ and log$(\bm{\sigma}^2)$, from which $\bm{z}$ is sampled. The log of the variance is used as this leads to improved numerical stability when training the network, a common approach when training VAEs. Tanh activation functions are used for all convolutional and fully connected layers except the latent space, which uses no activation function, and the output layer, which uses a sigmoid activation function to constrain the data to the pre-processing range. In total, the model contains 25,993,050 parameters. PyTorch~\cite{paszke2019pytorch} is used to train the model in addition to the LSTM ensemble.

\begin{table}[!h]
  \centering
  \begin{tabular}{@{}lllll@{}}
  \toprule
  {Layer}                          & {Number of Filters} & {Kernel Size} & {Activation Function} & {Size of Output} \\ \midrule
  Input                          &                   &             &                     & 384 $\times$ 256 $\times$ 2  \\ 
  Convolutional                  & 16                 & 3 $\times$ 3       & Tanh        & 192 $\times$ 128 $\times$ 16  \\ 
  Batch Norm                      &                   &        &                         & 192 $\times$ 128 $\times$ 16  \\  
  Convolutional                  & 32                & 3 $\times$ 3       & Tanh         & 96 $\times$ 64 $\times$ 32   \\ 
  Batch Norm                      &                   &        &                         & 96 $\times$ 64 $\times$ 32   \\
  Convolutional                  & 64                & 3 $\times$ 3       & Tanh         & 48 $\times$ 32 $\times$ 64   \\ 
  Batch Norm                      &                   &        &                         & 48 $\times$ 32 $\times$ 64   \\ 
  Convolutional                  & 128                & 3 $\times$ 3       & Tanh        & 24 $\times$ 16 $\times$ 128   \\ 
  Batch Norm                      &                   &        &                         & 24 $\times$ 16 $\times$ 128   \\ 
  Convolutional                  & 256                & 3 $\times$ 3       & Tanh        & 12 $\times$ 8 $\times$ 256   \\ 
  Batch Norm                      &                   &        &                         & 12 $\times$ 8 $\times$ 256     \\ 
  Reshape                        &                   &             &                     & 24576           \\ 
  Fully Connected                &                   &             & Tanh                & 512           \\ 
  Fully Connected (Latent Space) &                   &             &                     & 10           \\ 
  Fully Connected                &                   &             & Tanh                & 512            \\ 
  Fully Connected                &                   &             & Tanh                & 24576           \\ 
  Reshape                        &  256                 &             &                     & 12 $\times$ 8 $\times$ 256     \\
  Convolutional Transpose        & 128               & 3 $\times$ 3       & Tanh         & 24 $\times$ 16 $\times$ 128   \\ 
  Batch Norm                     &                   &             &                     & 24 $\times$ 16 $\times$ 128   \\
  Convolutional Transpose        & 64               & 3 $\times$ 3       & Tanh         & 48 $\times$ 32 $\times$ 64   \\ 
  Batch Norm                     &                   &             &                     & 48 $\times$ 32 $\times$ 64   \\
  Convolutional Transpose        & 32                & 3 $\times$ 3       & Tanh         & 96 $\times$ 64 $\times$ 32   \\ 
  Batch Norm                     &                   &             &                     & 96 $\times$ 64 $\times$ 32   \\
  Convolutional Transpose        & 16                & 3 $\times$ 3       & Tanh         & 192 $\times$ 128 $\times$ 16  \\ 
  Batch Norm                     &                   &             &                     & 192 $\times$ 128 $\times$ 16  \\
  Convolutional Transpose        & 2                 & 3 $\times$ 3       & Sigmoid      & 384 $\times$ 256 $\times$ 2  \\ \bottomrule
  
  \end{tabular}
  \caption{Koopman $\beta$-VAE architecture.}
  \label{table:arch}
  \end{table}

\section{Effect of Pre-training}
\label{appendix:pre}
\begin{figure}[!htpb]
  \centering
  \includegraphics[width=0.6\textwidth]{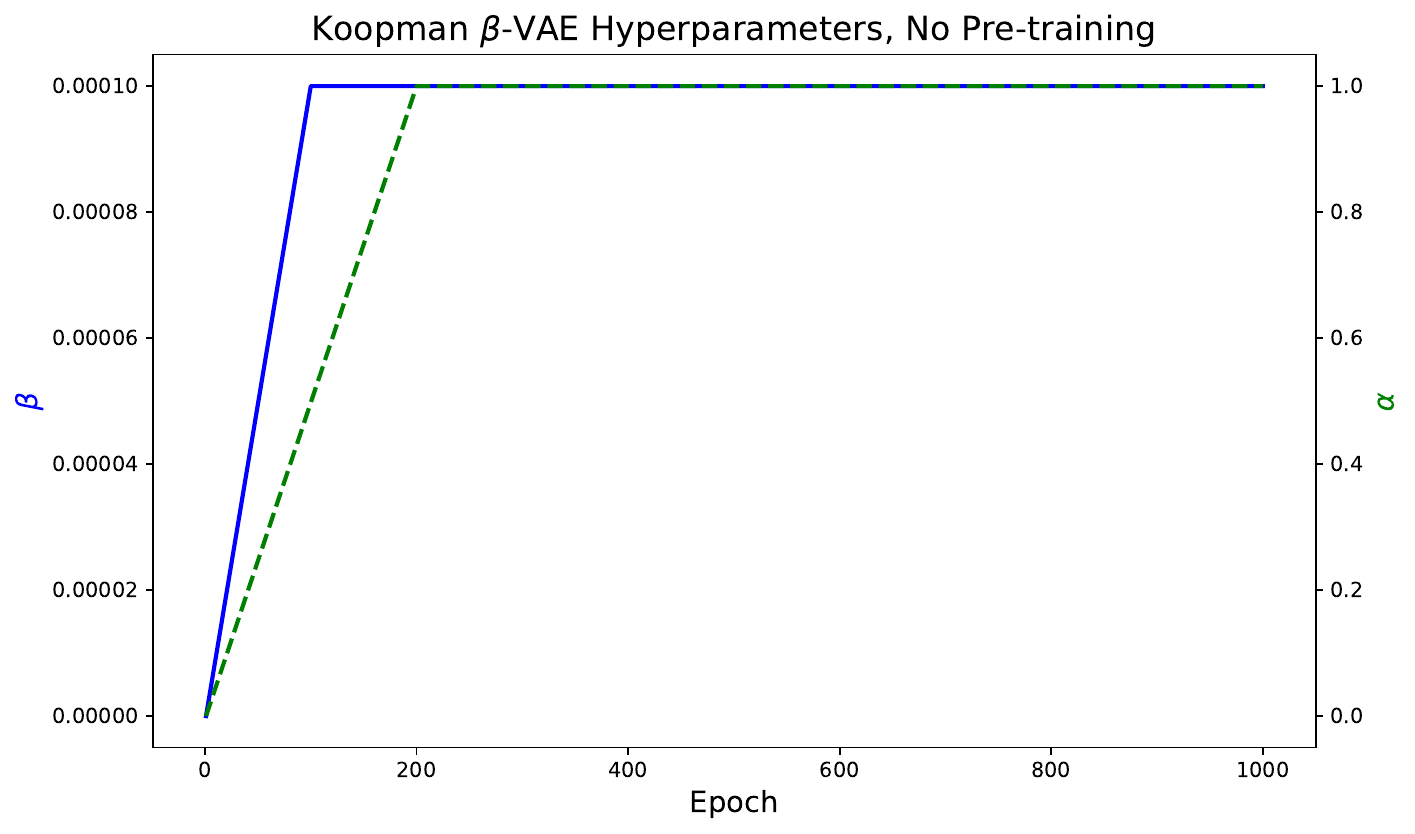}
  \caption{Koopman $\beta$-VAE hyperparameters for a non pre-trained model.}
  \label{fig:param_pre}
  \end{figure}

\begin{figure}[!htpb]
    \centering
    \includegraphics[width=0.49\textwidth]{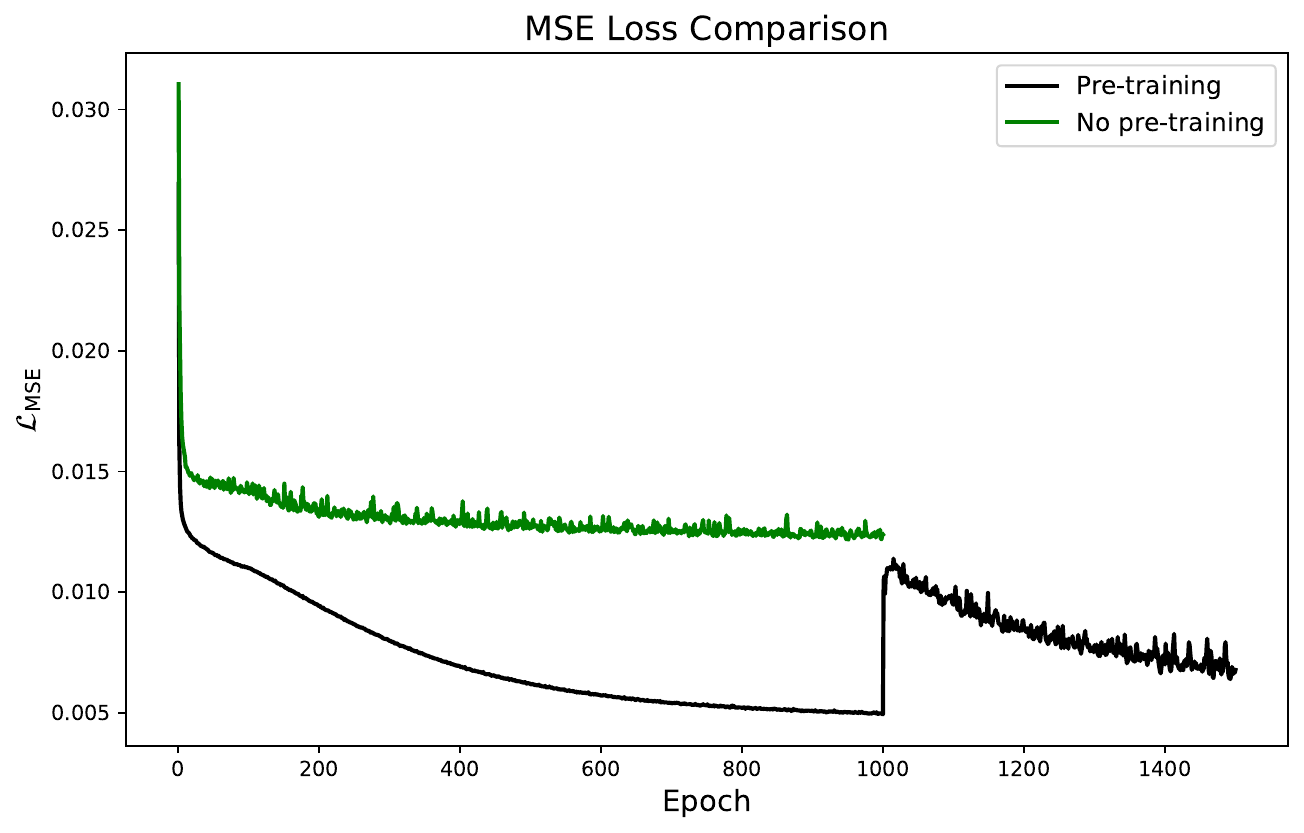}
    \includegraphics[width=0.49\textwidth]{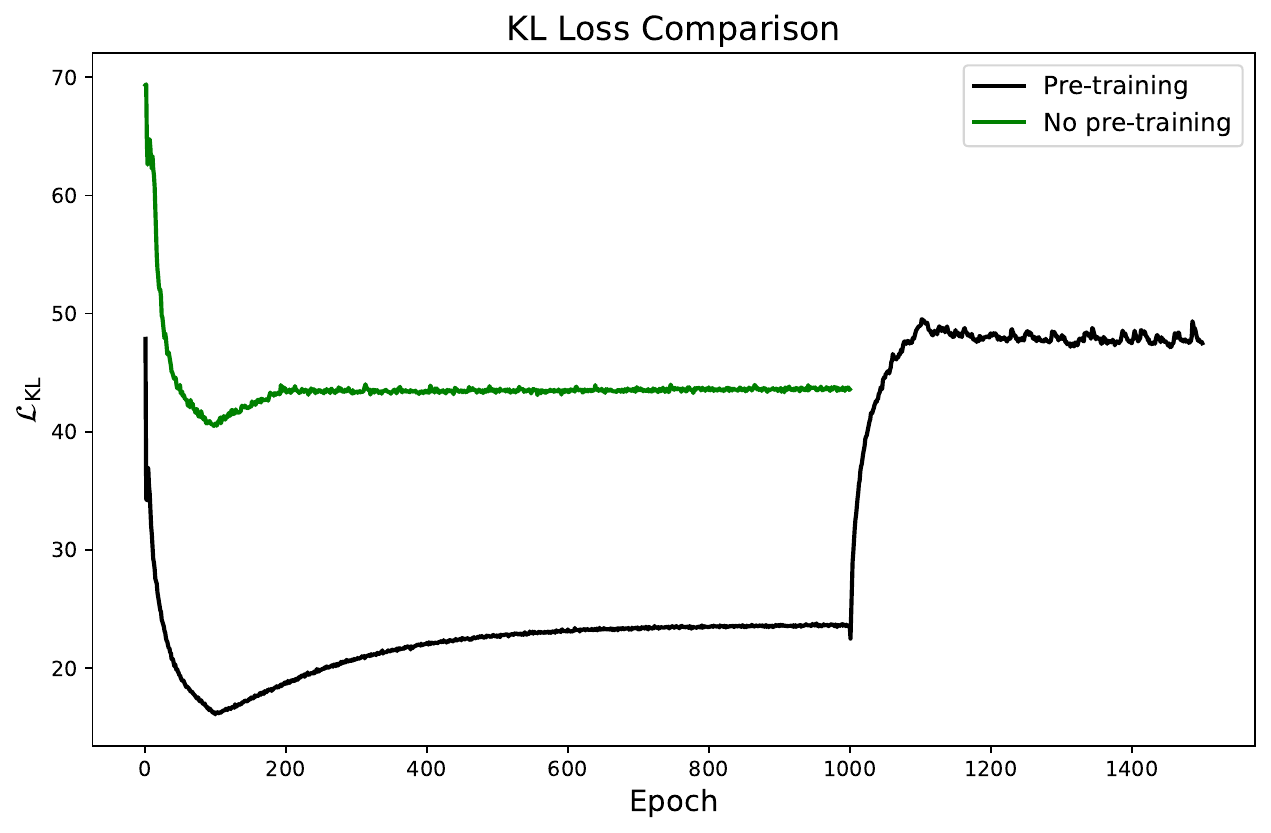}
    \includegraphics[width=0.49\textwidth]{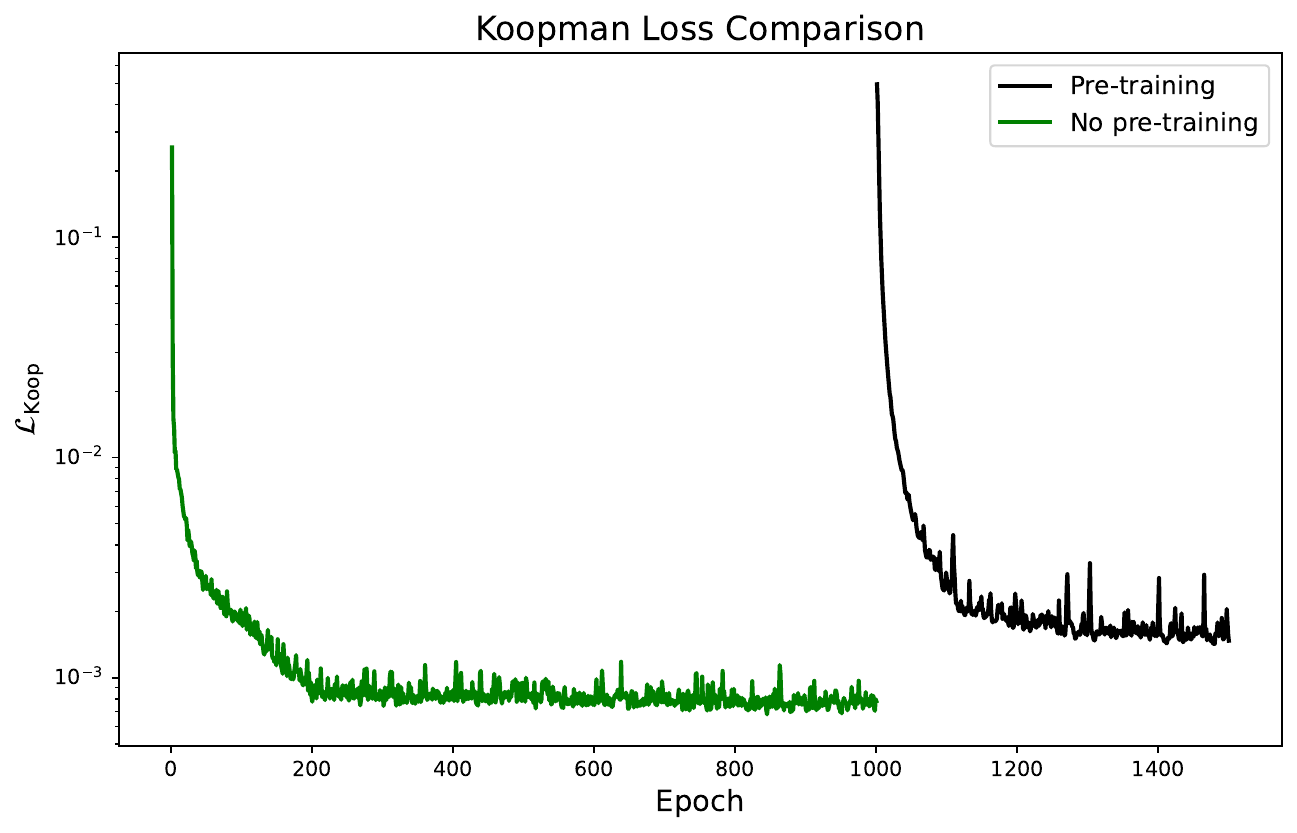}
    \caption{Training loss comparisons with a non pre-trained model.}
    \label{fig:pre_loss}
    \end{figure}

\begin{figure}[!htpb]
    \centering
    \includegraphics[width=0.49\textwidth]{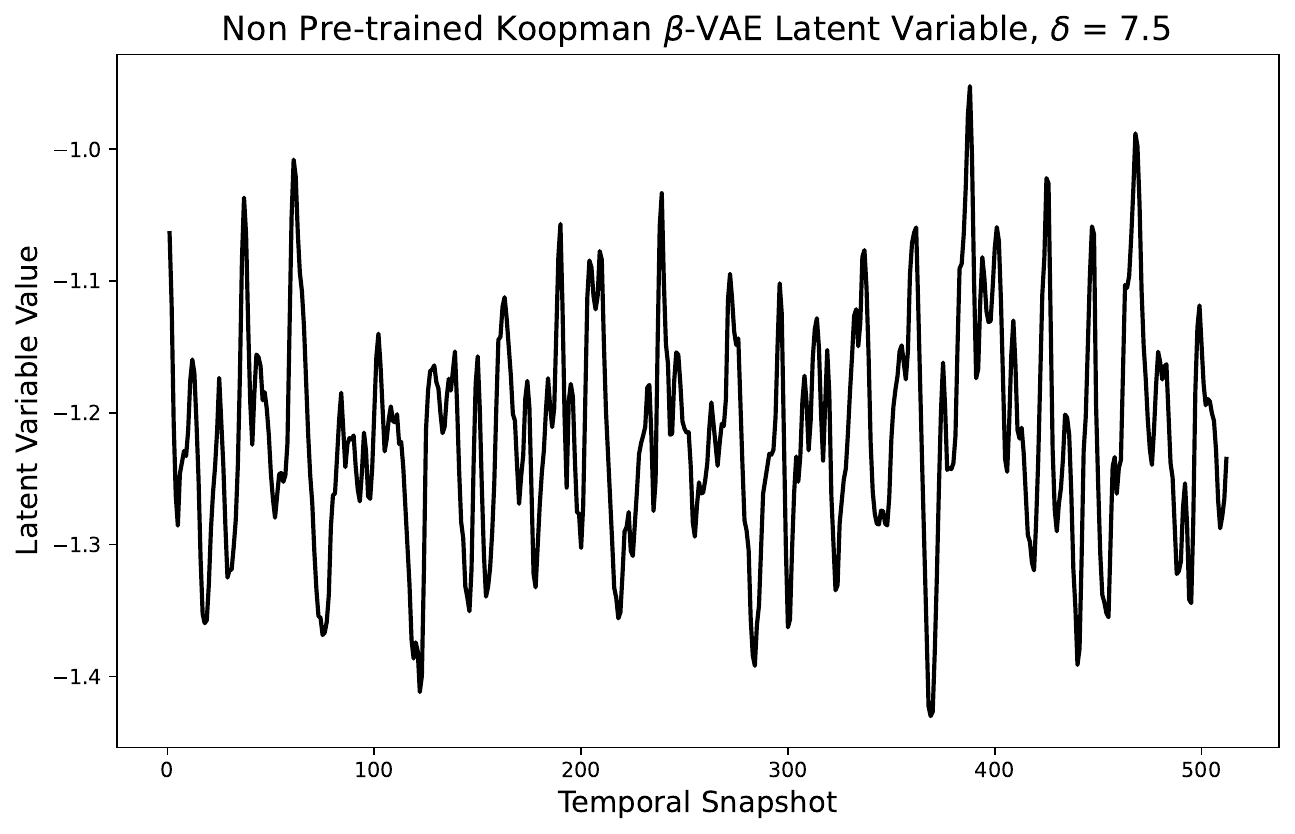}
    \includegraphics[width=0.49\textwidth]{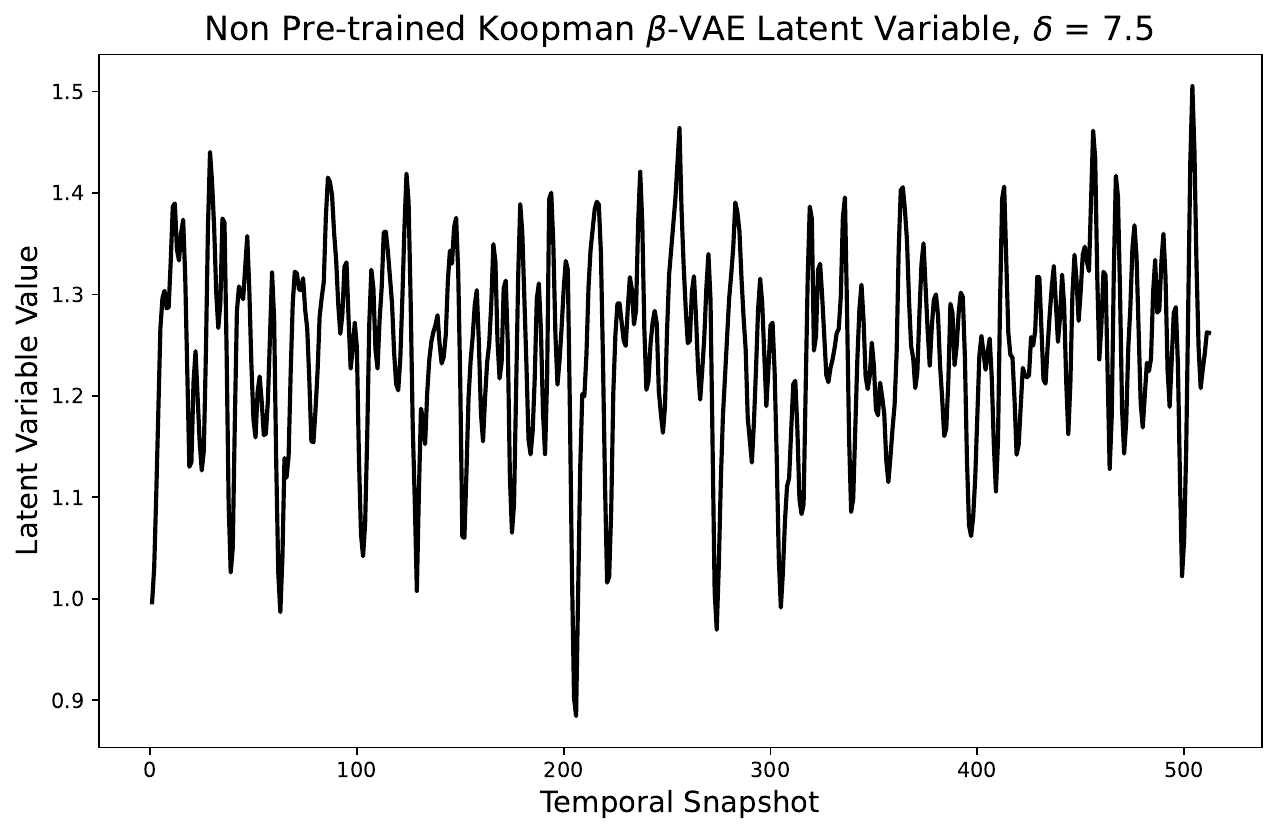}
    \caption{Non pre-trained Koopman $\beta$-VAE latent variable samples at $\delta$ = 7.5.}
    \label{fig:pre_lat}
    \end{figure}

To avoid overfitting to temporal patterns present within the training data and provide a well-structured latent space before using the Koopman loss function, the model is pre-trained on randomly shuffled mini-batches using only the $\beta$-VAE loss. The hyperparameter scheduling for a model trained only on temporally ordered mini-batches without a pre-training stage is shown in Figure~\ref{fig:param_pre}. Again, $\beta$ is linearly increased to its maximum value over the first 10\% of the total number of epochs, while $\alpha$ is over the first 20\%. Figure~\ref{fig:pre_loss} shows a comparison for each component of the training loss against the non pre-trained model. Both $\mathcal{L}_{\text{MSE}}$ are $\mathcal{L}_{\text{KL}}$ are significantly higher and plateau quickly, while $\mathcal{L}_{\text{Koop}}$ decreases rapidly and also plateaus early in the training procedure. This suggests that the model quickly overfits to temporal patterns in the training data due to the use of temporally ordered mini-batches. This also poses a problem for $\mathcal{L}_{\text{KL}}$, as a key assumption of VAEs is that samples are i.i.d. Additionally, implementing the Koopman loss function immediately further hinders the models ability to learn a well-structured latent space as there are two competing regularizing loss functions. Table~\ref{table:pre} shows the overall and component-wise reconstruction errors for the training data from both models, where the pre-trained model offers markedly better performance in all metrics. Figure~\ref{fig:pre_lat} shows latent variable samples from the non pre-trained model at $\delta = 7.5$. Large amounts of noise are retained in the latent variables and the scales do not match those of a standard Gaussian, highlighting the importance of allowing the latent variables to become well-structured before implementing the Koopman loss function.

  \begin{table}[!htbp]
    \centering
    \begin{tabular}{|l|l|l|l|l|l|}
    \hline
    \textbf{Model} & \textbf{$\epsilon$} & \textbf{$\epsilon_u$} & \textbf{$\epsilon_v$}  \\ \hline
    Pre-trained    &           0.252  & 0.188  & 0.443  \\ \hline
    Non pre-trained    & 0.332                      & 0.248                   & 0.582                                           \\ \hline
    \end{tabular}
    \caption{Effect of pre-training on reconstruction errors for the training data.}
    \label{table:pre}  
  \end{table}

\section{Comparison to a Koopman CAE}
\label{appendix:cae}
\begin{figure}[!htpb]
  \centering
  \includegraphics[width=0.49\textwidth]{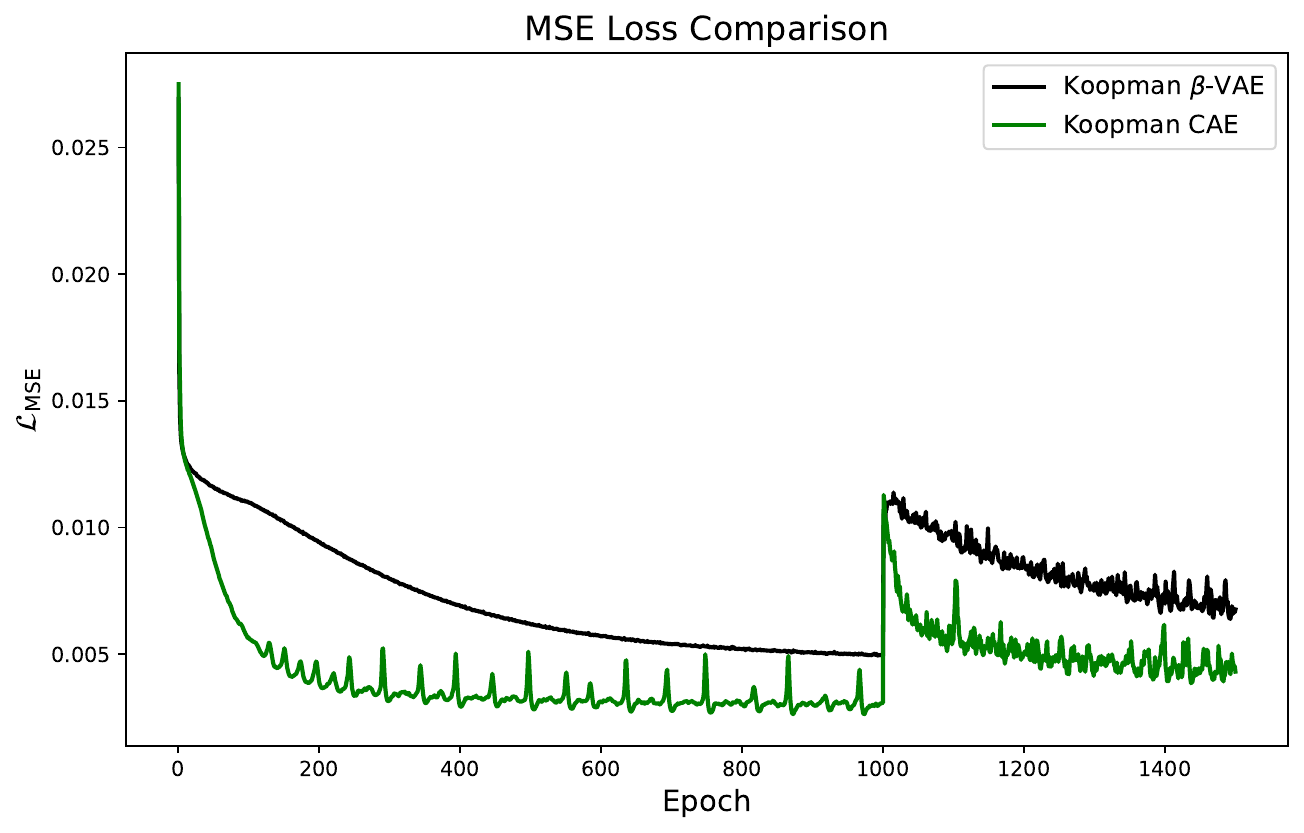}
  \includegraphics[width=0.49\textwidth]{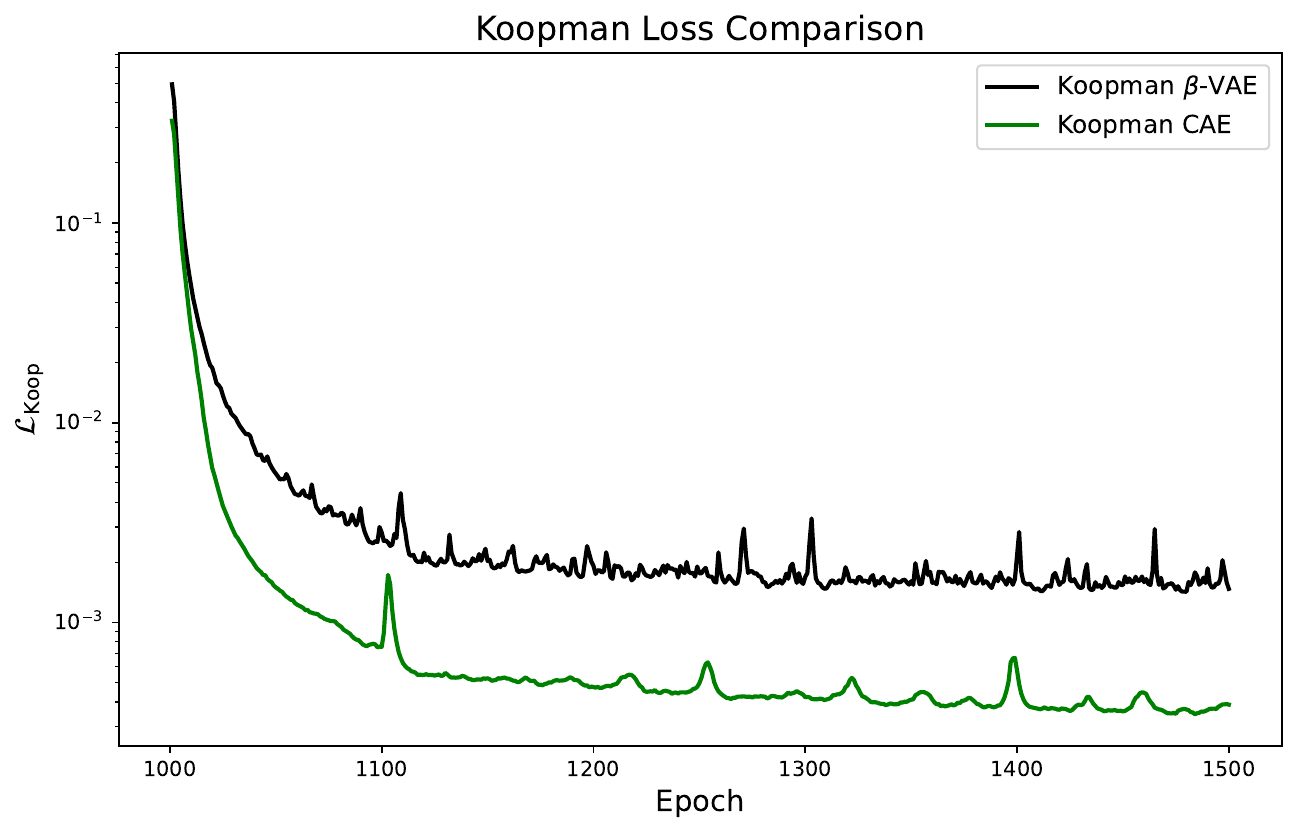}
  \caption{Training loss comparison between Koopman $\beta$-VAE and CAE.}
  \label{fig:koop_cae_loss}
  \end{figure}

\begin{figure}[!htpb]
  \centering
  \includegraphics[width=0.49\textwidth]{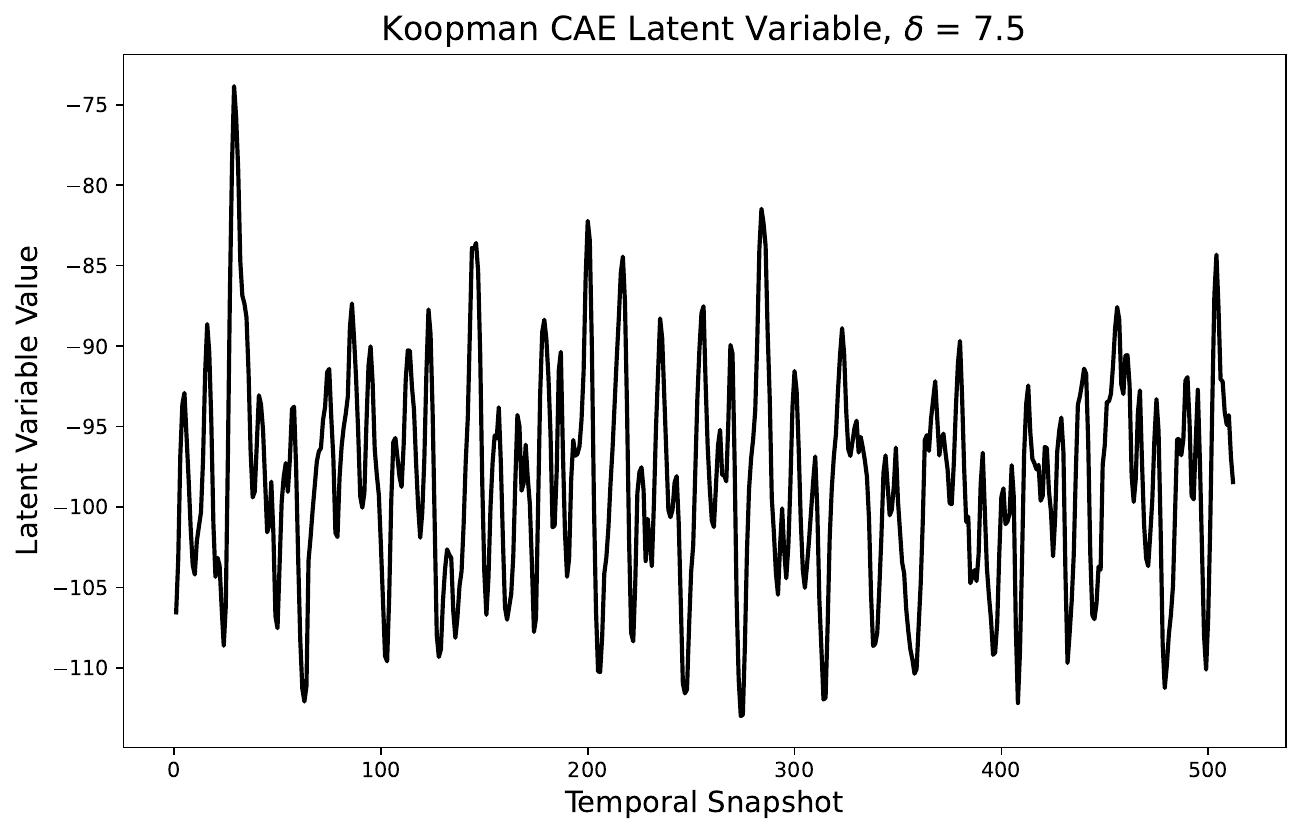}
  \includegraphics[width=0.49\textwidth]{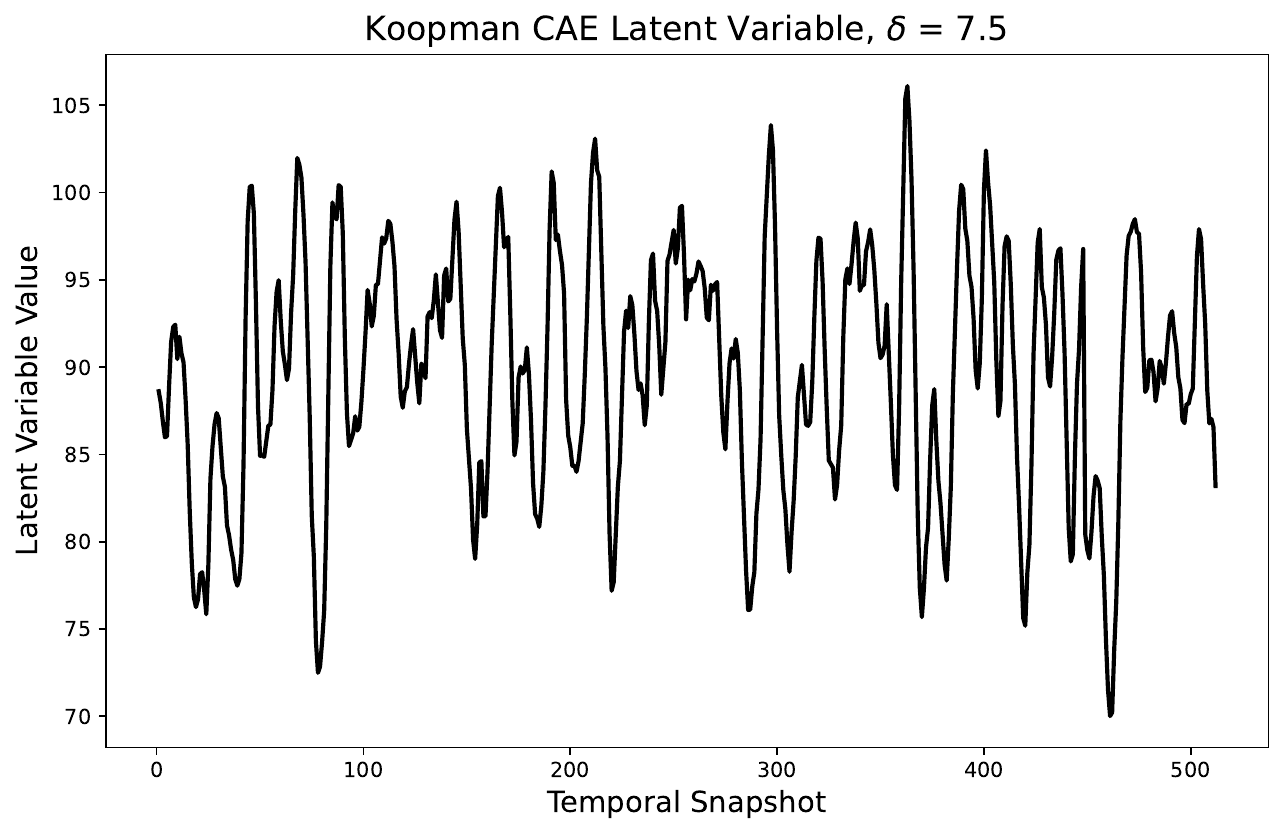}
  \caption{Koopman CAE latent variable samples at $\delta$ = 7.5.}
  \label{fig:koop_cae_lat}
  \end{figure}

\begin{figure}[!htpb]
    \centering
    \includegraphics[width=0.49\textwidth]{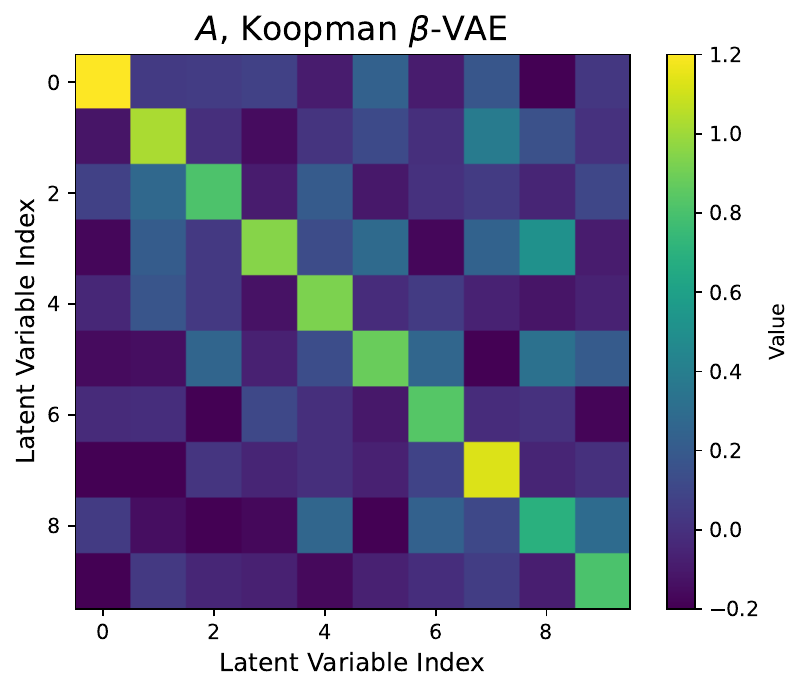}
    \includegraphics[width=0.49\textwidth]{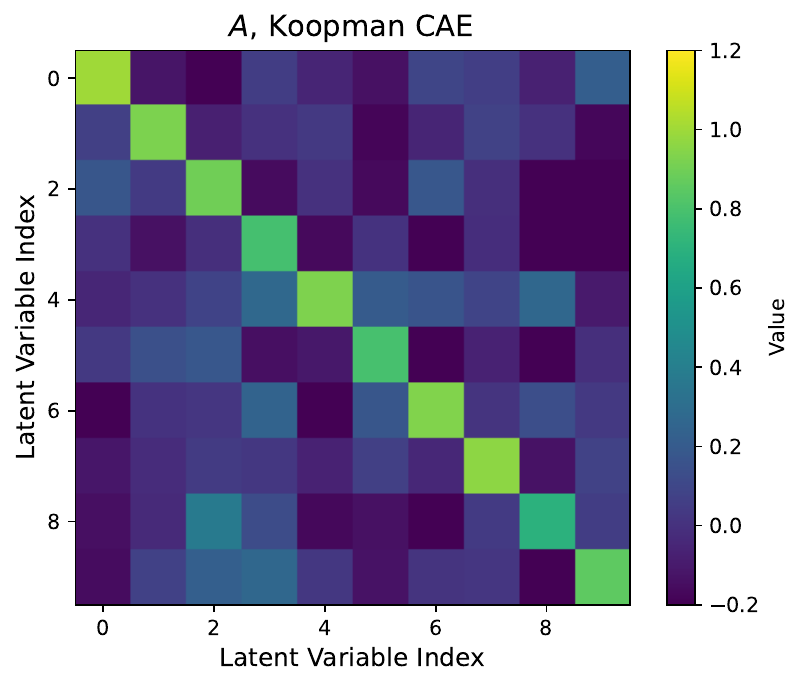}
    \caption{Koopman operator $\bm{A}$ from both models.}
    \label{fig:koop_oper}
    \end{figure}

\begin{figure}[!htpb]
    \centering
    \includegraphics[width=1.0\textwidth]{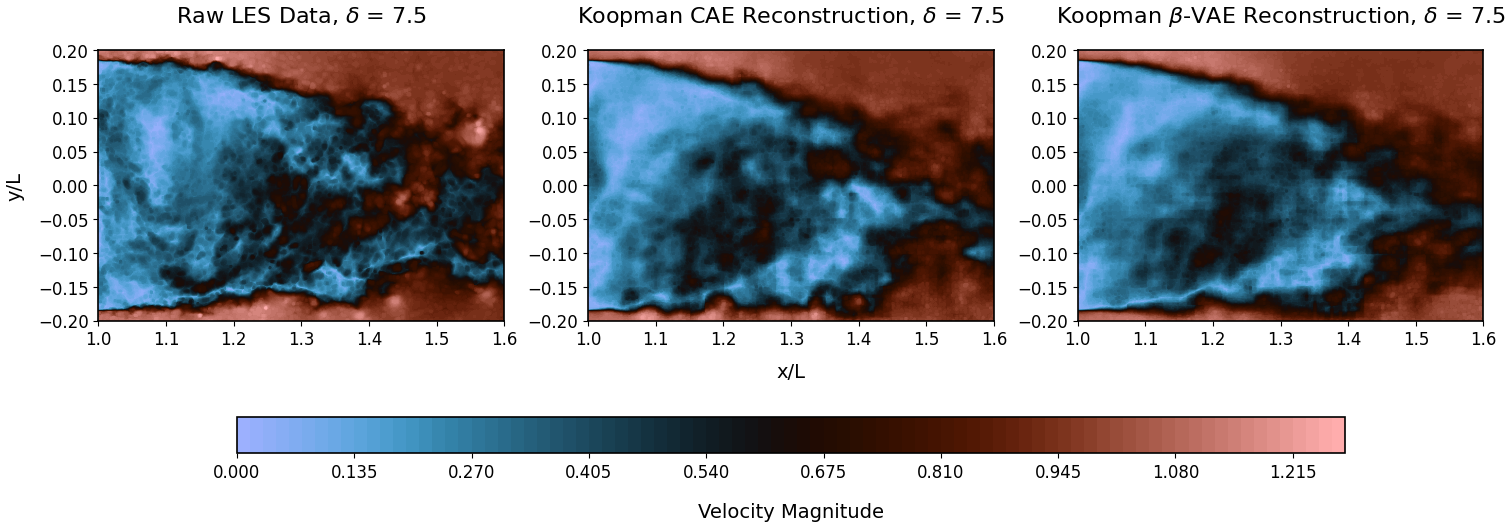}
    \caption{Velocity magnitude comparison at $t = 256$ and $\delta = 7.5$ with a Koopman CAE.}
    \label{fig:recon_7.5_cae}
    \end{figure}

The rationale behind using a $\beta$-VAE is that it results in a well-structured latent space with latent variables being similar in magnitude and distribution, leading to a more stable implementation of the Koopman loss function and better denoising. Using the same hyperparameter setup and number of epochs as in Figure~\ref{fig:params} except with $\beta = 0$, results are compared to a Koopman CAE using the same architecture. A comparison of the training losses are shown in Figure~\ref{fig:koop_cae_loss}. As expected, $\mathcal{L}_{\text{MSE}}$ is lower for the Koopman CAE in both training stages. $\mathcal{L}_{\text{Koop}}$ is also lower for the Koopman CAE, which can be attributed to the lack of any constraint being placed on the latent space. Figure~\ref{fig:koop_cae_lat} shows two latent variable samples at $\delta = 7.5$ from the Koopman CAE; in spite of attaining a significantly lower final value of $\mathcal{L}_{\text{Koop}}$, the latent variables fail to be adequately denoised. When applied to a CAE, where the latent variables are unconstrained and exhibit high variance in scale, the Koopman operator $\bm{A}$ is forced to reduce the loss while fitting to spurious patterns within the latent space. While the latent variables do follow an approximately linear evolution in time and small-scale structures are filtered out, their mismatched scales result in a non-smooth variation. The Koopman operators from both models are shown in Figure~\ref{fig:koop_oper}. There are no significant differences between them; as expected, the diagonal elements are close to 1, while the off-diagonal elements are much smaller and similar in scale. Figure~\ref{fig:recon_7.5_cae} shows velocity magnitude comparisons between the raw LES data, Koopman CAE, and Koopman $\beta$-VAE at $t = 256$ and $\delta = 7.5$. While the reconstructions from both models are similar, the latent variables produced by the Koopman CAE are much more difficult to predict over time due to the amount of noise present.

\section{Impact of $\alpha$}
\label{appendix:alpha}
\begin{figure}[!htpb]
  \centering
  \includegraphics[width=0.49\textwidth]{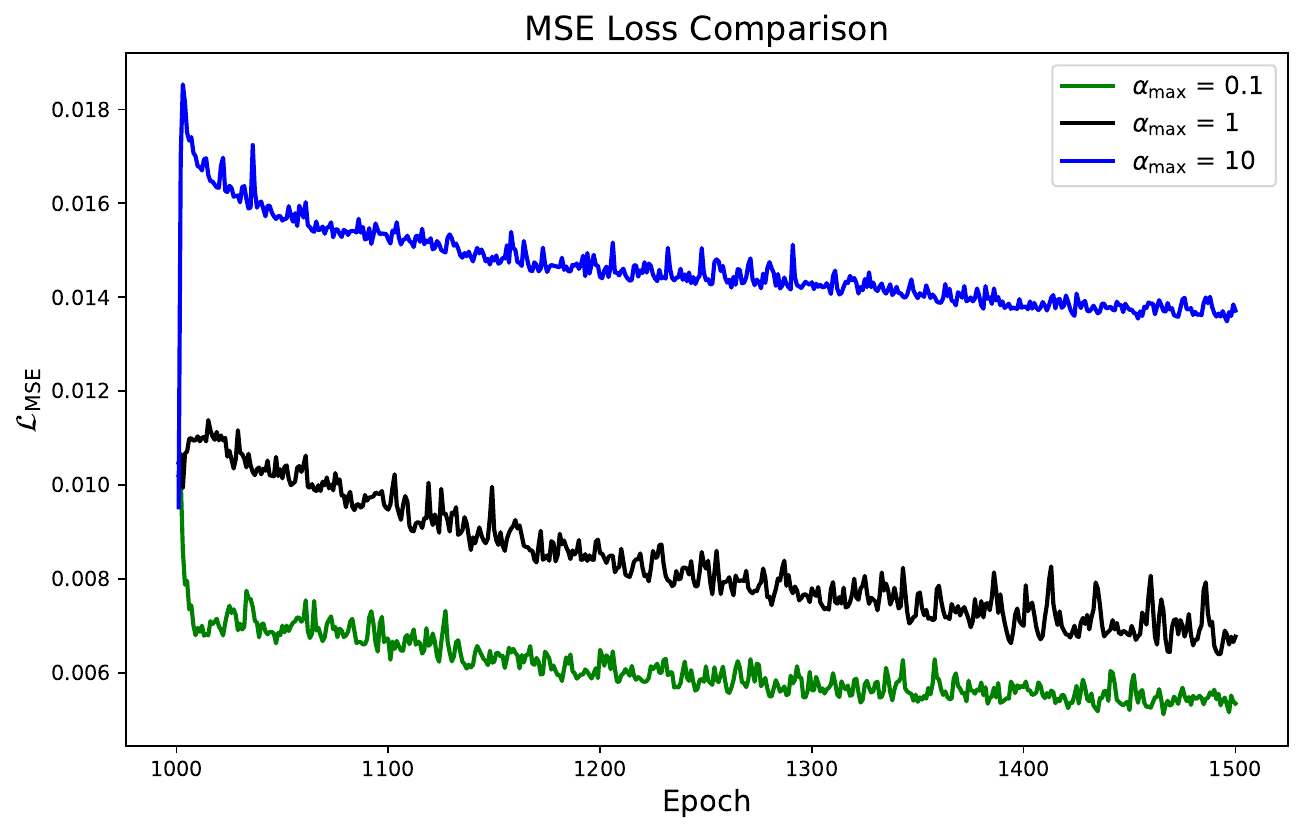}
  \includegraphics[width=0.49\textwidth]{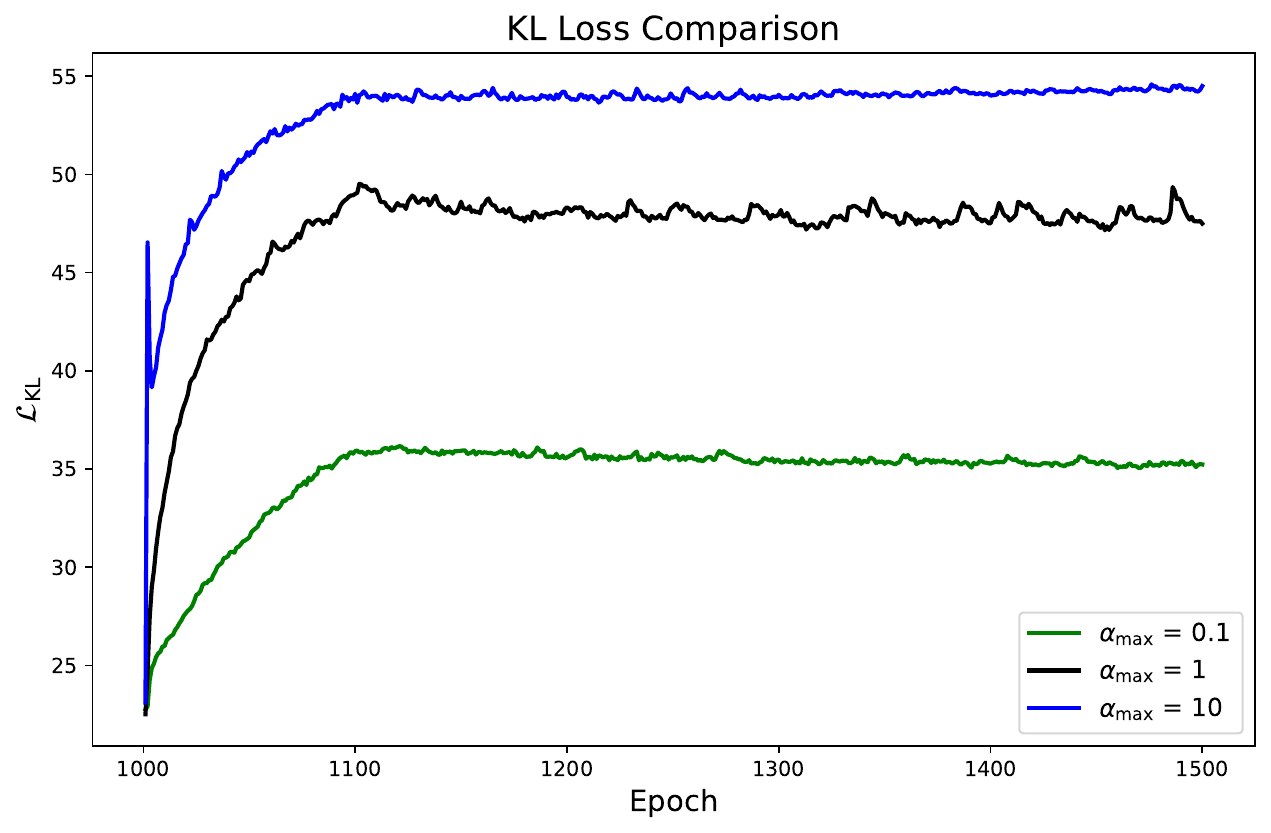}
  \includegraphics[width=0.49\textwidth]{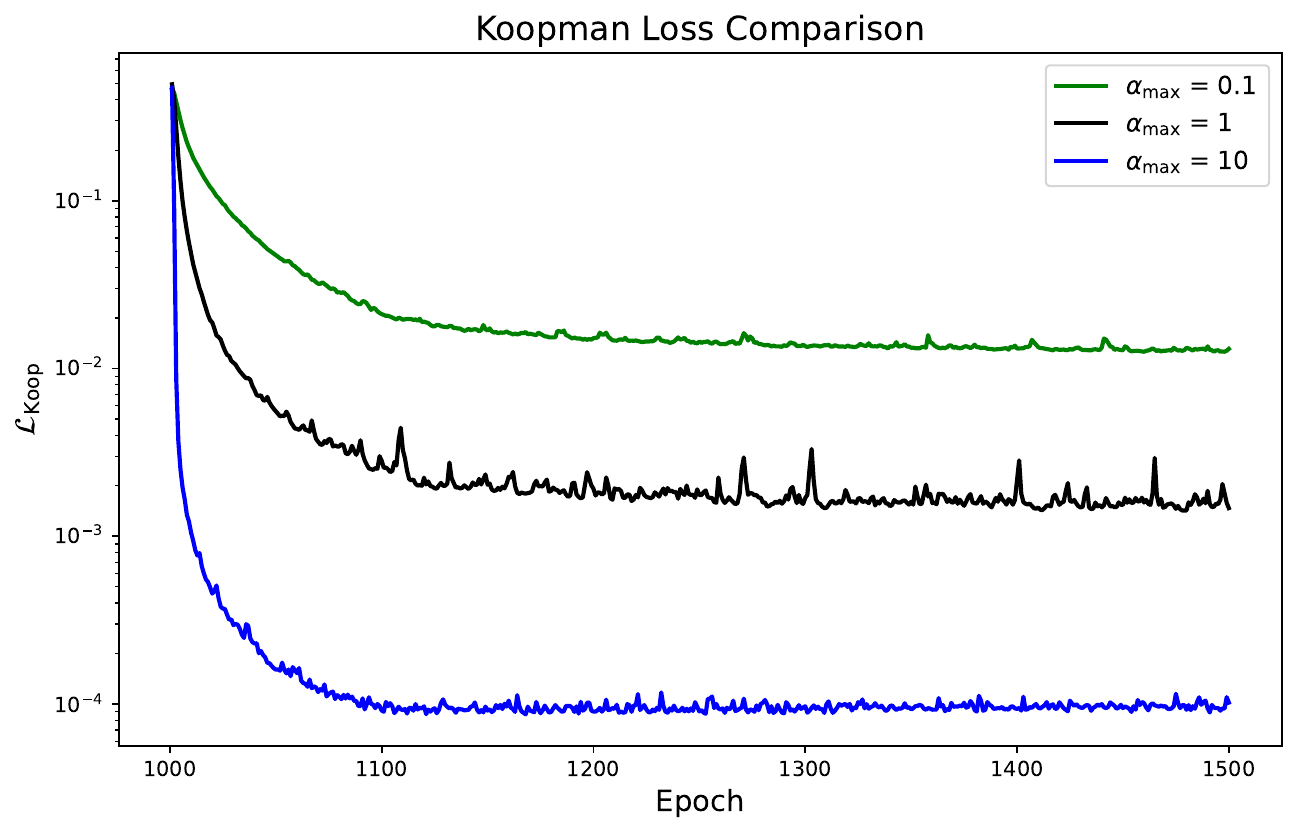}
  \caption{Training losses with different values of $\alpha_{\text{max}}$.}
  \label{fig:alpha_loss}
  \end{figure}

\begin{figure}[!htpb]
  \centering
  \includegraphics[width=0.49\textwidth]{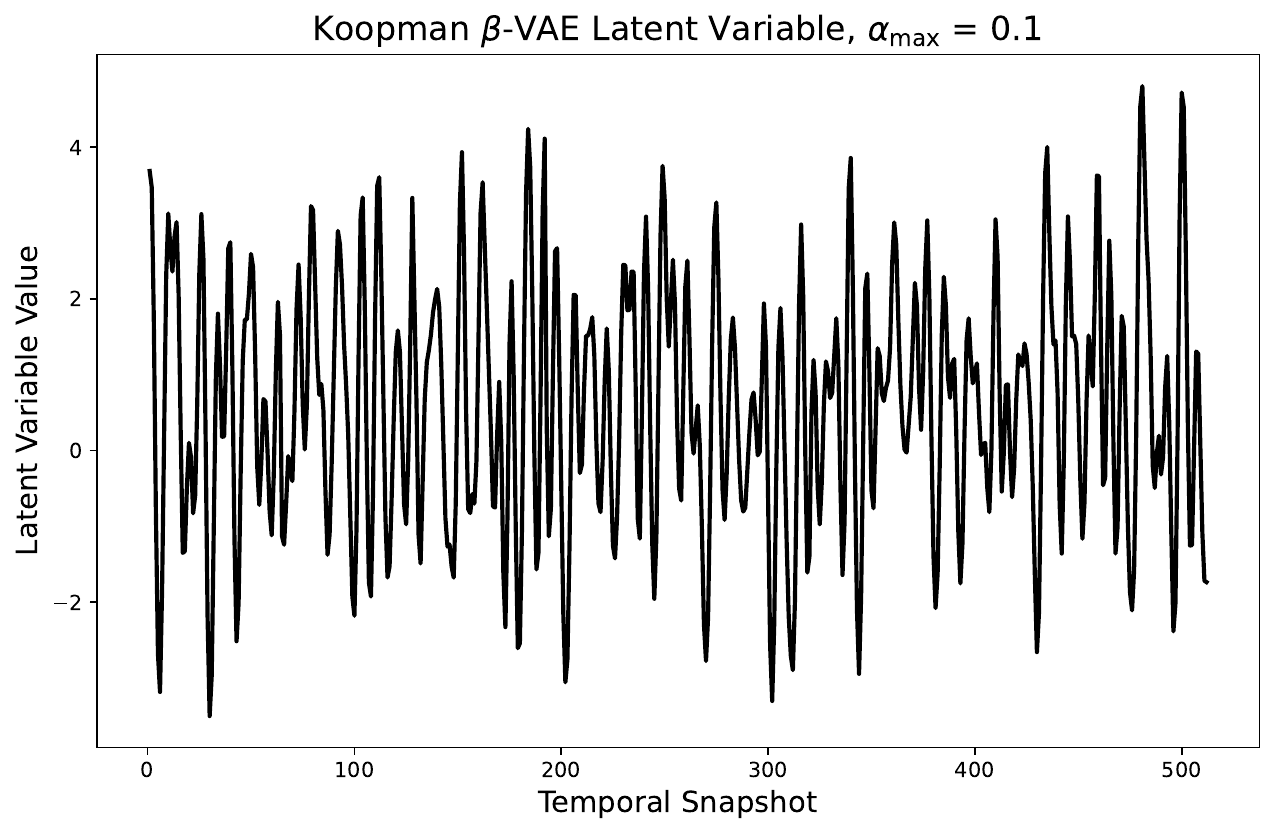}
  \includegraphics[width=0.49\textwidth]{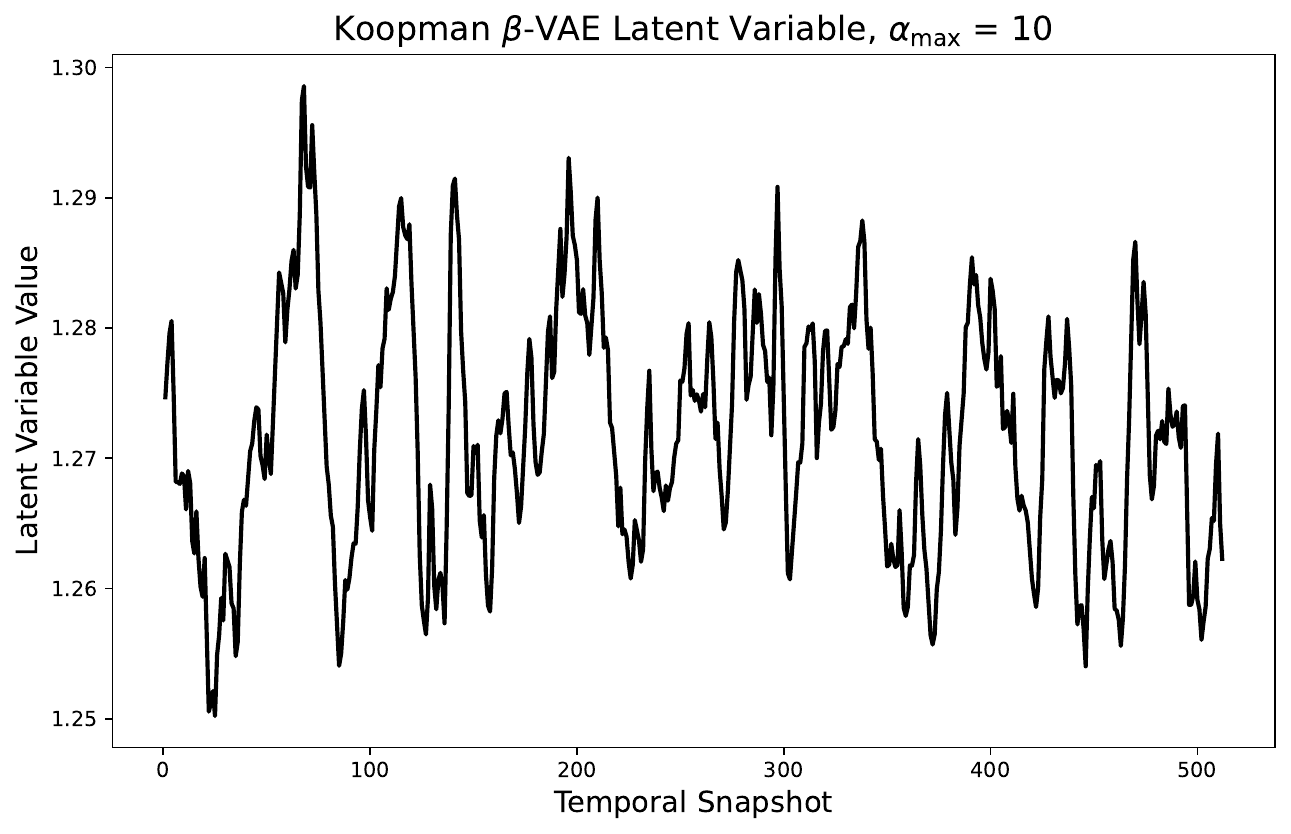}
  \caption{Latent variable samples at $\delta$ = 7.5 and $\alpha_{\text{max}}$ = 0.1 and 10.}
  \label{fig:alpha_lat}
  \end{figure}

\begin{figure}[!htpb]
    \centering
    \includegraphics[width=1.0\textwidth]{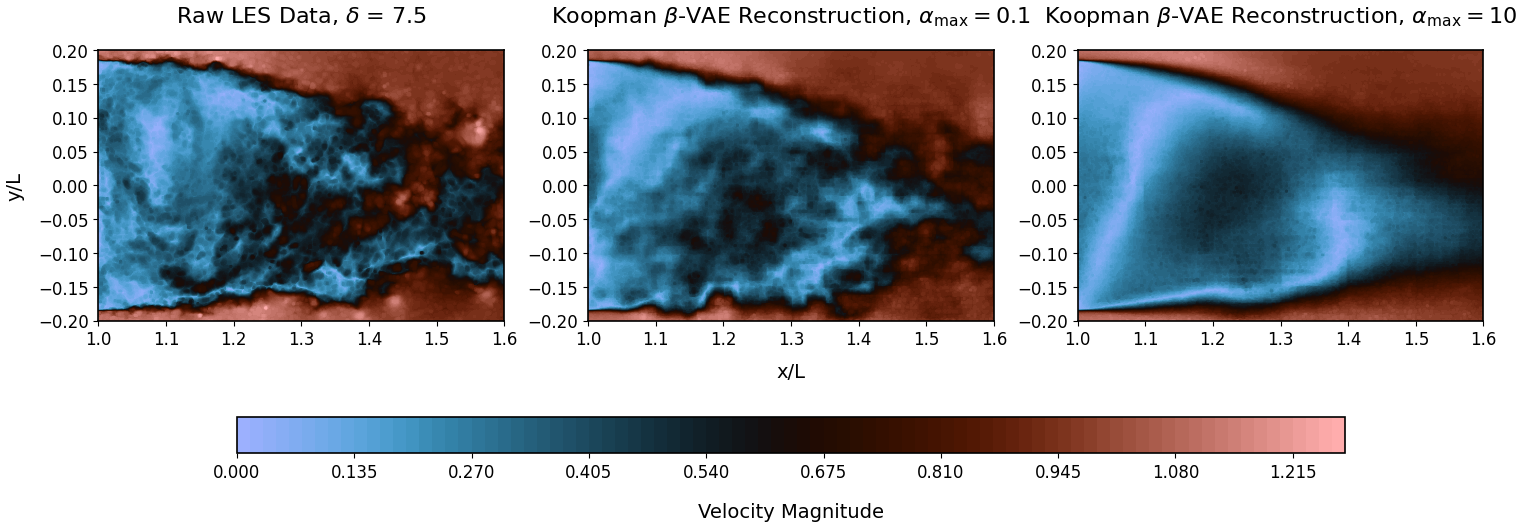}
    \caption{Velocity magnitude comparison at $t = 256$ and $\delta = 7.5$ for different values of $\alpha_{\text{max}}$.}
    \label{fig:recon_7.5_alpha}
    \end{figure}

While the Koopman $\beta$-VAE efficiently removes small-scale structures from reconstructions of turbulent input data, a limitation is that a threshold for which flow scales are retained cannot be set explicitly. However, $\alpha_{\text{max}}$ can be altered to control $\mathcal{L}_{\text{Koop}}$ and subsequently the degree of linearity. Figure~\ref{fig:alpha_loss} shows a comparison between the three different components of the loss function with values of $\alpha_{\text{max}} = [0.1, 1, 10]$ using the same hyperparameter scheduling in Figure~\ref{fig:params}. Increasing $\alpha_{\text{max}}$ imposes a stricter constraint on the evolution of the latent space in time, and higher values result in higher values of $\mathcal{L}_{\text{MSE}}$ and $\mathcal{L}_{\text{KL}}$. As expected, $\mathcal{L}_{\text{Koop}} $ significantly decreases as $\alpha_{\text{max}}$ is increased. Figure~\ref{fig:alpha_lat} shows latent variable samples at $\delta = 7.5$ and $\alpha_{\text{max}} = 0.1$ and 10. A value of 0.1 results in under-regularization of the latent space and a failure to denoise. Raising $\alpha_{\text{max}}$ to 10 results in over-regularization; the range of the latent variables no longer follow that of a standard Gaussian, implying that $\mathcal{L}_{\text{KL}}$ rises too much. While the latent variables do lower in frequency, they still exhibit high levels of noise and non-smoothness, suggesting that coherent structures are also filtered out. This is shown in velocity magnitude comparisons at $t = 256$ and $\delta = 7.5$ in Figure~\ref{fig:recon_7.5_alpha}; using $\alpha_{\text{max}} = 0.1$ does filter out small-scale structures but retains noise in the latent space, while using $\alpha_{\text{max}} = 10$ filters out both large and small-scale structures, resulting in flow resembling the mean being reconstructed. These results show that the ability to both filter out small-scale structures and denoise the latent variables is highly sensitive to $\alpha_{\text{max}}$ and requires a careful choice, which is likely problem dependent.

\FloatBarrier

\section{CAE-LSTM Results}
\label{appendix:caelstm}
\begin{figure}[!htpb]
  \centering
  \includegraphics[width=0.6\textwidth]{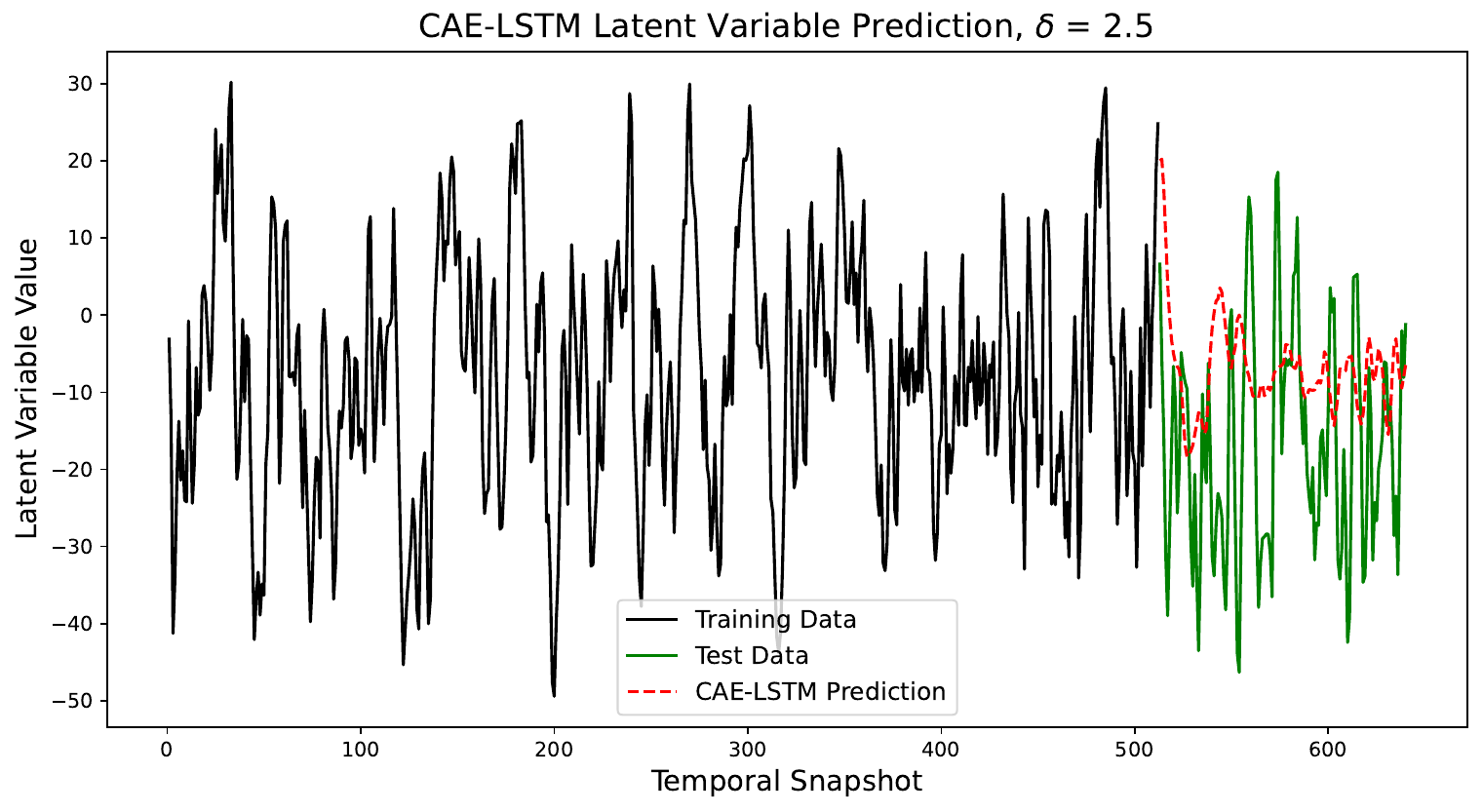}
  \caption{CAE-LSTM latent variable prediction at $\delta = 2.5$.}
  \label{fig:cae_lstm_2.5}
\end{figure}

\begin{figure}[!htpb]
  \centering
  \includegraphics[width=0.6\textwidth]{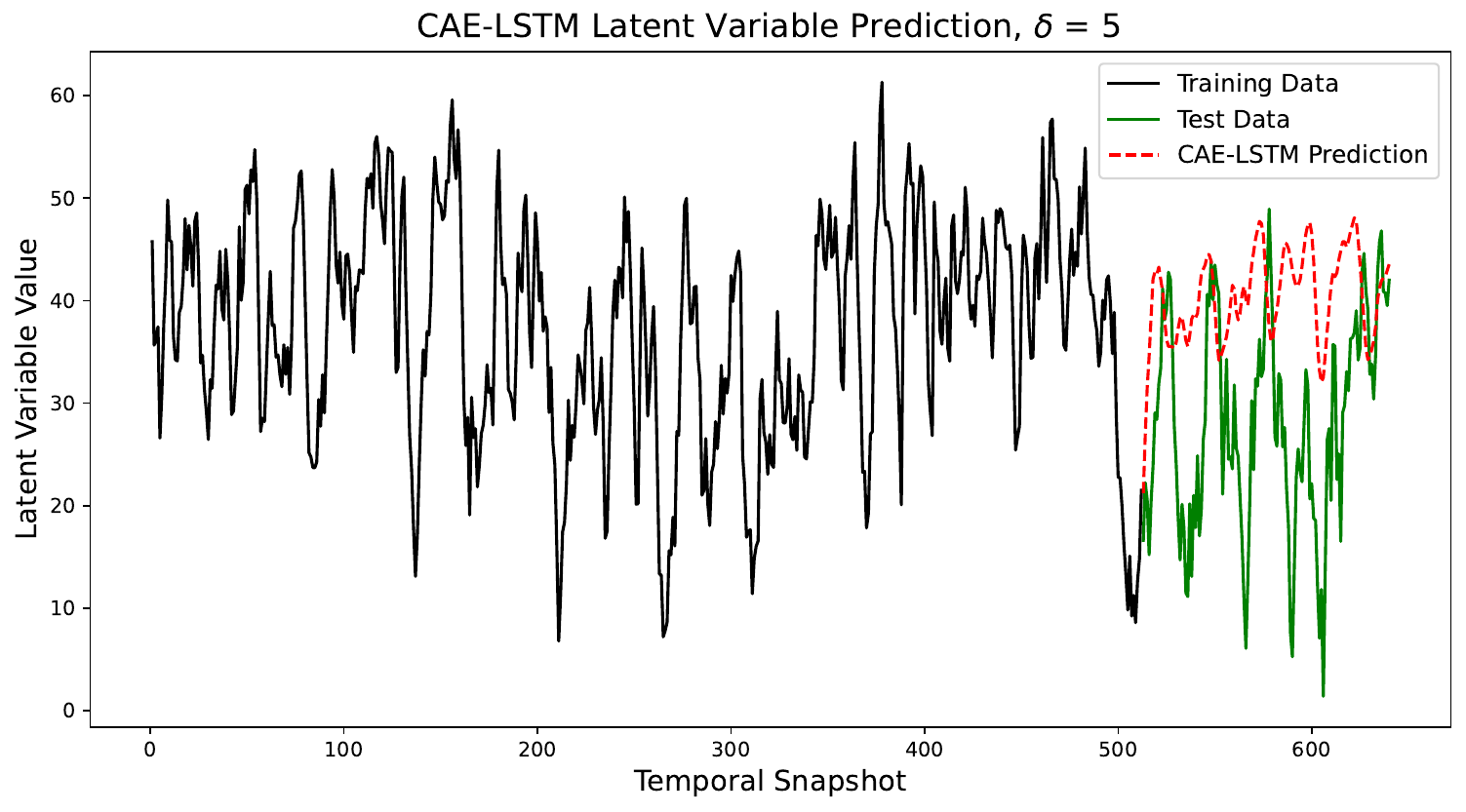}
  \caption{CAE-LSTM latent variable prediction at $\delta = 5$.}
  \label{fig:cae_lstm_5}
\end{figure}

\begin{figure}[!htpb]
  \centering
  \includegraphics[width=0.6\textwidth]{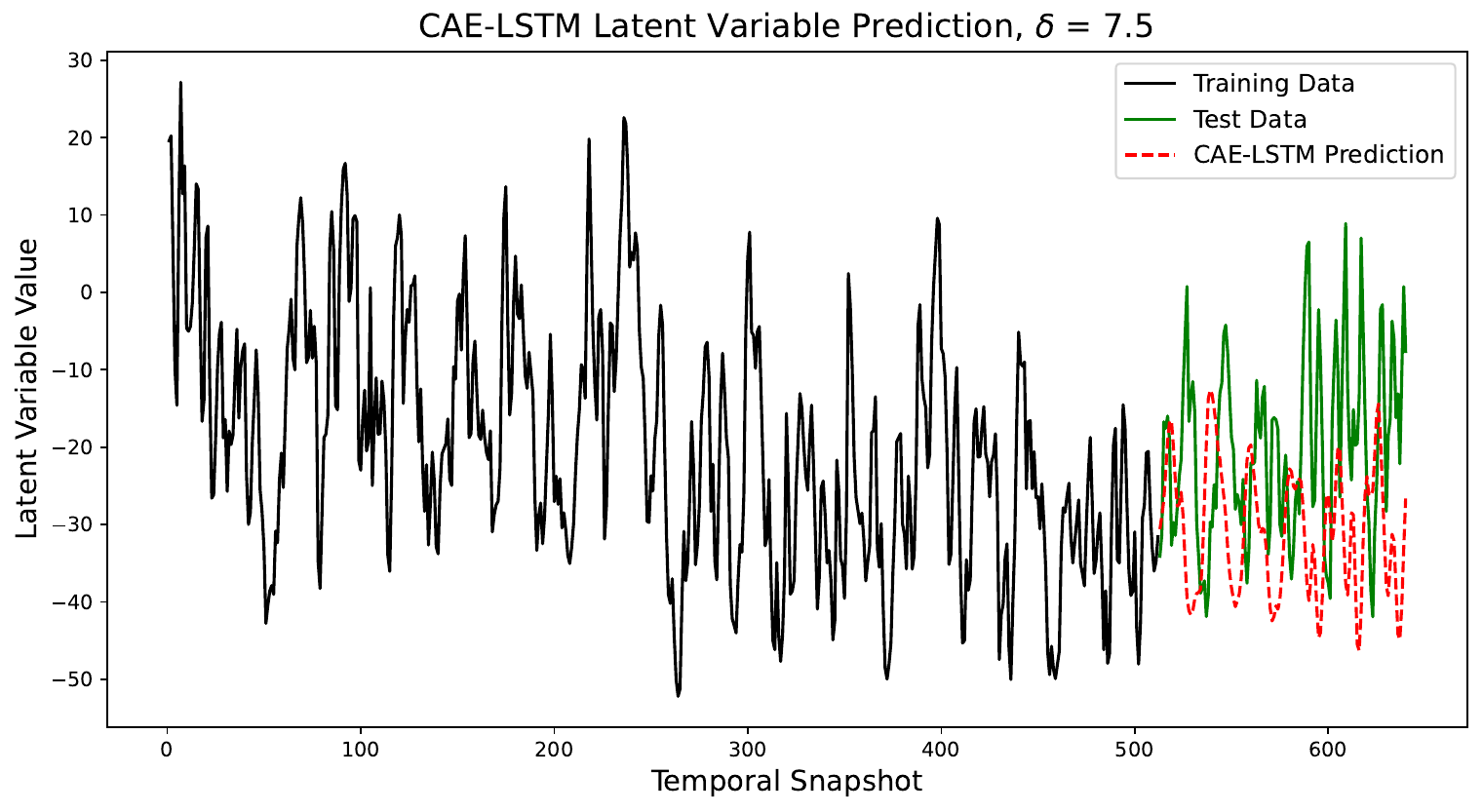}
  \caption{CAE-LSTM latent variable prediction at $\delta = 7.5$.}
  \label{fig:cae_lstm_7.5}
\end{figure}

\begin{figure}[!htpb]
  \centering
  \includegraphics[width=0.6\textwidth]{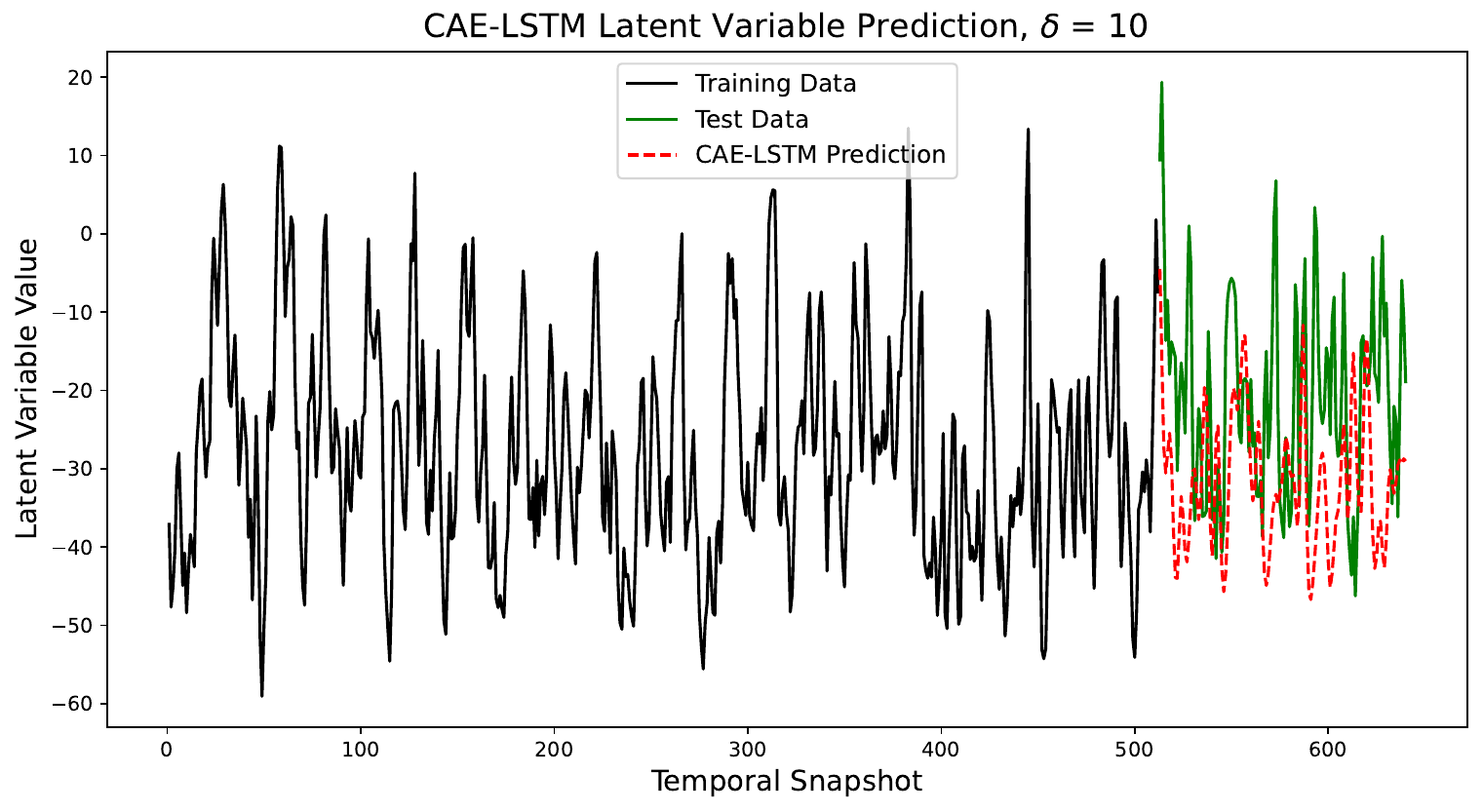}
  \caption{CAE-LSTM latent variable prediction at $\delta = 10$.}
  \label{fig:cae_lstm_10}
\end{figure}

\begin{figure}[!htpb]
  \centering
  \includegraphics[width=0.6\textwidth]{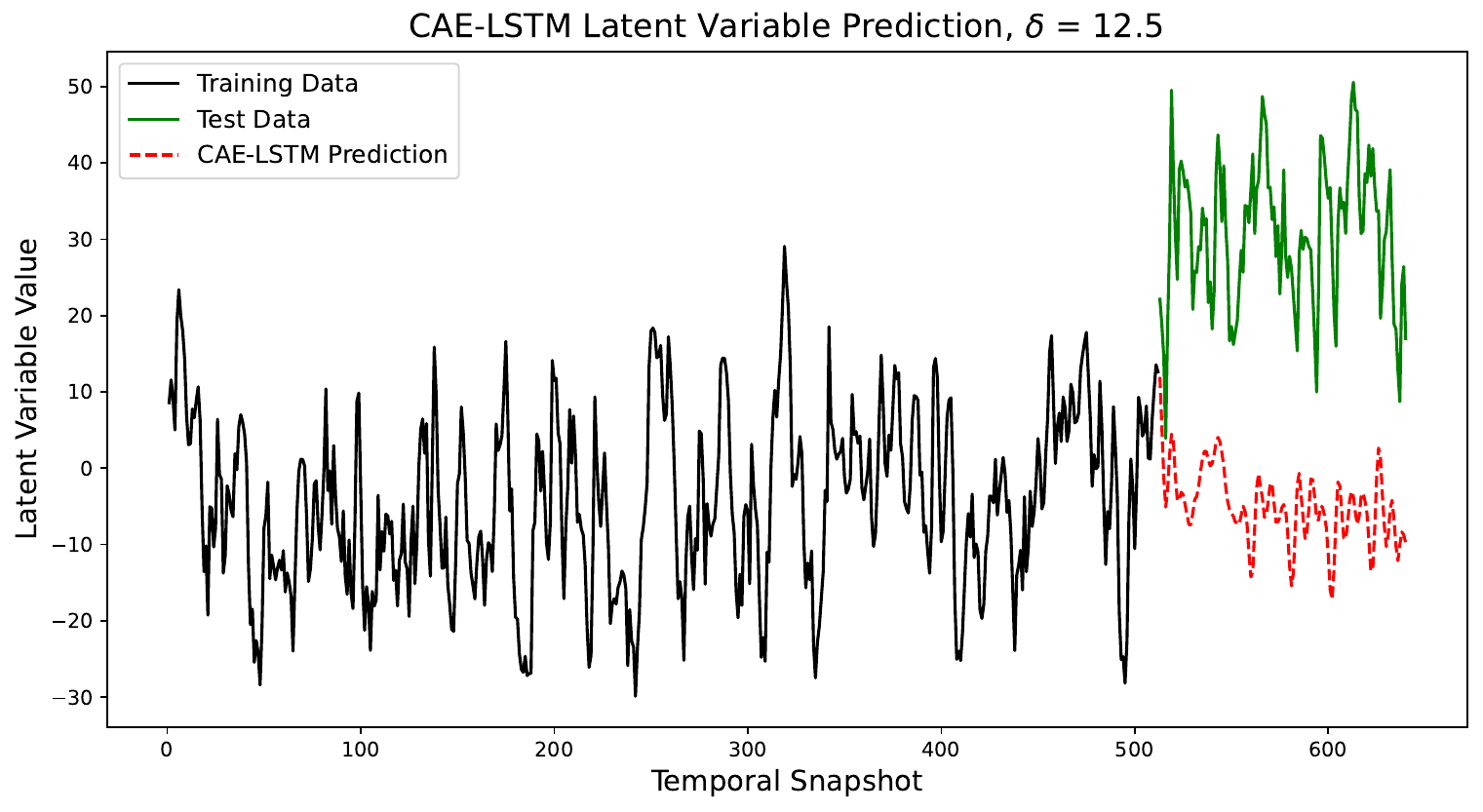}
  \caption{CAE-LSTM latent variable prediction at $\delta = 12.5$.}
  \label{fig:cae_lstm_12.5}
\end{figure}

To provide a baseline for comparison, results from a vanilla CAE-LSTM ROM are provided. The LSTM ensemble is the same one used for the Koopman $\beta$-VAE ROM. 
Figures~\ref{fig:cae_lstm_2.5}-\ref{fig:cae_lstm_12.5} show examples of latent variable predictions given by CAE-LSTM at each yaw angle. Due to the large amounts of noise present in the training data, the CAE-LSTM predictions are poor and fail to accurately forecast the latent variables over time.

Table~\ref{table:cae_lstm} shows the reconstruction and prediction errors when using CAE-LSTM. Although the vanilla CAE attains significantly better reconstruction errors on the training data, they are very similar to those of the Koopman $\beta$-VAE for the test data. The CAE-LSTM ROM produces larger prediction errors, with a greater defect from the accuracy of the reconstructions when compared to the Koopman $\beta$-VAE. 

\begin{table}[!htbp]
  \centering
  \begin{tabular}{|l|l|l|l|l|l|}
  \hline
  \textbf{Prediction} & \textbf{$\epsilon$} & \textbf{$\epsilon_u$} & \textbf{$\epsilon_v$}  \\ \hline
  CAE Reconstruction    &          0.383              & 0.284               & 0.671                                \\ \hline
  Koopman $\beta$-VAE Reconstruction    &          0.392              & 0.286               & 0.704                                \\ \hline
  CAE-LSTM ROM    & 0.489                      & 0.362                   & 0.861                                           \\ \hline
  Koopman $\beta$-VAE ROM    & 0.442                      & 0.323                   & 0.787                                           \\ \hline
  \end{tabular}
  \caption{Prediction error comparison for test data with CAE-LSTM.}
  \label{table:cae_lstm}  
\end{table}

\begin{figure}[!htpb]
  \centering
  \includegraphics[width=0.84\textwidth]{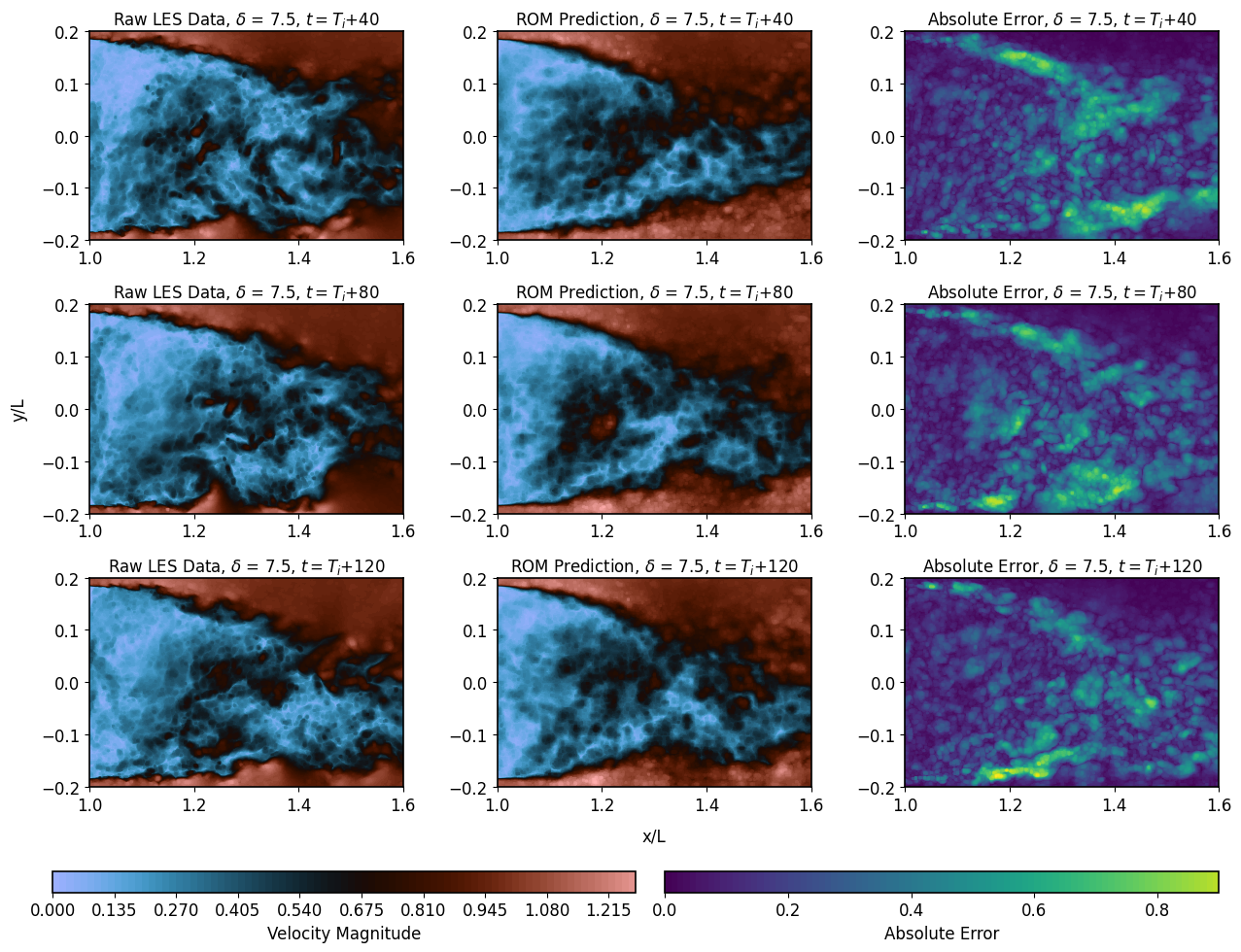}
  \caption{CAE-LSTM predictions and absolute errors at $\delta = 7.5$.}
  \label{fig:cae_7.5}
  \end{figure}

\begin{figure}[!htpb]
    \centering
    \includegraphics[width=0.6\textwidth]{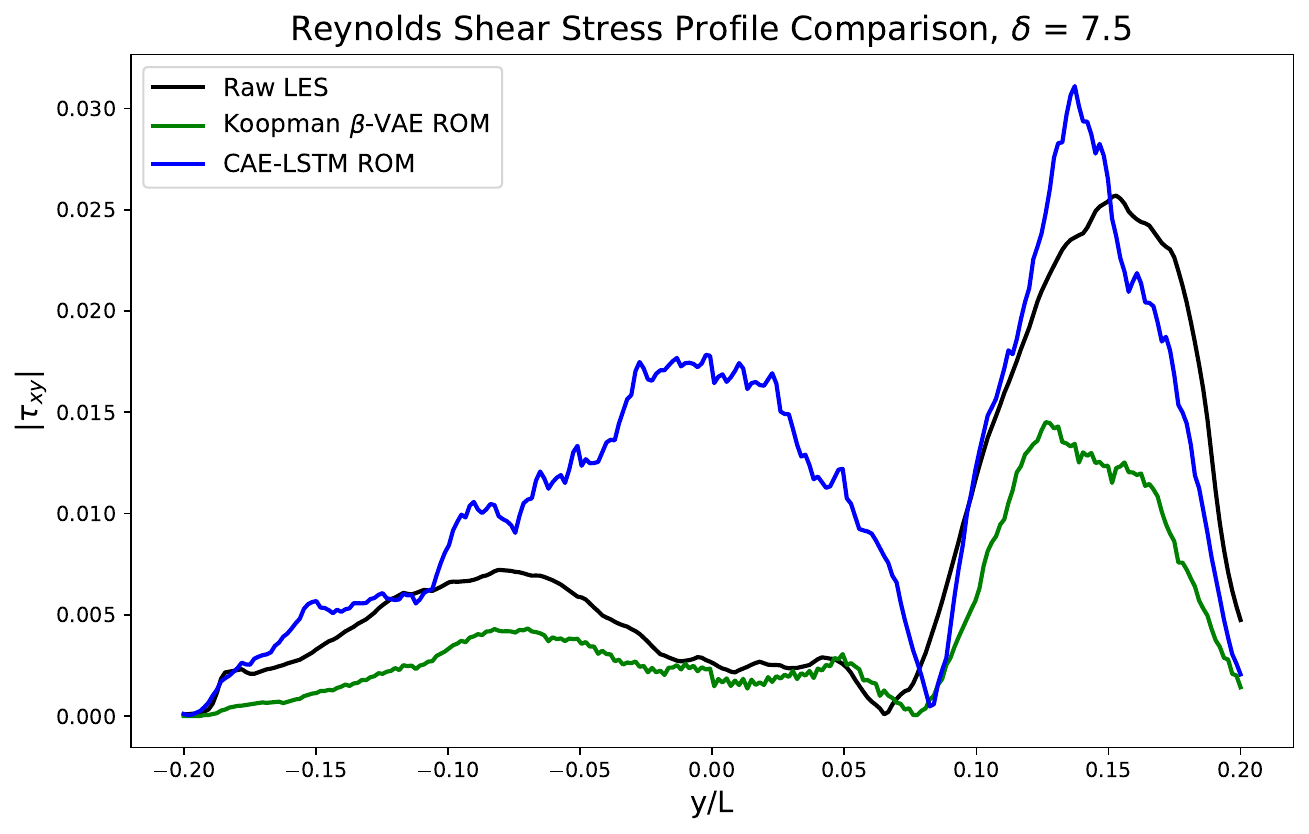}
    \caption{Reynolds shear stress comparison for the test data at $\delta = 7.5$ with CAE-LSTM.}
    \label{fig:tau_cae}
  \end{figure}

Figure~\ref{fig:cae_7.5} shows velocity magnitude contours of the raw LES data, ROM predictions, and absolute errors at $\delta = 7.5$. The predictions retain small-scale structures but fail to accurately predict the test data, with error magnitudes being higher than those from the Koopman $\beta$-VAE ROM. Figure~\ref{fig:tau_cae} shows a comparison of the Reynolds shear stress profiles between the raw test data and CAE-LSTM predictions at $\delta = 7.5$. $\tau_{xy}$ exhibits significantly higher values for CAE-LSTM along the y-axis, implying that noise is amplified in ROM predictions, resulting in increased turbulent mixing. The highly noisy nature of the latent variables make them difficult to predict over time, leading to unstable predictions that amplify the presence of fluctuations, also leading to increased prediction errors. The Koopman $\beta$-VAE ROM results in significantly lower turbulent mixing as a result of reduced fluctuations; this implies that the prediction errors are mostly a result of the absence of small-scale structures in predictions. Together, these results show that CAE-LSTM provides poor predictions of latent variables and velocity fields while amplifying fluctuations.

\end{appendices}

\end{document}